\newtheorem{subsub}{}
\newcounter{so}
\addtocounter{so}{1}

\documentstyle[12pt]{report}
\begin{document}
\parskip=4pt plus 1pt
\textheight=8.7in
\newcommand{\beq}{\begin{equation}}
\newcommand{\eeq}{\end{equation}}
\newcommand{\beqa}{\begin{eqnarray}}
\newcommand{\eeqa}{\end{eqnarray}}
\newcommand{\no}{\nonumber}
\newcommand{\grts}{\greaterthansquiggle}
\newcommand{\lets}{\lessthansquiggle}
\newcommand{\ul}{\underline}
\newcommand{\ol}{\overline}
\newcommand{\ra}{\rightarrow}
\newcommand{\Ra}{\Rightarrow}
\newcommand{\ve}{\varepsilon}
\newcommand{\vp}{\varphi}
\newcommand{\vt}{\vartheta}
\newcommand{\dg}{\dagger}
\newcommand{\wt}{\widetilde}
\newcommand{\wh}{\widehat}
\newcommand{\dfrac}{\displaystyle \frac}
\newcommand{\fsl}{\not\!}
\newcommand{\ben}{\begin{enumerate}}
\newcommand{\een}{\end{enumerate}}
\newcommand{\bfl}{\begin{flushleft}}
\newcommand{\efl}{\end{flushleft}}
\newcommand{\ba}{\begin{array}}
\newcommand{\ea}{\end{array}}
\newcommand{\btab}{\begin{tabular}}
\newcommand{\etab}{\end{tabular}}
\newcommand{\bit}{\begin{itemize}}
\newcommand{\eit}{\end{itemize}}

\newcommand{\be}{\begin{equation}}
\newcommand{\ee}{\end{equation}}
\newcommand{\bearr}{\begin{eqnarray}}
\newcommand{\eearr}{\end{eqnarray}}
\newcommand{\bea}{\begin{eqnarray}}
\newcommand{\eea}{\end{eqnarray}}

\newcommand{\per}{\;\;.}
\newcommand{\mtiny}[1]{{\mbox{\tiny #1}}}
\newcommand{\MS}{\mtiny{MS}}
\newcommand{\GeV}{\mbox{GeV}}
\newcommand{\MeV}{\mbox{MeV}}
\newcommand{\keV}{\mbox{keV}}
\newcommand{\ren}{\mtiny{ren}}
\newcommand{\kin}{\mtiny{kin}}
\newcommand{\hint}{\mtiny{int}}
\newcommand{\tot}{\mtiny{tot}}
\newcommand{\CHPT}{\mtiny{CHPT}}
\newcommand{\QED}{\mtiny{QED}}
\newcommand{\syst}{\mbox{syst.}}
\newcommand{\stat}{\mbox{stat.}}
\newcommand{\wave}{\mbox{wave}}
\newcommand{\co}{\; \; ,}
\newcommand{\nn}{\nonumber \\}
\newcommand{\fff}{\bar{f}}
\newcommand{\ffg}{\bar{g}}
\renewcommand{\theequation}{\arabic{equation}}
\renewcommand{\thetable}{\arabic{table}}

\newcommand{\fs}{\; \; .}

\newcommand{\bdm}{\begin{displaymath}}
\newcommand{\edm}{\end{displaymath}}

\newcommand{\nl}{\nonumber \\}

\thispagestyle{empty}

\renewcommand{\theequation}{\arabic{equation}}
\renewcommand{\thetable}{\arabic{table}}

\newcommand{\pl}{PL}
\newcommand{\spi}{s_\pi}
\newcommand{\ql}{QL}
\newcommand{\qn}{QN}
\newcommand{\pn}{PN}
\newcommand{\qq}{Q^2}
\newcommand{\mee}{m_l^2}
\newcommand{\thp}{\theta_\pi}

\chapter*{}

\begin{center}

{\Large \bf SEMILEPTONIC KAON DECAYS}
\refstepcounter{chapter}
\label{semi}
\thispagestyle{empty}

\vspace{2cm}
{\bf{J. Bijnens$\;^1$, G. Colangelo$\;^{2,3}$, G. Ecker$\;^{4}$ and  J.
Gasser$\;^2$}}

\end{center}

\vspace{8cm}
\noindent
{\underline{\hspace{5cm} }}\\

\noindent
${}^{1)}$ Nordita, Blegdamsvej 17, DK--2000 Copenhagen

\noindent
${}^{2)}$ Inst. Theor. Physik, Univ. Bern, Sidlerstrasse 5, CH--3012 Bern

\noindent
${}^{3)}$ Dipartimento di Fisica,
Universit\`{a} di Roma II - "Tor Vergata"

\hspace{-.34cm} Via della Ricerca Scientifica 1, I--00173 Roma

\noindent
${}^{4)}$ Inst. Theor. Physik, Univ. Wien, Boltzmanngasse 5, A--1090 Wien

\vspace{2cm}

\tableofcontents
\vspace{1cm}
\newpage
\begin{center}
 {\underline{Note:}}
\end{center}
\begin{itemize}
\item
The number of events quoted for DA$\Phi$NE are based on a
luminosity of $5\cdot 10^{32}~ cm^{-2}s^{-1}$,
which is equivalent \cite{franzini}
to an annual rate of $9\cdot 10^9$ $(1.1\cdot 10^9)$ tagged
$K^{\pm}$ $(K_L)$ (1 year = $10^7~ s$ assumed).
\item
Whenever we quote a branching ratio for a semileptonic
$K^0$ decay, it stands for the branching ratio of the corresponding
$K_L$ decay, e.g.,
\bearr
BR(K^0 \to \pi^- l^+ \nu ) \equiv BR(K_L \to \pi^{\pm} l^{\mp} \nu )~.
\nonumber
\eearr
\item
We use the data from the Particle Data Group edition 1990 \cite{pdg}
throughout. Please contact one of the authors in case that very high precision
is needed for a particular
matrix element. We would then  convert the relevant quantity to the
newest data compilation available.
 \item
If not stated explicitly, we always use for the low-energy constants
$L_1,\ldots,L_{10}$ the values displayed in table 1 in ref. \cite{Intro}.
 \item
 More
notation is provided in appendix \ref{notation}.
\end{itemize}

\newpage

\renewcommand{\thesection}{\arabic{section}}
\renewcommand{\thesubsection}{\arabic{section}.\arabic{subsection}}
\renewcommand{\theequation}{\arabic{section}.\arabic{equation}}
\renewcommand{\thefigure}{\arabic{section}.\arabic{figure}}
\renewcommand{\thetable}{\arabic{section}.\arabic{table}}

\setcounter{equation}{0}
\setcounter{table}{0}
\setcounter{figure}{0}
\setcounter{subsection}{0}
\setcounter{so}{1}

\section{Radiative $K_{l2}$ decays}

We consider the $K_{l2\gamma}$ decay
\be
K^+ (p) \rightarrow l^+ (p_l) \nu_l (p_\nu) \gamma (q) \hspace{1cm}
[K_{l2\gamma}] \label{k1}
\ee
where $l$ stands for $e$ or $\mu$, and $\gamma$ is a real photon with $q^2 =
0$. Processes where the (virtual) photon converts into a $e^+ e^-$ or
$\mu^+\mu^-$ pair are considered in the next subsection. The $K^-$ mode is
obtained from (\ref{k1}) by charge conjugation.

\subsection{Matrix elements and kinematics}

The matrix element for $K^+\rightarrow l^+ \nu_l \gamma$ has the
structure
\be
T = -iG_F eV_{us}^\star \epsilon^\star_\mu \left \{ F_K L^\mu - H^{\mu \nu}
l_\nu \right \}
\label{k3}
\ee
with
\bearr
L^\mu &=& m_l \bar{u}(p_\nu) (1 + \gamma_5) \left ( \frac{2p^\mu}{2pq}
- \frac{2 p^\mu_l + \not \!{q} \gamma^\mu}{2 p_l q} \right ) v (p_l)
\nonumber \\
l^\mu &=& \bar{u} (p_\nu)\gamma^\mu  (1 -\gamma_5) v (p_l)
\nonumber \\
H^{\mu \nu} &=& i V (W^2) \epsilon^{\mu \nu \alpha \beta} q_\alpha p_\beta -
A(W^2) (q W g^{\mu \nu} - W^\mu q^\nu)
\nonumber \\
W^\mu &=& (p-q)^\mu = (p_l + p_\nu)^\mu.
\label{k4}
\eearr

Here, $\epsilon_\mu$ denotes the polarization vector of the photon with
$q^\mu \epsilon_\mu= 0$, whereas $A$,
$V$ stand for two Lorentz
invariant amplitudes which occur in the general decomposition of the
tensors

\be
I^{\mu \nu} =
\int dx e^{iqx+iWy} < 0 \mid T V^\mu_{em} (x) I^\nu_{4-i5}(y) \mid K^+(p)>
\; \;, \; \; I=V,A \; \; .
\label{k5}
\ee
The form factor $A$  $(V)$ is related to the matrix element of
the axial (vector)
current in (\ref{k5}).
 In appendix \ref{kl2g} we display the general
decomposition of $A^{\mu \nu}$,
$V^{\mu\nu}$ for $q^2 \neq 0$ and provide also the link with the notation
used by the PDG \cite{pdg} and in \cite{ke22,km21}.

The  term proportional to $L^\mu$ in (\ref{k3}) does not contain
unknown quantities -- it
is determined by the amplitude of the nonradiative decay $K^+ \rightarrow l^+
\nu_l$. This part of the amplitude is usually referred to as ``in\-ner
Brems\-strahlung (IB) contribution", whereas the term proportional to $H^{\mu
\nu}$ is called ``structure dependent (SD) part" .

The form factors are analytic functions in the complex $W^2$-plane cut
along the positive real axis. The cut starts at $W^2 = (M_K + 2 M_\pi)^2$ for
$A$ (at $W^2 = (M_K + M_\pi)^2$ for $V$). In our phase convention, $A$ and
$V$ are real in the physical region of $K_{l 2 \gamma}$ decays,
\be
m_l^2 \leq W^2 \leq M^2_{K}.
\label{k6}
\ee

The kinematics of (spin averaged) $K_{l 2 \gamma}$ decays needs two
variables, for which we choose the conventional quantities
\be
x = 2 p q/M^2_{K} \hspace{0.2cm} , \hspace{0.2cm} y = 2 p  p_l/
M_{K}^2 \; \; .
\label{k15}
\ee
In the $K$ rest frame, the variable $x$ ($y$) is proportional to the photon
(charged lepton) energy,
\be
x=2 E_\gamma /M_K \; \; , \; \; y=2 E_l/M_K \; \; ,
\label{k15a}
\ee
and the angle $\theta_{l\gamma}$ between the photon and the charged lepton is
related to $x$ and $y$ by
\be
x=\frac{ (1-y/2+A/2)(1-y/2-A/2)}{1-y/2+A/2 {\mbox{cos}} \theta_{l \gamma}} \;
\; ; A=\sqrt{y^2-4 r_l} \; \; .
\label{k15b}
\ee
In terms of these quantities, one has
\be
W^2 = M^2_{K} (1-x) \hspace{0.2cm} ; \hspace{0.2cm} (q^2 = 0) \hspace{0.2cm} .
\label{k16}
\ee

We write the physical  region for $x$ and $y$ as
$$
2 \sqrt{r_l} \leq y \leq 1 + r_l
$$
\be
1 - \frac{1}{2} (y + A) \leq x \leq 1 - \frac{1}{2} (y - A)
\label{k17}
\ee
or, equivalently, as
\bearr
0 \leq & x &\leq 1-r_l \nonumber \\
1-x +\frac{r_l}{(1-x)} \leq & y & \leq 1+r_l
\label{k17a}
\eearr
where
\be
r_l = m^2_l/ M^2_{K} = \left \{ \begin{array}{ll}
1.1 \cdot 10^{-6} (l = e) \\
4.6 \cdot 10^{-2} (l = \mu) \; \; .
\end{array} \right .
\label{k18}
\ee

  \subsection{Decay rates}

The partial decay rate is
\be
d\Gamma = \frac{1}{2M_K (2\pi)^5} \sum_{spins} |T|^2 d_{LIPS}(p;p_l,p_\nu,q).
\ee
The Dalitz plot density
\be
\rho(x,y) = \frac{d^2\Gamma}{dx dy} = \frac{M_K}{256\pi^3} \sum_{spins} |T|^2
\ee
is a Lorentz invariant function which contains $V$ and $A$ in the following
form \cite{brym},
\bearr
\rho(x,y)&=&
\rho_{\mbox{\tiny{IB}}}(x,y)  +  \rho_{\mbox{\tiny{SD}}}(x,y)
+  \rho_{\mbox{\tiny{INT}}}(x,y)
\nonumber \\
 \rho_{\mbox{\tiny{IB}}}(x,y)& =& A_{\mbox{\tiny{IB}}}
f_{\mbox{\tiny{IB}}}(x,y)
\nonumber \\
 \rho_{\mbox{\tiny{SD}}}(x,y)& =& A_{\mbox{\tiny{SD}}}
      M_K^2   \left[ (V+A)^2
f_{{\mbox{\tiny{SD}}}^+}(x,y) + (V-A)^2 f_{{\mbox{\tiny{SD}}}^-} (x,y) \right]
\nonumber \\
 \rho_{\mbox{\tiny{INT}}}(x,y)& =&
 A_{\mbox{\tiny{INT}}} M_K
\left [ (V+A) f_{{\mbox{\tiny{INT}}}^+} (x,y) + (V-A)
f_{{\mbox{\tiny{INT}}}^-} (x,y) \right]
\label{k19}
\eearr
where
\bearr
f_{\mbox{\tiny{IB}}}(x,y)& =&\left[ \frac{1-y+r_l}{x^2(x+y-1-r_l)}\right]
\left[x^2 +2(1-x)(1-r_l) -\frac{2x r_l (1-r_l)}{x+y-1-r_l} \right]
\nonumber \\
 f_{{\mbox{\tiny{SD}}^+}}(x,y)& =& \left[ x+y-1-r_l\right]
\left[ (x+y-1)(1-x)-r_l \right]
\nonumber \\
 f_{{\mbox{\tiny{SD}}^-}}(x,y)& =& \left[1-y+r_l \right]
 \left[ (1-x) (1-y) +r_l\right]
\nonumber \\
 f_{{\mbox{\tiny{INT}}^+}}(x,y)& =& \left[ \frac{1-y+r_l}{x(x+y-1-r_l)}\right]
\left[ (1-x)(1-x-y)+r_l \right]
\nonumber \\
 f_{{\mbox{\tiny{INT}}^-}}(x,y)& =& \left[ \frac{1-y+r_l}{x(x+y-1-r_l)}\right]
\left[ x^2 -(1-x)(1-x-y)-r_l\right]
\label{k20}
\eearr

and
\bearr
A_{\mbox{\tiny{IB}}}& =& {4r_l} \left ( \frac{F_K}{M_K} \right)^2
A_{\mbox{\tiny{SD}}}
\nonumber \\
A_{\mbox{\tiny{SD}}}& =& \frac{G_F^2 |V_{us}|^2 \alpha}{32 \pi^2} M_K^5
\nonumber \\
A_{\mbox{\tiny{INT}}}& =&{4r_l} \left ( \frac{F_K}{M_K} \right)
A_{\mbox{\tiny{SD}}} \; \; .
\label{k20a}
\eearr
For later convenience, we note that
\be
A_{\mbox{\tiny{SD}}}
                    = \frac{\alpha}{8\pi}
\frac{1}{r_l(1-r_l)^2} {\left( \frac{M_K}{F_K} \right) }^2
\Gamma (K\rightarrow l\nu_l) \; .
\label{k20b}
\ee
The indices IB, SD and INT stand respectively for the contribution from
in\-ner Brems\-strahlung, from the structure dependent part and from the
interference term between the IB and the SD part in the amplitude.

To get a feeling for the magnitude of the various contributions
IB,$\mbox{SD}^\pm$ and ${\mbox{INT}}^\pm$ to the decay rate, we consider
the integrated  rates
\be
\Gamma_I = \int_{R_I} dxdy  \rho_I (x,y) \; \; ; \; \; I =
\mbox{SD}^\pm, \mbox{INT}^\pm ,\mbox{IB} \; \; ,
\label{dr1}
\ee
where $\rho_{SD}=\rho_{SD^+} + \rho_{SD^-}$ etc.
 For the region $R_I$ we take the full phase space
 for $I \neq \mbox{IB}$, and
\be
R_{{\mbox{{\tiny{IB}}}}} =214.5 {\mbox{MeV/c}} \leq p_l \leq 231.5
{\mbox{MeV/c}}\; . \label{dr2}
\ee
for the Bremsstrahlung contribution. Here $p_l$ stands for the modulus of the
lepton three momentum in the kaon rest system \footnote{
This cut has been used in \cite{km21} for $K_{\mu 2\gamma}$, because this
kinematical region is free from $K_{\mu3}$ background.
We apply it here for illustration also to the electron mode $K_{e2\gamma}$.
                                 }.
  We consider constant
form factors
$V$, $A$ and write for the rates and for the corresponding branching ratios
\bearr
\Gamma_I & =& A_{\mbox{\tiny{SD}}} \left \{ M_K (V \pm A) \right \}^{N_I}X_I \;
\;
\nonumber \\
{\mbox{BR}}_I &\doteq& \Gamma_I / \Gamma_{\mbox{\tiny{tot}}} = N \left \{
M_K (V \pm A) \right \}^{N_I} X_I
\label{dr3}
\eearr
with
\be
N = A_{\mbox{\tiny{SD}}}/{\Gamma_{\mbox{\tiny{tot}}}}
 = 8.348 \cdot 10^{-2}.
\label{dr3a}
\ee
The values for $N_I$ and $X_I$ are listed in table \ref{t:pskl2}.

\begin{table}[t]
\protect
\begin{center}
\caption{
 The quantities
 $X_I,N_I$.
 SD$^\pm$ and INT$^\pm$ are evaluated with full phase space,
 IB with
restricted kinematics (\protect\ref{dr2}).
\label{t:pskl2}         }
\vspace{1em}
\begin{tabular}{|c||c|c|c|c|c||c|} \hline
\multicolumn{1}{|c||}{}&{${\mbox{SD}}^+$}&{${\mbox{SD}}^-$}&
{${\mbox{INT}}^+$}&{INT$^-$}
&{${\mbox{IB}}$}& \multicolumn{1}{|c|}{}\\ \hline
{$X_I$} & {$1.67\cdot 10^{-2}$}&{$1.67\cdot 10^{-2}$}&
{$-8.22 \cdot 10^{-8}$}&{$3.67\cdot 10^{-6}$}&{$3.58\cdot 10^{-6}$}&
{$K_{e2\gamma}$} \\ \hline
{$X_I$} & {$1.18\cdot 10^{-2}$}&{$1.18\cdot 10^{-2}$}&
{$-1.78\cdot 10^{-3}$}&{$1.23\cdot 10^{-2}$}&{$3.68\cdot 10^{-2}$}&
{$K_{\mu 2 \gamma}$} \\ \hline
{$N_I$} &2&2&1&1&0&\multicolumn{1}{|c|}{} \\ \hline
  \end{tabular}
\end{center}
\end{table}

To estimate $\Gamma_I$ and ${\mbox{BR}}_I$, we note that the form factors $V,A$
 are of
 order
\be
M_K(V+A) \simeq -10^{-1} \; \; , \; \; M_K(V-A) \simeq -4 \cdot 10^{-2} \; \; .
\ee
{}From this and from the entries in the table
one concludes that for the above regions $R_I$, the interference terms
$\mbox{INT}^\pm$ are negligible in
$K_{e2\gamma}$, whereas they are important in $K_{\mu 2 \gamma}$. Furthermore,
IB is negligible for $K_{e2\gamma}$, because it is helicity suppressed as can
be seen from the factor $m_l^2$ in $A_{\mbox{\tiny{IB}}}$. This term
dominates however in $K_{\mu 2 \gamma}$.

\subsection{Determination of $A(W^2)$ and $V(W^2)$}

The decay rate  contains two real functions
\be
F^\pm(W^2) = V(W^2) \pm A(W^2)
\label{k21}
\ee
as the only unknowns. In Figs. (\ref{psp1},\ref{psp2}) we display
contour plots for the density distributions
$f_{\mbox{\tiny{IB}}}, \ldots,$ $f_{{\mbox{\tiny{INT}}^\pm}}$ for $l = \mu,
e$. These
five terms have obviously very different Dalitz plots. Therefore, in principle,
one can determine the strength of each term by choosing a suitable kinematical
region of observation. To pin down $F^\pm$, it would   be sufficient to
measure
at each photon energy the interference term INT$^\pm$. This has not yet
\begin{figure}[p]
\vspace{7.5in}
\caption{Contour plots for
 $f_{{\mbox{\protect\tiny{IB}}}}$,\ldots,
 $f_{{\mbox{\protect\tiny{INT}}}^\pm}$
 [$K_{\mu 2 \gamma}$]. The numbering on the
lines points towards increasing modulus. The normalization is arbitrary.
\label{psp1} }
\end{figure}
\begin{figure}[p]
\vspace{7.5in}
\caption{Contour plots for
 $f_{{\mbox{\protect\tiny{IB}}}}$,\ldots,
 $f_{{\mbox{\protect\tiny{INT}}}^\pm}$
 [$K_{e 2 \gamma}$]. The numbering on the
lines points towards increasing modulus. The normalization is arbitrary.
 \label{psp2}}
\end{figure}
been achieved so far, either because the contribution of INT$^\pm$
is too small (in $K_{e 2 \gamma}$)
, or because too few events have been collected  (in
$K_{\mu 2 \gamma}$). On the other hand, from a
measurement
of {\mbox{SD}}$^\pm$ alone one can determine $A, V$ only up to a fourfold
 ambiguity:
\be
\mbox{SD}^\pm \rightarrow \left \{ (V,A); - (V,A); (A,V); - (A,V) \right \}.
\label{k22}
\ee
In terms of the ratio
\be
\gamma_K = A/V
\label{k23}
\ee
this ambiguity  amounts to
\be
\mbox{SD}^\pm \rightarrow \left \{ \gamma_K ; 1/ \gamma_K \right \}.
\label{k24}
\ee
Therefore, in order to pin down the amplitudes $A$ and $V$ uniquely, one must
measure the interference terms INT$^\pm$ as well.

\subsection{Previous experiments}

 \vspace{.5cm}

{\underline{{\bf{$K^+ \rightarrow e^+ \nu_e  \gamma$}}}}

 \vspace{.5cm}

The PDG uses data from two experiments \cite{ke22,ke21}, both of which have
been
sensitive mainly to the {\mbox{SD}}$^+$ term in (\ref{k19}). In \cite{ke21}, 56
events
with $E_\gamma > 100$ MeV, $E_{e^+} > 236$ MeV and $\theta_{e^+\gamma}> 120^0$
have been identified, whereas the later experiment \cite{ke22} has
collected 51 events with $E_\gamma > 48$ MeV, $E_{e^+} > 235$ MeV and
$\theta_{e^+ \gamma} > 140^0$. In these kinematical regions, background from
$K^+ \rightarrow e^+ \nu_e \pi^0$ is absent because $E_{e}^{\mbox{\tiny{max}}}
(K_{e3}) = 228$ MeV. The combined result of both experiments is
\footnote{In all four experiments \cite{ke21,ke22,km21,km22} discussed here
and below , the form factors $A$ and $V$ have been treated as constants.}
\cite{ke22}
\be
\Gamma(\mbox{SD}^+) / \Gamma(K_{\mu2}) = (2.4 \pm 0.36) \cdot 10^{-5}.
\label{k25}
\ee
For {\mbox{SD}}$^-$, the bound
\be
\Gamma(\mbox{SD}^-) / \Gamma_{\mbox{\tiny{total}}} < 1.6 \cdot 10^{-4}
\label{k26}
\ee
has been obtained from a sample of electrons with energies 220 MeV $\leq
E_{e} \leq 230$ MeV \cite{ke22}. Using (\ref{dr3},\ref{dr3a}), the
result (\ref{k25}) leads to
\be
 M_K \mid V+A \mid = 0.105 \pm 0.008 \; \; .
\label{k27}
\ee

The bound (\ref{k26}) on the other hand implies \cite{ke22}
\be
\mid V -A \mid / \mid V+A \mid < \sqrt{11},
\label{k28}
\ee
from where one concludes \cite{ke22} that $\gamma_K$ is outside the range
$-1.86$ to $-0.54$,
\be
\gamma_K  \not \in [-1.86, - 0.54] \; \; .
\label{k29}
\ee
As we already mentioned, the interference terms INT$^\pm$ in $K \rightarrow
e \nu_e \gamma$ are small and can hardly ever be measured.
As a result of this,
the amplitudes $A,V$ and the ratio $\gamma_K$ determined from $K_{e2\gamma}$
are
subject to the ambiguities (\ref{k22}), (\ref{k24}).

\vspace{.5cm}

{\underline{{\bf{$K^+ \rightarrow \mu^+ \nu_\mu \gamma$}}}}

\vspace{.5cm}

Here, the interference terms INT$^\pm$ are nonnegligible in appropriate regions
of phase space (see Figs. (\ref{psp1},\ref{psp2})). Therefore, this
decay allows one in principle to
pin down $V$ and $A$. The PDG uses data from two experiments
\cite{km21,km22}. In \cite{km21},
the momentum spectrum of the muon was measured in the region (\ref{dr2}).
 In total $2 \pm 3.44$ {\mbox{SD}}$^+$ events
have been found with 216 MeV/c  $< p_\mu< $  230 MeV/c and $E_\gamma > 100$
 MeV, which leads to
\be
M_K \mid V+ A \mid < 0.16\; \;.
\label{k30}
\ee
In order to identify
the effect of the {\mbox{SD}}$^-$ terms, the region 120 MeV/c
$<p_\mu < $150 MeV/c was searched. Here, the background from $K_{\mu3}$ decays
was very serious. The authors found 142 $K_{\mu \nu \gamma}$ candidates and
conclude that
\be
- 1.77 < M_K (V-A) < 0.21.
\label{k31}
\ee

The result (\ref{k30}) is consistent with
(\ref{k27}), and the bound  (\ref{k31}) is worse than the
result (\ref{k28}) obtained from   $K_{e 2 \gamma}$. The
branching ratios which follow  \cite{km21} from
(\ref{k30},\ref{k31}) are displayed in table
\ref{t:erkl2}, where we also show the $K_{e2\gamma}$ results \cite{ke21,ke22}.
The entry SD$^-$+INT$^-$
for $K_{\mu 2 \gamma}$ is based on additional
 constraints from $K_{e2\gamma}$ \cite{km21}.

\begin{table}[t]
\begin{center}
\protect
\caption{ Measured
 branching ratios $\Gamma (K\rightarrow l \nu_l \gamma)
/\Gamma_{\mbox{{\protect\tiny{total}}}}$. The $K_{e2\gamma}$ data are from
\protect\cite{ke21,ke22}, the $K_{\mu 2\gamma}$ data from
\protect\cite{km21,km22}. The last column corresponds \protect\cite{km21}
 to the cut
(\protect\ref{dr2}).
\label{t:erkl2}         }
\vspace{1em}
{\footnotesize{
\begin{tabular}{|c|c|c|c|c|c|} \hline
&{${\mbox{SD}}^+$}&{${\mbox{SD}}^-$}&{${\mbox{INT}}^+$}&
{${\mbox{SD}}^- + {\mbox{INT}}^-$}&{total}
\\ \hline
{$K_{e2\gamma}$} & {$(1.52 \pm 0.23)\cdot 10^{-5}$}&{$<1.6\cdot 10^{-4}$}&
{}&{}&{ }
\\ \hline
{$K_{\mu 2 \gamma}$} & {$< 3\cdot 10^{-5}$}&{}&
{$<2.7\cdot 10^{-5}$}& {$<2.6\cdot 10^{-4}$}&
{$(3.02\pm 0.10)\cdot 10^{-3}$}
\\
\multicolumn{1}{|c|}{}&{}&{}&{(modulus)}&{(modulus)}&{}
\\ \hline
\end{tabular}
}}
\end{center}
\end{table}

\subsection{Theory}

The amplitudes $A(W^2)$ and $V(W^2)$ have been worked out in the framework of
various approaches, viz., current algebra, PCAC, resonance exchange,
dispersion relations, \ldots . For a rather detailed review together with an
extensive list of references up to 1976 see \cite{BARDIN}. Here, we concentrate
on the predictions of $V, A$ in the framework of CHPT.

\vspace{.5cm}
\begin{subsub}{\bf Chiral expansion to one loop}\end{subsub}

The amplitudes $A$ and $V$ have been evaluated \cite{donog,beg} in the
framework
of CHPT to one loop. At leading order in the low-energy expansion, one has
\be
A=V=0 .
\label{k36}
\ee
As a consequence of this, the rate is
entirely
given by the IB contribution
at leading order. At the one-loop level, one finds
\bearr
A &=& -\frac{4}{F} (L_9^{r} + L_{10}^{r})
\nonumber \\
V &=& - \frac{1}{8 \pi^2} \frac{1}{F}
\nonumber \\
\gamma_K &=& 32 \pi^2 (L_9^r + L_{10}^r)\; ,
\label{k37}
\eearr
where $L_9^r$ and $L_{10}^r$ are the renormalized low-energy couplings
evaluated at the scale $\mu$ (the combination $L_9^r + L_{10}^r$ is scale
independent). The vector form factor stems from the Wess-Zumino term
\cite{wessz} which enters the low-energy expansion at order $p^4$,
 see Ref. \cite{Intro}.

{\underline {Remarks}:}

\begin{description}

\item(i)
 At this order in the low-energy expansion, the form factors $A,V$ do
not
exhibit any $W^2$-dependence. A nontrivial $W^2$-dependence only occurs at the
next order in the energy expansion (two-loop effect, see  the discussion
below). Note that the available analyses of experimental data of $K \rightarrow
l \nu_l \gamma$ decays \cite{ke21,ke22,km21,km22} use constant form factors
throughout.

\item (ii) Once the combination $L_9 + L_{10}$ has been pinned down
from other processes, Eq. (\ref{k37}) allows one to
evaluate
$A,V$ unambiguously at this order in the low-energy expansion. Using $L_9 +
L_{10} = 1.4 \cdot 10^{-3}$  and $F=F_\pi$, one has
\bearr
 M_K (A+V) &=& -0.097
\nonumber \\
M_K(V-A) &=& -0.037
\nonumber \\
\gamma_K &=& 0.45 \; \; .
\label{k38}
\eearr
The result for the combination $(A+V)$ agrees with (\ref{k27}) within the
errors,
while $\gamma_K$ is consistent with (\ref{k29}).

\end{description}

We display in table \ref{t:chpr} the branching ratios ${\mbox{BR}}_I$
 (\ref{dr3})
 which follow from
the prediction (\ref{k38}). These predictions satisfy of course the
inequalities found from experimental data (see table \ref{t:erkl2}).

\begin{table}[t]
\protect
\begin{center}
\caption{Chiral
prediction at order $p^4$ for the branching ratios
$\Gamma(K\rightarrow l \nu_l \gamma)/\Gamma_{\mbox{\protect\tiny{total}}}$. The
cut used in the last column is  given in Eq. (\protect\ref{dr2}).
\label{t:chpr}
}
\vspace{1em}
\begin{tabular}{|c|c|c|c|c|c|} \hline
{}&{{\mbox{SD}}$^+$}&{{\mbox{SD}}$^-$}&{{\mbox{INT}}$^+$}&{{\mbox{INT}}$^-$}&{
total} \\ \hline
{$K_{e2\gamma}$} & {$1.30\cdot 10^{-5}$}&{$1.95\cdot 10^{-6}$}&
{$6.64\cdot 10^{-10}$}&{$-1.15\cdot 10^{-8}$}&{$2.34 \cdot 10^{-6}$}
\\ \hline
{$K_{\mu 2\gamma}$} & {$9.24\cdot 10^{-6}$}&{$1.38\cdot 10^{-6}$}&
{$1.44\cdot 10^{-5}$}&{$-3.83\cdot 10^{-5}$}&{$3.08 \cdot 10^{-3}$}
\\ \hline
\end{tabular}
\end{center}
\end{table}

\newpage

\vspace{.5cm}

\begin{subsub}{\bf
$W^2$-dependence of the form factors}
\end{subsub}

The chiral prediction gives constant form factors at order $p^4$. Terms of
order $p^6$ have not yet been calculated. They would, however, generate a
nontrivial $W^2$ - dependence both in $V$ and $A$. In order to estimate the
magnitude of these corrections, we consider one class of $p^6$ - contributions:
terms which are generated
by vector and axial-vector resonance exchange with
strangeness \cite{BARDIN,egpr},
\be
V(W^2) = \frac{V}{1-W^2/{M_{K^\star}}^2}\; \; ,\; \; A(W^2) =
\frac{A}{1-W^2/{M_{K_1}}^2} \label{k40}
\ee
where $V,A$ are given in (\ref{k37}). We now examine the effect of the
denominators in (\ref{k40}) in the region $y \geq 0.95, x \geq 0.2$
which has been explored in $K^+ \rightarrow e^+ \nu_e \gamma$ \cite{ke22}.
We put $m_e = 0$ and evaluate the rate
\be
\frac{dP(x)}{dx} = \frac{N_{\mbox{\tiny{tot}}}}{\Gamma_{\mbox{\tiny{tot}}}}
\int^1_{y=0.95}  \rho_{{\mbox{\tiny{SD}}}^+}(x,y) dy
\label{k41}
\ee
where $N_{\mbox{\tiny{tot}}}$ denotes the total number of $K^+$ decays
considered, and $\Gamma^{-1}_{\mbox{\tiny{tot}}} = 1.24 \cdot 10^{-8}$
 sec.

\begin{figure}[t]
\vspace{9cm}
\caption{The rate $dP(x)/dx$ in (\protect\ref{k41}), evaluated with the form
factors
(\protect\ref{k40}) and
$N_{\mbox{\protect\tiny{tot}}} = 9
\cdot 10^9$.
The solid line corresponds to $M_{K^\star}
= 890$ MeV, $M_{K_1} = 1.3$ GeV. The dashed  line is evaluated with
$M_{K^\star} = 890$ MeV, $ M_{K_1} = \infty$ and the dotted line corresponds to
$M_{K^\star} = M_{K_1} = \infty$.
 The total number of events is also indicated in each case.
 \label{figform}
        }
\end{figure}

The function $\frac{dP(x)}{dx}$ is displayed in Fig. (\ref{figform}) for
three different values of $M_{K^\star}$ and $M_{K_1}$, with
 $N_{\mbox{\protect\tiny{tot}}} = 9
\cdot 10^9$.
 The total number of events
\be
N_P = \int^{1}_{x= 0.2} dP(x)
\label{k42}
\ee
is also indicated in each case. The difference between the dashed and the
dotted line shows  that
the nearby singularity in the anomaly form factor influences the decay rate
substantially at low photon energies.
The effect disappears at $x \rightarrow 1$, where $W^2 = M_K^2 (1-x)
\rightarrow 0$. To minimize the effect of resonance exchange, the
 large-$x$ region should thus be
considered. The low-$x$ region, on the other hand, may be used to explore the
$W^2$-dependence of $V$ and of $A$. For a rather exhaustive discussion
of the relevance of this $W^2$ - dependence
for the analysis of $K_{l2\gamma}$ decays we refer the reader to Ref.
\cite{BARDIN}.

\subsection{Comment on tensor couplings}
\label{tensor1}
Bolotov et al. \cite{bolotov} have analyzed radiative pion decays
$\pi^-\rightarrow e^-\bar{\nu}_e\gamma$ in flight ($\simeq 80$ events) in a
wider kinematical region than was explored in the
high-statistics experiment
of Bay et al. \cite{bay} (where $\simeq 700\;\; \pi^+\rightarrow
e^+\nu_e\gamma$  events had been observed).
The theoretical branching ratio, calculated with the standard $V-A$ coupling,
differs from the measured one by more than three standard deviations.
This discrepancy may be avoided by adding to the standard matrix element the
amplitude of a tensorial interaction \cite{poblaguev}.
 Belyaev and Kogan \cite{belyaev} and Voloshin \cite{voloshin} have pointed
out, however, that in the standard model the induced
tensor coupling is too small to generate the rate observed in
ref. \cite{bolotov}.

Gabrielli \cite{gabrielli} has  worked out the effect of tensor
couplings for $K^+\rightarrow l^+\nu_l\gamma$ decays. Using the above
quoted
values for the form factors $A$ and $V$ and a tensor coupling of a size
suggested to explain the data in Ref.
\cite{bolotov}, he finds a $\leq 30\%$ effect
in the partial decay rates (the exact size depends on the chosen coupling,
channel, decay region,...). The author then suggests that these effects may be
accessible to detection at high precision experiments carried out at
DA$\Phi$NE.

We wish to point out that this may be difficult for the
following reason. The calculation of the decays $K^+\rightarrow l^+\nu_l\gamma$
presented in this section is based on the one-loop formulae for the decay
matrix elements.
Higher-order effects may well be sizeable, see e.g.
figure \ref{figform}.
There, it is explicitly seen that the effect of resonance exchange is
$\simeq30\%$ in particular regions of phase space. Therefore, in order to
identify effects due to  tensor couplings, one first has to pin down
the contribution from higher-order effects in CHPT. This is not
an easy task to achieve to the accuracy required. On the other hand, it is of
course needless to say that the finding of a tensorial coupling of the size
suggested in Ref. \cite{poblaguev} would be spectacular.

\subsection{Improvements at DA$\Phi$NE}

Previous experiments have used various cuts in phase space in order (i)
to identify the individual contributions IB, {\mbox{SD}}$^\pm$, INT$^\pm$ as
far
 as
possible, and (ii) to reduce the background from $K_{l3}$ decays. This
background has in fact forced so severe cuts that only the upper end of the
lepton spectrum remained.

The experimental possibilities to reduce background from $K_{l3}$ decays are
presumably more favourable with today's techniques. Furthermore, the
annual yield of $9\cdot 10^9 K^+$ decays at DA$\Phi$NE is more than two orders
of
magnitude higher than the samples which were available in
 \cite{ke22,km21,ke21,km22}.
This allows for a big improvement in the determination
of the amplitudes $A$ and $V$, in particular in $K_{\mu 2 \gamma}$ decays.
 It would  be very interesting to pin
down the
combination $L_9 + L_{10}$ of the low-energy constants which occur
in the chiral representation of the amplitude $A$  and to
investigate the $W^2$-dependence of the form factors.

\newpage

\newpage
\setcounter{equation}{0}
\setcounter{subsection}{0}
\setcounter{so}{1}
\setcounter{table}{0}
\setcounter{figure}{0}
\clearpage

\section{The decays $K^{\pm} \rightarrow l^{\pm} \nu l'^{+}l'^{-}$}

Here we consider decays where the photon turns into a
lepton-anti-lepton pair,
\begin{eqnarray}
K^+ &\to&e^+ \nu \mu^+\mu^- \label{S21}\\
K^+ &\to&\mu^+ \nu e^+ e^- \label{S22}\\
K^+ &\to&e^+ \nu e^+ e^- \label{S23}\\
K^+ &\to&\mu^+ \nu \mu^+ \mu^- \ .\label{S24}
\end{eqnarray}

\subsection{Matrix elements}

We start with the processes (\ref{S21}) and (\ref{S22}),
\begin{eqnarray}
K^+(p) &\to&l^+(p_l) \nu(p_\nu) l'^{+}(p_1) l'^{-}(p_2)\nonumber\\
(l,l')&=& \ (e,\mu)\mbox{ or } (\mu,e) .
\end{eqnarray}
The matrix element is
\begin{equation}
\label{S27}
T = -i G_F e V_{us}^* \overline{\epsilon}_\rho   \left\{
F_K \overline{L}^\rho - \overline{H}^{\rho\mu}l_{\mu} \right\}
\end{equation}
where
\begin{eqnarray}
\overline{L}^\mu & = & m_l \overline{u}(p_\nu) (1+\gamma_5)
\left\{
\frac{2 p^\mu - q^\mu}{2 p q - q^2} -\frac{2p^\mu_l +\not\!q\gamma^\mu}
{2 p_l q + q^2}\right\}
v(p_l)
\nonumber\\
l^\mu &=& \overline{u}(p_\nu)\gamma^\mu (1-\gamma_5) v(p_l)
\nonumber\\
\overline{H}^{\rho\mu} & = &i V_1 \epsilon^{\rho\mu\alpha\beta}
 q_\alpha p_\beta - A_1 ( qW g^{\rho\mu} - W^\rho q^\mu )
\nonumber\\&&
-A_2 (q^2 g^{\rho\mu} - q^\rho q^\mu )
-A_4 (qW q^\rho - q^2 W^\rho) W^\mu
\end{eqnarray}
with
\begin{equation}
A_4 = \frac{2 F_K}{M_K^2 - W^2} \frac{F_V^{K}   (q^2) - 1 }{q^2} + A_3
\ .
\end{equation}
The form factors $A_i(q^2,W^2),\ V_1(q^2,W^2)$ are the ones defined
in appendix \ref{kl2g}.
                $F_V^{K}  (q^2)$ is the electromagnetic form factor
of the $K^+$. Finally the quantity $\overline{\epsilon}^{ \mu}$ stands
for
\begin{equation}
\overline{\epsilon}^{\mu} =   \frac{e}{q^2} \overline{u}(p_2)
\gamma^\mu v(p_1)\ ,
\end{equation}
and the four-momenta are
\begin{equation}
q = p_1 + p_2 ,\ W = p_l + p_\nu = p - q
\end{equation}
such that $q_\mu \overline{\epsilon}^{\mu} = 0$.

In order to obtain the matrix element for (\ref{S23}) and (\ref{S24}),
\begin{equation}
K^+ (p) \to l^+(p_l) \nu(p_\nu) l^+(p_1) l^-(p_2)\ ,
\end{equation}
one identifies $m_l$ and $m_l'$ in (\ref{S27}) and subtracts the
contribution obtained from interchanging $p_1 \leftrightarrow p_l$ :
\begin{eqnarray}
(p_1 , p_l) &\to& (p_l,p_1) \nonumber\\
q &\to& p_l + p_2\nonumber\\
W &\to& p- q=p_\nu + p_1\ .
\end{eqnarray}

\subsection{Decay distributions}

The decay width is given by
\begin{equation}
d\Gamma  =  \frac{1}{2M_K (2\pi)^8 }
\sum_{spins} |T^2| d_{LIPS}(p;p_l,p_\nu,p_1,p_2)
\end{equation}
and the total rate is the integral over this for the case $l\ne l'$.
For the case $l = l'$ the integral has to be divided by the factor 2
for two identical particles in the final state.

We first consider the case where
$l\ne l'^{}$ and introduce the dimensionless variables
\begin{eqnarray}
x = \frac{2 pq}{M_K^2},\;
y = \frac{2p_l p}{M_K^2},\;
z = \frac{q^2}{M_K^2},\;
r_l = \frac{m_l^2}{M_K^2},\;
r_l' =\frac{m_{l'}^2}{M_K^2}\fs
\end{eqnarray}
Then one obtains, after integrating over $p_1$ and $p_2$ at fixed $q^2$
\cite{KRISHNA},
\begin{eqnarray}
\label{S16}
d\Gamma_{K^+ \to l^+ \nu l'^{+}l'^{-}} &=&
\alpha^2 G_F^2 |V_{us}|^2 M_K^5 F(z,r'_l) \left\{
   -\sum_{spins} \overline{T}_\mu^* \overline{T}^\mu \right\}
    dx dy dz
\nonumber\\
F(z,r'_l)&=& \frac{1}{192 \pi^3 z} \left\{
     1 + \frac{2r'_{l}}{z}\right\}
  \sqrt{1 - \frac{4r'_{l}}{z}}
\nonumber\\
{\overline{T}}^\mu &=& M_K^{-2} \left\{
F_K \overline{L}^\mu -
\overline{H}^{\mu\nu}l_\nu\right\} .
\end{eqnarray}
This result allows one to evaluate, e.g.,
the distribution $d\Gamma/dz$ of produced $l'^+ l'^-$ pairs rather
easily. The kinematically allowed region is
\begin{eqnarray}
4 r'_l \le&z&\le 1 + r_l - 2 \sqrt{r_l}
\nonumber\\
2 \sqrt{z} \le &x& \le 1 + z - r_l
\nonumber\\
A - B \le & y & \le A+B
\end{eqnarray}
with
\begin{eqnarray}
A&=&\frac{(2-x)(1+z+r_l -x)}{2 ( 1+z-x)}\nonumber\\
B&=&\frac{(1+z-x-r_l)\sqrt{x^2 - 4 z}}{2(1+z-x)} \; \; .
\end{eqnarray}
The case $l = l'$ is slightly more elaborate. We feel that it does not
make sense to display the term $\sum_{spins} |T|^2$ because it is of
considerable complexity in the general case when all the form factors
$A_i$, $V_1$ and $F_V^{K}$ are $q^2$ and $W^2$ dependent. The expression
together with the Monte Carlo program to do the phase space integrals
is available on request from the authors.

\subsection{Theory}

The form factors $A_i,\ V_1$ and $F_V^K$ have been discussed in all
kinds of models, Vector Meson Dominance, hard meson, etc.. For a
discussion see Ref. \cite{BARDIN}. We will restrict ourselves to the
predictions in the framework of CHPT.

To leading order we have
\begin{eqnarray}
V_1&=&0 \nonumber\\
A_1\ =\ A_2&=&A_3\ = 0\ .
\end{eqnarray}
We also have $F_V^K = 1$. The rate here is entirely given by the
in\-ner Brems\-strahlung contribution. At the one-loop level
several form factors get non-zero values \cite{beg}
\begin{eqnarray}
V_1&=&- \frac{1}{8\pi^2 F}\nonumber\\
A_1&=&- \frac{4}{F}\left( L_9^r + L_{10}^r \right) \nonumber\\
A_2&=&- \frac{2 F_K ( F_V^K(q^2) -1 ) }{q^2} \nonumber\\
A_3&=&0\nonumber\\
F_V^K(q^2)&=&1+H_{\pi\pi}(q^2)+2 H_{KK}(q^2) \; \; .
\label{ourresult}
\end{eqnarray}
These results obey the current algebra relation of Ref. \cite{BARDIN}.
The function $F_V^K(q^2)$ does, however, deviate somewhat from the
linear parametrization often used.
The function $H(t)$ is defined in appendix \ref{loop}.

The fact that the form factors at
next-to-leading order could be written in terms of the kaon
electromagnetic form factor
       in a simple way is          not true anymore at the $p^6$
level.
The Lagrangian at order $p^6$ contains a term of the form
\begin{equation}
\mbox{\rm tr}\left\{D_\alpha F_L^{\alpha\mu} U^{\dag} D^{\beta}
F_{R\beta\mu} U \right\}
\end{equation}
that contributes to $A_2$ and $A_3$ but not to the kaon electromagnetic
form factor,  $F_V^K (q^2)$.

\subsection{Numerical results}

We have
calculated the rates for a few cuts, including those
given in the literature.
For the case of \underline{unequal} leptons, the results are given
in table \ref{KLLL1} for the decay
$K^+ \to \mu^+ \nu e^+ e^-$. These include the cuts used in Refs.
\cite{KRISHNA} and \cite{DIAMANT},
$x \ge 40\ MeV /M_K$ and $ z \ge (140\ MeV/M_K)^2$,
respectively.
\begin{table}
\caption{\label{KLLL1}
         Theoretical values for the branching ratios for the decay
$K^+ \to \mu^+ \nu e^+ e^-$ for various cuts.}
\begin{center}
\begin{tabular}{|c|c|c|}
\hline
               &  tree level & form factors as given by CHPT\\
\hline
full phase space &$2.49 \cdot 10^{-5}$  &         $2.49 \cdot 10^{-5}$\\
\hline
$z\le 10^{-3}$ & $ 2.07\cdot 10^{-5} $ & $ 2.07 \cdot 10^{-5}$\\
\hline
$z\ge 10^{-3}$ &$4.12\cdot 10^{-6}$& $4.20\cdot 10^{-6}$\\
\hline
$z \ge ( 20\ MeV/M_K)^2$ & $ 3.15\cdot 10^{-6}$ & $3.23\cdot 10^{-6}$\\
\hline
$z \ge (140\ MeV/M_K)^2$ & $ 4.98\cdot 10^{-8}$ & $8.51\cdot 10^{-8}$\\
\hline
$ x\ge 40\ MeV/M_K$ & $ 1.58\cdot 10^{-5} $ & $1.58\cdot 10^{-5}$ \\
\hline
\end{tabular}
\end{center}
\end{table}
It can be seen that for this decay most of the branching ratio is
generated at very low electron-positron invariant masses. As can be seen
from the result for the cuts used in Ref. \cite{DIAMANT}, the effect
of the structure dependent terms is most visible at high invariant
electron-positron invariant mass. Our calculation, including the effect
of the form factors agrees well with their data. We disagree, however,
with the numerical result obtained by Ref. \cite{KRISHNA} by about an
order of magnitude.

For the decay $K^+ \to e^+ \nu \mu^+\mu^-$, we obtain for the tree level
or IB contribution a branching ratio
\begin{equation}
BR_{IB}(K^+ \to e^+ \nu \mu^+ \mu^- ) =
     3.06\cdot 10^{-12}
\end{equation}
      and,
including the form factors,
\begin{equation}
BR_{total}(K^+ \to e^+ \nu \mu^+ \mu^- ) =
 1.12\cdot 10^{-8}.
\end{equation}
 Here the structure
dependent terms are the leading-contribution since the in\-ner
Brems\-strahlung  contribution is helicity suppressed as can be seen
from the factor $m_l$ in $\overline{L_\mu}$.

For the decays with \underline{identical}
leptons we obtain for the muon case a
branching ratio  of
\begin{equation}
BR_{total}(
K^+ \to \mu^+\nu\mu^+\mu^-) =
1.35\cdot10^{-8}
\label{predicted}
\end{equation}
for the full phase space including the effects of the
form factors.
The in\-ner Brems\-strahlung or the tree level branching
ratio for this decay is
\begin{equation}
BR_{IB}(
K^+ \to \mu^+\nu\mu^+\mu^-) =
3.79\cdot10^{-9}.
\end{equation}
For the decay with two positrons and one electron the integration
over full phase space
for the tree level results
is very sensitive to the behaviour for small
pair masses.
 We have given the tree level and the full prediction,
including form factor effects in table \ref{KLLLP}. The cuts are
always on both invariant masses :
\begin{eqnarray}
z&=&(p_1 + p_2)^2 / M_K^2 \nonumber\\
z_1 &=& (p_l + p_2)^2 / M_K^2\ .
\end{eqnarray}
\begin{table}
\caption{\label{KLLLP}
         Theoretical values for the branching ratios for the decay
$K^+ \to   e^+ \nu e^+ e^-$ for various cuts.}
\begin{center}
\begin{tabular}{|c|c|c|}
\hline
               &  tree level & form factors as given by CHPT\\
\hline
full phase space & $\approx 4 \cdot 10^{-9}$ & $ 1.8\cdot 10^{-7}$\\
\hline
$z,\ z_1 \ge 10^{-3}$&  $3.0\cdot 10^{-10}$&      $1.22 \cdot 10^{-7}$\\
\hline
$z,\ z_1 \ge (50\ MeV/M_K)^2 $
&  $5.2\cdot 10^{-11}$&      $8.88 \cdot 10^{-8}$\\
\hline
$z,\ z_1 \ge (140\ MeV/M_K)^2 $
&  $2.1\cdot 10^{-12}$&      $3.39 \cdot 10^{-8}$\\
\hline
\end{tabular}
\end{center}
\end{table}
The values for the masses used are those of     $K^+$ and $\pi^+$.
For $L_9$ and $L_{10}$ we used the values given in table 1 in Ref.
\cite{Intro},

\begin{eqnarray}
L_9^r (M_\rho ) & = & 6.9 \cdot 10^{-3} \nonumber\\
L_{10}^r (M_\rho ) &=& -5.5 \cdot 10^{-3} \; \; .
\end{eqnarray}

\subsection{Present experimental status}

Only decays with an electron positron pair in the final state,
decays (\ref{S22}) and (\ref{S23}),
have been observed.

 Both have been measured
in the same experiment       \cite{DIAMANT}.
The decay $K^+ \to \mu^+ \nu e^+ e^-$ was measured with a branching
ratio of $(1.23 \pm 0.32)\cdot 10^{-7}$ with a lower cut on the
electron positron invariant mass of $140\ MeV$.
The measurement is compatible with our calculation including the
form factor effects for the relevant region of phase space.
This measurement
was then  extrapolated \cite{DIAMANT}
using the result of \cite{KRISHNA} to the
full phase space. Since we  disagree with that calculation,
we also disagree with the extrapolation.

In the same experiment,  4 events of the type
                                            $K^+ \to e^+\nu e^+ e^-$
were
observed where both electron positron pair invariant masses were
above 140 $MeV$. This corresponds to a branching ratio for this
region of phase space of $(2.8^{+2.8}_{-1.4})\cdot 10^{-8}$.
This result is compatible within errors  with our calculation, see
table \ref{KLLLP}.
The matrix element of Ref. \cite{KRISHNA} was again used for the
extrapolation to full phase space\cite{DIAMANT}.
Apart from our numerical disagreement,
the calculation of Ref. \cite{KRISHNA} was for the case of
non-identical leptons and cannot be applied here.

For the decay $K^+ \to \mu^+\nu\mu^+\mu^-$ an upper limit
of $4.1 \cdot 10^{-7}$ exists \cite{ATIYA}. This upper limit is
compatible with our theoretical result, Eq. (\ref{predicted}).

The decay $K^+ \to e^+\nu\mu^+\mu^-$ has not been looked for so far
and should be within the capabilities of DA$\Phi$NE given the
branching ratio predicted in the previous subsection.
This decay proceeds
almost entirely through the structure dependent terms and is as such a
good test of our calculation.

\subsection{Improvements at DA$\Phi$NE}

The decays discussed in this subsection, $K^+ \to l^+\nu
l^+l^-$, are
complementary to the decays $K^+ \to l^+\nu\gamma$. As was the case
for the analogous decay, $\pi^+ \to e^+\nu e^+ e^-$ \cite{egli},
 it may be
possible to explore phase space
more easily with this process than with
$K^+ \to l^+ \nu \gamma$ to resolve ambiguities in the form factors.

As can be seen from our predictions, tables \ref{KLLL1} and
\ref{KLLLP}, all the decays considered in this subsection should be
observable at DA$\Phi$NE. Large improvements in statistics are
possible since less severe cuts than those used in the past
experiments should be possible. In the decays with a $\mu^+ \mu^-$
pair and the decay $K^+ \to e^+ \nu e^+ e^-$
the effects of the form factors are already large in the total
rates and should be easily visible at DA$\Phi$NE. In the decay
$K^+ \to \mu^+ \nu e^+ e^-$
most of the total rate is for small invariant
mass of the pair and is given by the inner Bremsstrahlung
contribution.  There are, however, regions of phase space where the
form factor effects are large and DA$\Phi$NE should have enough
statistics to be able to study these regions.

\newpage
\setcounter{equation}{0}
\setcounter{subsection}{0}
\setcounter{so}{1}
\setcounter{table}{0}
\setcounter{figure}{0}
\clearpage
\section{$K_{l3}$ decays}

The decay channels considered in this subsection are
\bearr
 K^+(p) &\rightarrow& \pi^0 (p') l^+ (p_l) \nu _l (p_\nu) \hspace{1cm}
[K_{l3}^+] \label{s31}
\\
K^0(p) &\rightarrow &\pi^- (p') l^+ (p_l) \nu_l (p_\nu) \hspace{1cm} [K_{l3}^0]
\label{s32}
\eearr
and their charge conjugate modes. The symbol  $l$ stands for $\mu$ or $e$. We
do
not consider electromagnetic corrections and correspondingly set $\alpha = 0$
throughout this subsection.

\subsection{Matrix elements and kinematics}

The matrix element for $K_{l3}^+$
has the general structure
\bearr
T& =& \frac{G_F} {\sqrt{2}} V_{us}^\star l^\mu { F_\mu}^+ (p',p)
\label{s34}
\eearr
with
\bearr
l^\mu& =& \bar{u} (p_\nu)\gamma^\mu  (1- \gamma_5) v (p_l)
\nonumber \\
{ F_\mu}^+ (p',p)& =& < \pi^0 (p') \mid V_\mu^{4-i5} (0)
\mid K^+(p)>
\nonumber \\
&=& \frac{1}{\sqrt{2}} [(p'+p)_\mu f^{K^+\pi^0}_+ (t) + (p-p')_\mu
f_-^{K^+\pi^0} (t)]. \label{s35}
\eearr
To obtain the matrix element for $K_{l3}^0$ , one replaces $F^+_\mu$ by
\bearr
{ F_\mu }^0 (p',p)& =& < \pi^- (p') \mid V_\mu^{4-i5} (0)
\mid K^0(p)>
\nonumber \\
&=&  (p'+p)_\mu f^{K^0\pi^-}_+ (t) + (p-p')_\mu f_-^{K^0\pi^-} (t).
\label{s35a}
\eearr
 The processes (\ref{s31})
and (\ref{s32}) thus involve the four $K_{l3}$ form factors
$f^{K^+\pi^0}_\pm (t)$, $f^{K^0 \pi^-}_\pm (t)$ which depend on
\be
t = (p'-p)^2 = (p_l + p_\nu)^2,
\label{s36}
\ee
the square of the four momentum transfer to the leptons.

Let $f_\pm^{K\pi}=f_\pm^{K^+\pi^0}$ or $f_\pm^{K^0\pi^-}$.
$f_+^{K\pi}$ is referred to as the vector form factor, because
it specifies the $P$-wave  projection of the crossed channel matrix elements
 $< 0 \mid V^{4-i5}_\mu(0) \mid K^+, \pi^0 \;\mbox{in} >$.
 The $S$-wave projection is described by the scalar form factor
\be
f^{K\pi}_0 (t) = f^{K\pi}_+ (t) + \frac{t}{M^2_K - M^2_\pi} f^{K\pi}_-(t) \; \;
{}.
 \label{s37}
\ee
Analyses of $K_{l3}$ data frequently assume a linear dependence
\be
f^{K\pi}_{+,0} (t) = f^{K\pi}_+ (0) \left[ 1 + \lambda_{+,0}
\frac{t}{M^2_{\pi^+}} \right] \; \; .
\label{s38}
\ee
For a discussion of the validity of this approximation see \cite{GL2,pdg}
and references cited therein. Eq. (\ref{s38}) leads to a constant
$f^{K\pi}_- (t)$ ,
\be
f^{K\pi}_- (t) = f^{K\pi}_- (0) = f^{K\pi}_+ (0) (\lambda_0 -
\lambda_+) \frac{M_K^2 - M_\pi^2}{M_{\pi^+}^2}.
\label{s381}
\ee
The form factors $f^{K\pi}_{\pm,0} (t)$ are analytic functions in the complex
$t$-plane cut along the positive real axis. The cut starts at $t=(M_K +
M_\pi)^2$. In our phase convention, the form factors are real in the physical
region
\be
m^2_l \leq t \leq (M_K - M_\pi)^2.
\label{s39}
\ee
The kinematics of (spin averaged) $K_{l3}$ decays needs two variables, for
which we choose
\be
y = 2 p p_l/M_K^2\;,\; \; z = 2 p p'/ M_K^2 =(-t+M_\pi^2 +M_K^2)/M_K^2 \; \;.
\label{s311}
\ee
In the $K$ rest frame, $y$ ($z$) is proportional to the charged lepton (pion)
energy,
\be
y=2E_l/M_K \; \; , \; \; z=2E_\pi/M_K \; \; .
\label{s311a}
\ee

The physical region for $y$ and $z$ is
\bearr
2 \sqrt{r_l} \leq & y& \leq 1 + r_l - r_\pi
\nonumber \\
A(y)-B(y) \leq& z& \leq A(y)+B(y)
\nonumber\\
A(y)&=&(2-y)(1+r_l+r_\pi-y)/[2(1+r_l-y)]
\nonumber \\
B(y)&=& \sqrt{y^2-4r_l} (1+r_l-r_\pi-y)/[2(1+r_l-y)]
\nonumber \\
r_l &=& m^2_l/M_K^2, r_\pi = M^2_\pi/ M_K^2.
\label{s3121}
\eearr
or, equivalently,
\bearr
2\sqrt{r_\pi} \leq &z& \leq 1+r_\pi -r_l
\nonumber\\
C(z)-D(z) \leq &y& \leq C(z)+D(z)
\nonumber \\
C(z)&=& (2-z)(1+r_\pi +r_l -z)/[2(1+r_\pi-z)]
\nonumber \\
D(z)&=&\sqrt{z^2-4r_\pi}(1+r_\pi -r_l -z)/[2(1+r_\pi -z)] \; \; .
\nonumber \\
\label{s312}
\eearr

  \subsection{Decay rates}

The differential decay rate for $K_{l3}^+$ is given by
\bearr
d\Gamma & =& \frac{1}{2M_K(2\pi)^5} \sum_{spins}|T|^2 d_{LIPS}(p;p_l,p_\nu,p')
\; \; .
 \label{s312a}
\eearr
The Dalitz plot density
\be
\rho(y,z)=\frac{d^2\Gamma}{dy dz} = \frac{M_K}{256\pi^3} \sum_{spins}|T|^2
\label{s312b}
\ee
is a Lorentz invariant function which contains $f_{\pm}^{K^+\pi^0}$ in the
following form,
\bearr
\rho(y,z) = \frac{M_K^5G^2_F \mid V_{us}\mid^2 }{256 \pi^3}
 \left [ A (f^{K^+\pi^0}_+)^2 + B f^{K^+\pi^0}_+ f^{K^+\pi^0}_- + C
(f^{K^+\pi^0}_-)^2 \right]
 \label{s313}
\eearr
with
\bearr
A(y,z)& =& 4 (z+y-1) (1-y) + r_l [4 y+ 3 z-3] - 4 r_\pi + r_l (r_\pi - r_l)
\nonumber \\
B(y,z)& =& 2 r_l (3-2y - z + r_l - r_\pi)
\nonumber \\
C(y,z)& =& r_l (1+ r_\pi - z - r_l).
\label{s314}
\eearr
The quantities $(A,B,C)$ are related to the ones quoted by the PDG \cite{pdg}
by
\be
(A,B,C) = \frac{8}{M_K^3} (A,B,C)_{\mbox{\tiny{PDG}}} \; \; .
\label{s315}
\ee
To obtain the rate for $K_{l3}^0$, one replaces in (\ref{s313})
$f^{K^+\pi^0}_\pm $ by $\sqrt{2} f^{K^0\pi^-}_\pm$.

For convenience we also display the $K_{\mu3}/K_{e3}$ rates evaluated in the
approximation (\ref{s38}) for the form factors,
\bearr
\Gamma (K^+_{\mu3}) / \Gamma (K^+_{e3}) &=&
\frac{
0.645 +2.087 \lambda_+ +1.464 \lambda_0 +3.375 \lambda_+^2 +2.573 \lambda_0^2}
{1 +3.457 \lambda_+ +4.783 \lambda_+^2}
\nonumber  \\
\Gamma (K^0_{\mu3}) / \Gamma (K^0_{e3})& = &
\frac{
0.645 +2.086 \lambda_+ +1.459 \lambda_0 +3.369 \lambda_+^2 +2.560\lambda_0^2}
{1+3.456\lambda_+ +4.776 \lambda_+^2} \; .
\label{s316b}
\eearr

We have used the physical masses \cite{pdg} in evaluating these ratios and
$M_{\pi^+}$ to scale the slope in both cases. The
terms linear and quadratic in $\lambda_0$ are proportional to $m_l^2$ and
 therefore strongly suppressed in the
electron case. We do not include them in the denominators,
 because these coefficients
are smaller than $10^{-4}$. The interference term $\lambda_0\lambda_+$ is
absent by angular momentum conservation.
Furthermore, one has
\be
\int dy\;dz A(y,z) = \left\{ \begin{array}{ll}
                             0.0623 & [K_{\mu 3}^+] \nonumber \\
                             0.0606 & [K_{\mu 3}^0] \; \; .\\
                             \end{array}
                         \right.
\ee
\subsection{Determination of
the $K_{l3}$ form factors}

Measurements of the Dalitz plot distribution (\ref{s313}) of $K_{\mu3}$ data
allow one in principle to pin down the form factors
(up to a sign) in the range
 $m_\mu^2 \leq t \leq
(M_K - M_\pi)^2$.
 Measuring the $K_{\mu3}/K_{e3}$ branching ratio and
then using
(\ref{s316b}) gives a relationship between $\lambda_+$ and
$\lambda_0$  which is valid
in the approximation (\ref{s38}). Furthermore, muon polarization experiments
measure the weighted average of the ratio $f_-^{K\pi}(t)/f_+^{K\pi}(t)$ over
the $t$
range of the experiment \cite{pdg,mupol}. On the other hand, the electron modes
$K_{e3}$ are sensitive to $f_+^{K\pi}$ only, because the other contributions
are suppressed by the factor $(m_e/M_K)^2 \simeq 10^{-6}$, see eqs.
(\ref{s313}), (\ref{s314}).

Isospin breaking effects in $f^{K^+\pi^0}_+ (0)$ and $f_+^{K^0 \pi^-}(0)$
play
a central role in the determination of the Kobayashi-Maskawa matrix element
$V_{us}$ from $K_{e3}$ data, see \cite{lroos} for a detailed discussion of this
point. In the following we concentrate on the measurement of the slopes
$\lambda_{+,0}$.

\subsection{Previous measurements}

We refer the reader to the 1982 version of the PDG \cite{pdg82}
\footnote{Please note that the most recent measurements of $\lambda_{+,0}$ go
back to 1981 \cite{pdg}!}
for a critical discussion of the wealth of experimental information on
$\lambda_{+,0}^{K\pi}$. Here we content ourselves with a short summary.

\underline{$K_{e3}$-experiments}

The $\lambda_+$ values obtained are fairly consistent. The average values are
\bearr
K^+_{e3} \;:\; \lambda_+ &=& 0.028 \pm 0.004 \; \; \; \; \mbox{Ref.} \cite{pdg}
\nonumber \\
K_{e3}^0 \ : \, \lambda_+& =& 0.030 \pm 0.0016 \; \; \mbox{Ref.} \cite{pdg} \;
\; .
\label{s318}
\eearr

\underline{$K_{\mu 3}$-experiments}

The result by Donaldson et al. \cite{donaldson}
\bearr
\lambda_+& =& 0.030 \pm 0.003
\nonumber \\
\lambda_0& =& 0.019 \pm 0.004
\label{s319}
\eearr
dominates the statistics in the $K^0_{\mu3}$ case. The $\lambda_+$ value
(\ref{s319}) is consistent with the $K_{e3}$ value (\ref{s318}).
 However, the situation concerning the slope $\lambda_0$
is rather unsatisfactory, as the following (chronological) list illustrates.
\be
\lambda_0 = \left\{\begin{array}{llll}
              0.0341 &\pm &0.0067 & \cite{birulev} \\
              0.050 &\pm &0.008  & \cite{cho} \\
              0.039 &\pm &0.010  &  \cite{hill} \\
              0.047 &\pm &0.009  & \cite{clark} \\
              0.025 &\pm &0.019  & \cite{buchanan} \\
              0.019 &\pm &0.004  & \cite{donaldson} \; \; .
                \end{array}
              \right.
\label{l:s324}
\ee
The $\chi^2$ fit to the $K^0_{\mu3}$ data yields $\lambda_+ = 0.034 \pm 0.005$,
$\lambda_0 = 0.025 \pm 0.006$ with a $\chi^2/DF = 88/16$ \cite[p.76]{pdg82}!
The situation in the charged mode $K^+_{\mu3}$ is slightly better \cite{pdg82}.

\subsection{Theory}

The theoretical prediction of $K_{l3}$ form factors has a long history,
starting
in the sixties with the current algebra evaluation of $f^{K^+\pi^0}_\pm$. For
an
early review of the subject and for references to work prior to CHPT
evaluations
of $f_\pm$ we refer the reader to \cite{chounet} (see also Ref.\cite{shabkl4}).
 Here we concentrate on the
evaluation of the form factors in the framework of CHPT. We restrict our
consideration to the isospin symmetry limit $m_u =m_d$, as a result of which
one has
\be
f^{K^0 \pi^-}_{\pm,0} (t) =  f^{K^+\pi^0}_{\pm,0} (t) \equiv f_{\pm,0}(t) \; ;
\; m_u = m_d \; \; . \label{s321}
\ee

\vspace{.5cm}

\begin{subsub}{\bf
Chiral prediction at one-loop order}
\end{subsub}

In Ref. \cite{GL2}, the vector current matrix elements $< M' \mid q
\gamma^\mu \frac{\lambda^a}{2} q \mid M>$ have been calculated up to and
including terms of order $t = (p'-p)^2$ and of order $m_u, m_d$ and $m_s$ in
the invariant form factors. For reasons which will become evident below, we
consider here, in addition to the $K_{l3}$ form factors, also
the electromagnetic form factor of the pion
\be
< \pi^+ (p') \mid V^\mu_{em} (0) \mid \pi^+ (p) > = (p'
+p)^\mu F^\pi_V (t).
\label{s322}
\ee
The low-energy representation for $F_V^\pi (t)$ \cite{GL2,bijnens} and
 $f_+ (t)$ \cite{GL2} reads

$$
F_V^\pi (t) = 1+ 2 H_{\pi \pi} (t) + H_{KK} (t)
$$
\be
f_+ (t) = 1 + \frac{3}{2} H_{K\pi} (t) + \frac{3}{2} H_{K\eta} (t).
\label{s323}
\ee
The quantity $H(t)$ is a loop function
displayed  in appendix \ref{loop}. It contains the low-energy constant $L_9$.
 The indices attached to $H(t)$ denote the masses running in the loop.

Since $L_9$ is  the only unknown occurring in $F_V^\pi (t)$ and
in $f_+
(t)$, we need experimental information on the {\em slope} of one of these two
form factors to obtain a parameter-free low-energy representation of the
other.

The analogous low-energy representation of the scalar form factor is
\bearr
f_0 (t) &=& 1 + \frac{1}{8F^2} \left( 5 t - 2 \Sigma_{K\pi} - 3
\frac{\Delta^2_{K\pi}}{t}  \right) \bar{J}_{K\pi} (t)
\nonumber \\
&+& \frac{1}{24F^2} \left( 3 t - 2 \Sigma_{K\pi} -
\frac{\Delta_{K\pi}^2}{t} \right) \bar{J}_{K\eta} (t)
\nonumber \\
&+& \frac{t}{\Delta_{K\pi}} \left( \frac{F_K}{F_\pi} - 1 \right).
\label{s325}
\eearr
The function $\bar{J} (t)$ is listed in appendix \ref{loop}, and
$\Sigma_{K\pi}$ and $\Delta_{K\pi}$ stand for
\bearr
\Sigma_{K\pi}& =& M_K^2 + M_\pi^2
\nonumber \\
\Delta_{K \pi}& =& M_K^2 - M_\pi^2 \; \; .
\label{s326}
\eearr
The measured value \cite{lroos} $F_K/F_\pi = 1.22 \pm 0.01$ may be used to
obtain a parameter-free prediction of the scalar form factor $f_0 (t)$.

\vspace{.5cm}

\begin{subsub}{\bf
Momentum dependence of the vector form factor
              }\end{subsub}

In the spacelike interval $\sqrt{-t} < 350$ MeV the low-energy representation
(\ref{s323}) for the electromagnetic form factor $F_V^\pi (t)$ is very well
approximated by the first two terms in the Taylor series expansion
around $t=0$,
\be
F_V^\pi (t) = 1 + \frac{1}{6} <r^2>^\pi_V t + \cdots \; \; .
\label{s324}
\ee
Likewise, the linear approximation
\be
f_+ (t) = f_+ (0) \left \{ 1 + \frac{1}{6} <r^2>_V^{K\pi}  t
+ \cdots \right\}
\label{s325a}
\ee
reproduces the low-energy representation (\ref{s323}) very
well, see Fig. \ref{fkl31}.
\begin{figure}[t]
\vspace{8cm}
\caption{The vector and scalar form factors $f_+(t)$ and $f_0(t)$.
\label{fkl31}
         }
\end{figure}
 This is  in agreement with the observed Dalitz plot distribution, which is
consistent with a form factor linear in $t$. The charge radii are
$$
<r^2>^\pi_V = \frac{12 L_9^{r}}{F^2}- \frac{1}{32 \pi^2 F^2} \left \{
2 \ln \frac{M_\pi^2}{\mu^2} + \ln \frac{M_K^2}{\mu^2} + 3 \right \}
$$
$$
< r^2>^{K\pi}_V = < r^2>_V^\pi - \frac{1}{64 \pi^2 F^2} \left \{ 3 h_1
\left (\frac{M_\pi^2}{M_K^2} \right) + 3 h_1 \left (\frac{M_\eta^2}{M_K^2}
\right ) \right.
$$
\be
\hspace{-1cm} \left. + \frac{5}{2} \ln \frac{M_K^2}{M_\pi^2} + \frac{3}{2} \ln
\frac{M_\eta^2}{M_K^2} - 6 \right \}
\label{s327}
\ee
where
\be
h_1(x) = \frac{1}{2} \frac{(x^3 - 3x^2 - 3x +1)}{(x-1)^3} \ln x + \frac{1}{2}
\left( \frac{x+1}{x-1} \right)^2 - \frac{1}{3} \; \; .
\label{s328}
\ee
To evaluate these relations numerically, we use the measured charge radius of
the pion:
\be
<r^2>_V^\pi = 0.439 \pm 0.008 \mbox{fm}^2 \; \;  \cite{r2pion}
\label{s329}
\ee
as input and obtain the prediction
\bearr
\lambda_+& =& \frac{1}{6} M^2_{\pi^+} <r^2>_V^{K\pi} = 0.031
\label{s330}
\eearr
in agreement with
the experimental results (\ref{s318}), (\ref{s319})
\footnote{
We do not quote an error for the result (\ref{s330}), because one should
estimate  higher-order chiral corrections for this purpose.}
                                .
{}From this (and from the considerably more detailed discussion in Ref.
\cite{GL2}), one concludes, in agreement with other
theoretical investigations \cite{kl3other}, that the measured charge radii
$<r^2>_V^{\pi}$ and $<r^2>_V^{K \pi}$ are consistent with the low-energy
prediction.

\vspace{.5cm}

\begin{subsub}{\bf
 Momentum dependence of the scalar form factor.\\
\hspace*{1.4cm} Dashen-Weinstein and
Callan-Treiman relations} \end{subsub}

In the physical region of $K_{l3}$ decay the low-energy representation
(\ref{s325}) for the scalar form factor is approximated by the linear
formula
\be
f_0 (t) = f_+ (0) \left \{ 1 + \frac{1}{6} <r^2>^{K\pi}_S t +
\cdots \right \}
\label{s331}
\ee
to within an accuracy of 1 \%. (See Fig. \ref{fkl31}). The curvature generated
by higher-order
 terms is also expected to be negligible in the physical region of the
decay \cite{GL2}. For the slope $<r^2>^{K\pi}_S$ one obtains
 \bearr
<r^2>^{K\pi}_S& = &\frac{6}{M_K^2 - M_\pi^2} \left ( \frac{F_K}{F_\pi}-1
\right) + \delta_2 + O(\hat{m},m_s)
\nonumber \\
\delta_2 &=& - \frac{1}{192 \pi^2F^2} \left \{ 15 h_2 \left(
\frac{M_\pi^2}{M_K^2} \right) + \frac{19 M_K^2 + 3 M^2_\eta}{M_K^2 + M_\eta^2}
h_2 \left( \frac{M_\eta^2}{M_K^2} \right) - 18 \right\}
\nonumber \\
\label{s332}
\eearr
where
\bearr
h_2(x)& =& \frac{3}{2} \left( \frac{1+x}{1-x}\right)^2 +
\frac{3x(1+x)}{(1-x)^3} \ln x,
\nonumber \\
h_2 (x) &=& h_2 \left(\frac{1}{x} \right)\; ,\;h_2(1) = 1,
\nonumber \\
\hat{m}&=&(m_u+m_d)/2\; \; .
\eearr
This (parameter-free) predic\-tion is a modi\-fied vers\-ion of the
Dashen-Wein\-stein
relation \cite{dawein}, which results if the nonanalytic contribution
$\delta_2$ is dropped. Dashen, Li, Pagels and Weinstein \cite{dalipa} were the
first to point out that the low-energy singularities generated by the Goldstone
bosons affect this relation. The modified relation is formulated as a
prediction for the slope of $f_0(t)$ at the unphysical point $t_1 = M^2_K +
M^2_\pi$. Their expression for this slope however has two shortcomings: (i) it
does not account for all corrections of order ${\cal M}$; (ii) The slope at
$t_1$ differs substantially from the slope in the physical region of the decay
\cite{GL2,slope}, see Fig. \ref{fkl32}.
\begin{figure}[t]
\vspace{8cm}
\caption{ The normalized slopes of the vector and the scalar form factors.
Curve 1: the normalized slope $M^2_{\pi^+} df_+(t)/dt$. Curve 2: the
normalized
slope $M^2_{\pi^+} df_0(t)/dt$. Near the $\pi K$ threshold $t_0=(M_K+M_\pi)^2$,
the vector form factor behaves as $f_+(t)=f_+(t_0) + O[(t-t_0)]$, whereas
$f_0(t)=f_0(t_0) + O[(\protect\sqrt{t-t_0})]$. The slope of the scalar form
factor is therefore singular at $t=(M_K+M_\pi)^2$.
\label{fkl32}
          }
\end{figure}

Algebraically, the correction $\delta_2$ is of the same order in the low-energy
expansion as the term involving $F_K/F_\pi -1$. Numerically, the correction is
however small: $\delta_2$ reduces the prediction by 11 \%. With $F_K/F_\pi =
1.22 \pm 0.01$ the low-energy theorem (\ref{s332}) implies
\bearr
<r^2>^{K\pi}_S &=& 0.20 \pm 0.05 \mbox{fm}^2
\nonumber \\
\lambda_0 &=& \frac{1}{6} M_{\pi^+}^2 <r^2>^{K\pi}_S = 0.017 \pm 0.004
\label{s333}
\eearr
where the error is an estimate of the uncertainties due to higher-order
contributions. The prediction (\ref{s333}) is in agreement with the
high-statistics experiment \cite{donaldson} quoted in (\ref{s319}) but in
flat disagreement with some of the more recent data listed in
(\ref{l:s324}).

In the formulation of Dashen and Weinstein \cite{dawein}, the Callan-Treiman
relation \cite{caltrei} states
that the scalar form factor evaluated at $t= M_K^2 - M^2_\pi$ differs from
$F_K/F_\pi$ only by terms of order $m_u, m_d$: the quantity
\be
\Delta_{\mbox{\tiny{CT}}} = f_0 (M_K^2 - M^2_\pi) -
\frac{F_K}{F_\pi} \label{s334}
\ee
is of order $\hat{m}$. Indeed,  the low-energy
representation (\ref{s325})  leads to
\be
\Delta_{\mbox{\tiny{CT}}} = - \frac{M^2_\pi}{2 F^2} \left \{ \bar{J}_{K\pi}
(M_K^2 - M_\pi^2) + \frac{1}{3} \bar{J}_{K\eta} (M^2_K - M_\pi^2) \right\} +
O(\hat{m}m_s) \; \; .
\label{s335}
\ee
Numerically, $\Delta_{\mbox{\tiny{CT}}} = - 3.5 \cdot 10^{-3}$. The
Callan-Treiman relation should therefore hold to a very high degree of
accuracy. If the form factor is linear from $t=0$ to $t = M_K^2 - M^2_\pi$ then
the slope must be very close to
\be
\lambda_0^{\mtiny{CT}} = \frac{ {{M_{\pi^+}}^{2}} }{M_K^2-M^2_\pi}
 \left( \frac{F_K}{F_\pi}-1 \right) = 0.019
,
\label{s336}
\ee
in agreement with (\ref{s333}) and with the experimental result of Ref.
\cite{donaldson}, but in disagreement with, e.g., the value found in Ref.
\cite{cho}. We see no way to reconcile the value $\lambda_0 = 0.050$ with
chiral symmetry.

\subsection{Comment on tensor couplings}

S.A. Akimenko et al. \cite{akimenko} have investigated
the general form of the matrix element for $K_{e3}$ decays,
  obtained by adding scalar and tensor
couplings to
the standard $V-A$ interaction. Analyzing the
Dalitz-plot
distribution of $3.2\cdot 10^4$  $K^+\rightarrow \pi^0e^+\nu_e$ events, they
find that the presence of scalar and tensor couplings or  nonlinearities in the
form factor $f_+$  cannot be excluded.
 DA$\Phi$NE may be
an ideal place to check this claim ($\simeq 4\cdot 10^8$
 $K^+\rightarrow \pi^0e^+\nu_e$   events in
one year). However, the same proviso as in the case
of radiative $K_{l2}$ decays should be made
 (see subsection \ref{tensor1}): before any firm conclusion
can
be drawn, one has to estimate the effect of higher-order terms in the chiral
calculation. In the present case this may be less difficult to achieve than in
radiative kaon decays,
as only one form factor comes into play, which, in addition,  depends on
one kinematical variable only.

\subsection{Improvements at DA$\Phi$NE}

DA$\Phi$NE provides the opportunity to improve our knowledge of $K_{l3}$ decays
in a
very substantial manner - in particular, it should be possible to clarify the
issue of the slope $\lambda_0$ of the scalar form factor $f_0$. To illustrate,
we compare in table \ref{tkl3} the hitherto
obtained
number of events (third column)  with the expected ones at DA$\Phi$NE
(fourth column). The last column displays the remarkable increase in statistics
obtainable at DA$\Phi$NE.

\begin{table}
\protect
\begin{center}
\caption{Rates of $K_{l3}$ decays.
The number of events in the
third column
 corresponds to  those data which are of relevance for the
determination of the  slope $\lambda_0$ of the scalar
form factor.
\label{tkl3}
}
\vspace{1em}
\begin{tabular}{|c|c|c|c|c|} \hline
\multicolumn{2}{|c|}{} & \multicolumn{2}{c|}{$\sharp$ events} & \\ \hline
  & branching  & Particle Data
  & DA$\Phi$NE  & improve-                  \\
  & ratio & Group
  & {1 year} & ment \\ \hline
$K^+ \rightarrow \pi^0 \mu^+ \nu_\mu $ &
 $3.18 \cdot 10^{-2} $ & $ 10^5 $ & $3\cdot 10^8$ & $3\cdot 10^3$      \\
\hline  $K_L \rightarrow \pi^{\pm} \mu^{\mp} \nu $ &
 $27 \cdot 10^{-2} $ & $ 4\cdot 10^6 $ & $3\cdot 10^8$ & $70$  \\ \hline
\end{tabular}
\end{center}
\end{table}


\setcounter{equation}{0}
\setcounter{subsection}{0}
\setcounter{so}{1}
\setcounter{table}{0}
\setcounter{figure}{0}
\clearpage

\section{Radiative $K_{l3}$ decays} \label{section4}
The decay channels considered in this subsection are
\beqa
K^+(p) & \ra & \pi^0(p') l^+(p_l) \nu_l(p_{\nu}) \gamma(q) \qquad
[K^+_{l3\gamma}] \no \\*
K^0(p) & \ra & \pi^-(p') l^+(p_l) \nu_l(p_{\nu}) \gamma(q) \qquad
[K^0_{l3\gamma}] \no
\eeqa
and the charge conjugate modes. We only consider real photons ($q^2 = 0$).

\subsection{Matrix elements}
The matrix element for $K^+_{l3\gamma}$ has the general structure
\beqa
T & = & \left.\dfrac{G_F}{\sqrt{2}} e V^*_{us} \ve^{\mu}(q)^*
\right\{(V^+_{\mu\nu} -
A^+_{\mu\nu}) \ol{u}(p_{\nu}) \gamma^{\nu} (1 - \gamma_5)
v(p_l) \label{eq:T}  \\*
  &   & + \left. \dfrac{F^+_{\nu}}{2 p_lq} \ol{u}(p_{\nu}) \gamma^{\nu}
(1 - \gamma_5) (m_l - \fsl p_l - \fsl q) \gamma_{\mu} v(p_l)\right\}
\equiv \ve^{\mu *} A^+_{\mu}. \no
\eeqa
The diagram of Fig. \ref{fig41}.a corresponding to the first part of
Eq. (\ref{eq:T}) includes Brems\-strah\-lung off the $K^+$.
The lepton Brems\-strah\-lung diagram of Fig. \ref{fig41}.b is represented
by the second part of Eq. (\ref{eq:T}).
The hadronic tensors $V^+_{\mu\nu}, A^+_{\mu\nu}$ are defined as
\beq
I^+_{\mu\nu} = i \int d^4x e^{i q x} \langle \pi^0(p') \mid
T\{V^{em}_\mu(x) I^{4-i 5}_\nu(0)\} \mid K^+(p) \rangle ,
\qquad I = V,A .
\eeq
$F^+_\nu$ is the $K^+_{l3}$ matrix element
\beq
F^+_\nu = \langle \pi^0(p') \mid V^{4-i 5}_\nu(0) \mid K^+(p) \rangle.
\eeq
The tensors $V^+_{\mu\nu}$ and $A^+_{\mu\nu}$ satisfy the Ward identities
\beqa
q^\mu V^+_{\mu\nu} & = & F^+_\nu \label{eq:Ward} \\*
q^\mu A^+_{\mu\nu} & = & 0  \no
\eeqa
leading in turn to
\beq
q^\mu A^+_\mu = 0~,\label{eq:WI}
\eeq
as is required by gauge invariance.

For $K^0_{l3\gamma}$, one obtains the corresponding amplitudes and
hadronic tensors by making the replacements
\beqa
K^+ & \ra & K^0,\qquad \pi^0 \ra \pi^- \no \\*
V^+_{\mu\nu} & \ra & V^0_{\mu\nu},\qquad A^+_{\mu\nu} \ra A^0_{\mu\nu}\\*
F^+_\nu & \ra & F^0_\nu,\qquad A^+_\mu \ra A^0_\mu. \no
\eeqa

To make the infrared behaviour transparent,
it is convenient to separate the tensors $V^+_{\mu\nu}, V^0_{\mu\nu}$
into two parts:
\beqa
V^+_{\mu\nu} & = & \hat{V}^+_{\mu\nu} + \dfrac{p_\mu}{pq} F^+_\nu
\label{eq:Low} \\*
V^0_{\mu\nu} & = & \hat{V}^0_{\mu\nu} + \dfrac{p'_\mu}{p'q} F^0_\nu. \no
\eeqa
Due to Low's theorem, the amplitudes $\hat{V}^{+,0}_{\mu\nu}$
are finite for $q \ra 0$. The axial amplitudes
$A^{+,0}_{\mu\nu}$ are automatically infrared finite.
The Ward identity (\ref{eq:Ward}) implies that the vector amplitudes
$\hat{V}^{+,0}_{\mu\nu}$ are transverse:
\beq
q^\mu \hat{V}^{+,0}_{\mu\nu} = 0. \eeq

For on-shell photons, Lorentz and parity invariance together with gauge
invariance allow the general
decomposition (dropping the superscripts +,0 and terms that vanish
upon contraction with the photon polarization vector)
\beqa
\hat{V}_{\mu\nu} & = & V_1 \left(g_{\mu\nu} - \dfrac{W_\mu q_\nu}
{qW}\right) + V_2 \left(p'_\mu q_\nu - \dfrac{p'q}{qW} W_\mu q_\nu
\right) \no \\*
& & + V_3 \left( p'_\mu W_\nu - \dfrac{p'q}{qW} W_\mu W_\nu \right)
+ V_4\left( p'_\mu p'_\nu - \dfrac{p'q}{qW} W_\mu p'_\nu \right)
\label{eq:tensor} \\
A_{\mu\nu} & = & i \ve_{\mu\nu\rho\sigma} (A_1 p'^\rho q^\sigma +
A_2 q^\rho W^\sigma) + i \ve_{\mu\lambda\rho\sigma} p'^\lambda
q^\rho W^\sigma (A_3 W_\nu + A_4 p'_\nu) \no \\
F_\nu & = & C_1 p'_\nu + C_2 (p - p')_\nu  \no \\*
W & = &  p_l + p_\nu. \no
\eeqa
With the decomposition (\ref{eq:Low}) we can write the matrix element
for $K^+_{l3\gamma}$ in (\ref{eq:T}) in a form analogous to Eq.
(\ref{k3}) for $K_{l2\gamma}$:
\beqa
T & = & \left.\dfrac{G_F}{\sqrt{2}} e V^*_{us} \ve^{\mu}(q)^*
\right\{(\hat{V}^+_{\mu\nu} -
A^+_{\mu\nu}) \ol{u}(p_{\nu}) \gamma^{\nu} (1 - \gamma_5) v(p_l)
\label{eq:Tnew}  \\*
  &   & + \left. F^+_{\nu} \ol{u}(p_{\nu}) \gamma^{\nu}
(1 - \gamma_5) \left[ \dfrac{p_\mu}{pq} -
\dfrac{(\fsl p_l + \fsl q - m_l) \gamma_\mu}
{2 p_lq} \right] v(p_l)\right\}~.  \no
\eeqa

The four invariant vector amplitudes $V_1,\ldots,V_4$ and
the four axial amplitudes $A_1,\ldots,A_4$ are functions of three scalar
variables. A convenient choice for these variables is
\beq
E_\gamma = pq/M_K ,\; E_\pi = pp'/M_K ,\; W = \sqrt{W^2}
\label{eq:kin1} \eeq
where $W$ is the invariant mass of the lepton pair. The amplitudes
$C_1, C_2$ can be expressed in terms of the $K_{l3}$ form factors
and depend only on the variable
\begin{figure}[t]
\vspace{7cm}
\caption{Diagrammatic representation of the $K^+_{l3\gamma}$
amplitude.} \label{fig41}
\end{figure}
$ (p - p')^2 = M^2_K + M^2_\pi - 2 M_K E_\pi$.
For the full kinematics of $K_{l3\gamma}$ two more variables are
needed, e.g.
\beq E_l = pp_l/M_K ,\; x = p_lq/M^2_K. \label{eq:kin2} \eeq
The variable $x$ is related to the angle $\theta_{l\gamma}$ between the
photon and the charged lepton in the $K$ rest frame:
\beq
x M^2_K = E_\gamma (E_l - \sqrt{E^2_l - m^2_l} \cos{\theta_{l\gamma}}).
\eeq
T invariance implies that the vector amplitudes $V_1,\ldots,V_4$, the
axial amplitudes $A_1,\ldots,A_4$ and the $K_{l3}$ form factors
$C_1, C_2$ are (separately) relatively real in the physical region.
We choose the standard
phase convention in which all amplitudes are real.

For $\theta_{l\gamma} \ra 0$ (collinear lepton and photon), there is
a lepton mass singularity in (\ref{eq:T}) which is numerically
relevant for $l = e$.
The region of small $E_\gamma, \theta_{l\gamma}$ is dominated by the
$K_{l3}$ matrix elements. The new theoretical information of
$K_{l3\gamma}$ decays resides in the tensor amplitudes
$\hat{V}_{\mu\nu}$ and $A_{\mu\nu}$. The relative importance of these
contributions can be enhanced by cutting away the region of low
$E_\gamma, \theta_{l\gamma}$. It may turn out to be of advantage to
reduce the statistics by applying more severe cuts than necessary from
a purely experimental point of view.

\subsection{Decay rates} \label{subsec-tree}
The total decay rate is given by \beqa
\Gamma(K \ra \pi l \nu \gamma) & = & \dfrac{1}{2 M_K (2 \pi)^8}
\int d_{LIPS}(p;p',p_l,p_\nu,q)
\sum_{spins}\mid T \mid^2 \label{eq:PS}
\eeqa
in terms of the amplitude $T$ in (\ref{eq:T}). The square of the matrix
element, summed over photon and lepton polarizations, is a bilinear form
in the invariant amplitudes $V_1,\ldots,V_4$, $A_1,\ldots,A_4$,
$C_1,C_2$.
Pulling out common factors, we write (\ref{eq:PS}) in the form
\beq
\Gamma(K \ra \pi l \nu \gamma) = \dfrac{4 \alpha G^2_F \mid V_{us} \mid^2}
{(2 \pi)^7 M_K} \int d_{LIPS}(p;p',p_l,p_\nu,q)\ SM \co \label{eq:rate} \eeq
where $SM$ is the reduced matrix element. For the actual numerical
calculations, we have found it useful to employ a tensor decomposition
different from the one in Eqs. (\ref{eq:Low}) and (\ref{eq:tensor})
\beqa
V_{\mu\nu} & = & B_1 g_{\mu\nu} + B_2 W_\mu q_\nu + B_3 p'_\mu q_\nu
+ B_4 W_\mu p'_\nu \no \\
& & + B_5 W_\mu W_\nu + B_6 p'_\mu W_\nu + B_7 p'_\mu p'_\nu~.
\label{eq:basis} \eeqa
One advantage is that (\ref{eq:basis}) applies equally well to both
charge modes while (\ref{eq:Low}) does not.
 In the numerical evaluation of the amplitudes, gauge invariance
can of course be used to express three of the $B_i$ in terms of the
remaining ones and of $C_1, C_2$.

To get some feeling for the magnitude of the various decay rates, let
us first consider the tree level amplitudes to lowest order $p^2$ in
CHPT. With the sign conventions of Ref. \cite{galenp1},
 these amplitudes are
\cite{beg,Hol} : \begin{flushleft} $\ul{K^+_{l3\gamma}:}$ \end{flushleft}
\beqa
V^+_{\mu\nu} & = & \dfrac{1}{\sqrt{2}} \left [ g_{\mu\nu} + \dfrac{
(p'+W)_\mu(2 p'+W)_\nu}{pq} \right ] \no \\*
A^+_{\mu\nu} & = & 0 \label{eq:tree+} \\*
F^+_\nu & = & \dfrac{1}{\sqrt{2}} (p+p')_\nu \no \eeqa
\begin{flushleft} $\ul{K^0_{l3\gamma}:}$ \end{flushleft}
\beqa
V^0_{\mu\nu} & = & - g_{\mu\nu} + \dfrac{p'_\mu(2 p'+2 q+W)_\nu}
{p'q} \no \\*
A^0_{\mu\nu} & = & 0 \label{eq:tree0} \\*
F^0_\nu & = & (p+p')_\nu. \no \eeqa

In table~\ref{tab:tree} the corresponding branching ratios
are presented for the four decay modes for
$E_\gamma \geq 30 MeV$ and $\theta_{l\gamma} \geq 20^\circ$.
For $K^0_{l3\gamma}$, the rates are to be understood as
$\Gamma(K_L \ra \pi^{\pm} l^{\mp} \nu \gamma)$. The
number of events correspond to the design values for DA$\Phi$NE (cf.
App. \ref{notation} ).
\begin{table}[t]
\begin{center}
\caption{Branching ratios for tree level amplitudes for $E_\gamma
\geq 30 MeV$ and $\theta_{l\gamma} \geq 20^\circ$ in the $K$ rest
frame.} \label{tab:tree} \vspace{.5cm}
$\begin{array}{|c|c|c|}
\hline
$decay$  & $BR(tree)$
& \# $events/yr$ \\ \hline
K^+_{e3\gamma} & 2.8 \times 10^{-4} & 2.5 \times 10^6 \\
K^+_{\mu 3\gamma} & 1.9 \times 10^{-5} & 1.7 \times 10^5 \\
K^0_{e3\gamma} & 3.6 \times 10^{-3} & 4.0 \times 10^6 \\
K^0_{\mu 3\gamma} & 5.2 \times 10^{-4} & 5.7 \times 10^5 \\
\hline  \ea $
\end{center}
\end{table}

\subsection{Previous experiments}
The data sample for $K_{l3\gamma}$ decays is very limited and it
is obvious that DA$\Phi$NE will be able to make significant improvements.
The present experimental status is summarized in table~\ref{tab:exp}.
\begin{table} \centering
\caption{Experimental results for $K_{l3\gamma}$ decays}
\label{tab:exp} \vspace{.5cm}
$\ba{|c|c|c|r|r|c|}
\hline
$decay$ & $exp.$ & E_{\gamma,min} & $\# events$
 & \multicolumn{1}{|c|}{$BR$} & \\ \hline
K^+_{e3\gamma} & \cite{Bol} & 10 MeV & 192 & (2.7 \pm 0.2)
\times 10^{-4} & 0.6 < \cos{\theta_{e\gamma}} < 0.9 \\
K^+_{e3\gamma} & \cite{Rom} & 10 MeV & 13 & (3.7 \pm 1.3)
\times 10^{-4} &  - " - \\
K^+_{e3\gamma} & \cite{Ljung} & 30 MeV & 16 & (2.3 \pm 1.0)
\times 10^{-4} & \cos{\theta_{e\gamma}} < 0.9 \\
K^+_{\mu 3\gamma} & \cite{Ljung} & 30 MeV & 0 & < 6.1
\times 10^{-5} & 90 \% \; c.l. \\
K^0_{e3\gamma} & \cite{Peach} & 15 MeV & 10 & (1.3 \pm 0.8)
\times 10^{-2} & \\ \hline
\ea $ \end{table}

A comparison between tables \ref{tab:tree} and \ref{tab:exp} shows
the tremendous improvement in statistics to be expected at DA$\Phi$NE.
We shall come back to the question whether this improvement will
be sufficient to test the standard model at the next-to-leading order,
$O(p^4)$, in CHPT.

\subsection{Theory}
Prior to CHPT, the most detailed calculations of $K_{l3\gamma}$
amplitudes were performed by Fearing, Fischbach and Smith
\cite{Fischbach} using current algebra techniques.

In the framework of CHPT, the amplitudes are
given by (\ref{eq:tree+}) and (\ref{eq:tree0})
to leading order in the chiral expansion.

\vspace{.5cm}

\begin{subsub}{{\bf
CHPT to $O(p^4)$} {\rm \cite{beg}}}
\end{subsub}

There are in general three types of contributions~\cite{galenp1}:
anomaly, local contributions due to ${\cal L}_4$ and loop amplitudes.

\begin{figure}
\vspace{7.5cm}
\caption{Loop diagrams (without tadpoles) for $K_{l3}$ at $O(p^4)$. For
$K_{l3\gamma}$, the photon must be appended on all charged lines and
on all vertices.} \label{fig42}
\end{figure}

The anomaly contributes to the axial amplitudes \beqa
A^+_{\mu\nu} & = & \dfrac{i \sqrt{2}}{16 \pi^2 F^2} \left\{\ve_
{\mu\nu\rho\sigma}q^\rho(4 p'+W)^\sigma + \dfrac{4}{W^2-M^2_K}
\ve_{\mu\lambda\rho\sigma}W_\nu p'^\lambda q^\rho W^\sigma \right\}
\label{eq:anom} \\*
A^0_{\mu\nu} & = & - \dfrac{i}{8 \pi^2 F^2}\ve_{\mu\nu\rho\sigma}
q^\rho W^\sigma.\no \eeqa

The loop diagrams for $K_{l3\gamma}$ are shown in
Fig. \ref{fig42}. We first write the $K^+_{l3}$ matrix element
in terms of three functions $f^+_1, f^+_2,f^+_3$ which will also appear
in the invariant amplitudes $V^+_i$. Including the contributions
from the low-energy constants $L_5,L_9$ in ${\cal L}_4$, the
$K_{l3}$ matrix element $F^+_\nu$ is given by \beqa
F^+_\nu & = & f^+_1(t) p'_\nu + \left[\dfrac{1}{2}(M^2_K - M^2_\pi - t)
f^+_2(t) + f^+_3(t)\right](p - p')_\nu \no \\
f^+_1(t) & = & \sqrt{2} + \dfrac{4 \ol{L_9}}{\sqrt{2} F^2} t +
2 \sum_{I=1}^{3} (c^I_2 - c^I_1) B^I_2(t) \no \\*
f^+_2(t) & = & - \dfrac{4 \ol{L_9}}{\sqrt{2} F^2} + \dfrac{1}{t}
\sum_{I=1}^{3} \left\{ (c^I_1 - c^I_2)\left [2 B^I_2(t)
- \dfrac{(t+\Delta_I)
\Delta_I J_I(t)}{2 t}\right ] - c^I_2 \Delta_I J_I(t)
\right\} \no \\*
f^+_3(t) & = & \dfrac{F_K}{\sqrt{2} F_\pi} + \dfrac{1}{2 t}
\sum_{I=1}^{3} \left\{(c^I_1+c^I_2)(t+\Delta_I)
- 2 c^I_3\right\}\Delta_I J_I(t) \label{eq:fi} \\
\ol{L_9} & = & L^r_9(\mu) - \dfrac{1}{256 \pi^2} \ln{\dfrac
{M_\pi M^2_K M_\eta}{\mu^4}}  \no \\
\Delta_I & = & M^2_I-m^2_I\enskip ,\enskip t \enskip = \enskip (p-p')^2~.
\no \eeqa
$\ol{L_9}$ is a scale independent coupling constant and we have traded
the tadpole contribution together with $L_5$ for $F_K/F_\pi$ in
$f^+_3(t)$. The sum over I corresponds to the three loop diagrams
of Fig. \ref{fig42} with coefficients $c^I_1,c^I_2,c^I_3$ displayed in
table~\ref{tab:loop+}.
\begin{table} \centering
\caption{Coefficients for the $K^+_{l3\gamma}$ loop amplitudes
corresponding to the diagrams $I=1,2,3$ in Fig. \protect\ref{fig42}.
All coefficients $c^I_i$ must be divided by
$6 \protect\sqrt{2} F^2$.}
\label{tab:loop+} \vspace{.5cm}
$\ba{|l|l|l|r|r|c|}
\hline
I & M_I & m_I & c^I_1
& c^I_2  & c^I_3 \\ \hline
1 & M_K & M_\pi & 1 & -2 & - M^2_K - 2 M^2_\pi \\
2 & M_K & M_\eta & 3 & -6 & - M^2_K - 2 M^2_\pi \\
3 & M_\pi & M_K & 0 & -6 & -6 M^2_\pi \\
\hline
\ea$ \end{table}
We use the Gell-Mann--Okubo mass formula
throughout to express $M^2_{\eta}$ in terms of $M^2_K, M^2_\pi$.
The functions $J_I(t)$ and $B^I_2(t)$ can be found in App. \ref{loop}.

The standard $K_{l3}$ form factors $f_+(t), f_-(t)$
as given in the previous subsection~\cite{GL2} are
\beqa
f_+(t) & = & \dfrac{1}{\sqrt{2}}f^+_1(t) \\*
f_-(t) & = & \dfrac{1}{\sqrt{2}}\left[(M^2_K-M^2_\pi-t)f^+_2(t)
+2 f^+_3(t)-f^+_1(t)\right]. \no \eeqa

It remains to calculate the infrared finite tensor amplitude
$\hat{V}^+_{\mu\nu}$. The invariant amplitudes $V^+_i$ can be
expressed in terms of the previously defined functions $f^+_i$ and of
additional amplitudes $I_1,I_2,I_3$. Diagrammatically, the latter
amplitudes arise from those diagrams in Fig. \ref{fig42} where the
photon is not appended on the incoming $K^+$ (non-Bremsstrahlung
diagrams). The final expressions are
\beqa
V^+_1 & = & I_1 + p'W_q f^+_2(W^2_q) + f^+_3(W^2_q)
\no \\*
V^+_2 & = & I_2 -\dfrac{1}{pq}\left[p'W_q f^+_2(W^2_q)
+ f^+_3(W^2_q)\right]
\no \\*
V^+_3 & = & I_3 +\dfrac{1}{pq}\left[p'W f^+_2(W^2)+f^+_3(W^2)
-p'W_q f^+_2(W^2_q) - f^+_3(W^2_q)\right]  \label{eq:Vhat} \\*
V^+_4 & = & \dfrac{f^+_1(W^2)-f^+_1(W^2_q)}{pq}  \no \\*
W_q & = & W + q \enskip = \enskip p - p'~.
\no \eeqa

The amplitudes $I_1, I_2, I_3$ in Eq.(\ref{eq:Vhat}) are given by
\beqa
I_1 & = & \dfrac{4 qW}{\sqrt{2}F^2}(\ol{L_9}+\ol{L_{10}}) + \dfrac
{8 p'q}{\sqrt{2}F^2}\ol{L_9}  \no \\*
&  & +\sum_{I=1}^{3}\left\{\left [(W^2_q+\Delta_I)(c^I_1+c^I_2)-
2(c^I_2 p'W_q+c^I_3)\right ]
\left[\dfrac{(W^2_q-\Delta_I)\hat{J}_I}
{2 W^2_q} -2 G_I) \right] \right.\no \\*
& & +\dfrac{(c^I_2-c^I_1)}{2}
\left [\dfrac{p'W_q}{W^2_q}(\dfrac{(W^4_q-\Delta^2_I)
\hat{J}_I}{W^2_q}+4\hat{B}^I_2)+p'(W-q)L^I_m\right ] \no \\*
& & +\dfrac{2(c^I_2-c^I_1)}{qW}\left.\left [p'q(F_I-
(W^2_q+\Delta_I)G_I)+p'W(\hat{B}^I_2-B^I_2)\right ] \right\}
\no \\
I_2 & = & -\dfrac{8\ol{L_9}}{\sqrt{2}F^2}+\dfrac{2}{qW}\sum_{I=1}
^{3}(c^I_2-c^I_1)\left [F_I-(W^2+\Delta_I)G_I\right ]  \no \\
I_3 & = & -\dfrac{4\ol{L_9}}{\sqrt{2}F^2}+\sum_{I=1}^{3}
\left\{2(c^I_2-c^I_1)
\left [G_I+\dfrac{L^I_m}{4}+\dfrac{\hat{B}^I_2-B^I_2}{qW}
\right ] -c^I_1 \dfrac{\Delta_I J_I}{W^2} \right\} \label{eq:loop} \\
\ol{L_{10}} & = & L^r_{10}(\mu)+\dfrac{1}{256\pi^2}\ln{\dfrac{M_\pi
M^2_K M_\eta}{\mu^4}}  \no \\
L^I_m & = & \dfrac{\Sigma_I}{32\pi^2\Delta_I}
\ln{\dfrac{m^2_I}{M^2_I}} \no \\
F_I & = & \hat{B}^I_2-\dfrac{W^2}{4}L^I_m+\dfrac{1}{qW}
\left(W^2 B^I_2-W^2_q \hat{B}^I_2\right) \no \\
G_I & = & \dfrac{M^2_I}{2}C(W^2_q,W^2,M^2_I,m^2_I)+\dfrac{1}{8 qW}
\left[(W^2_q+\Delta_I)\hat{J}_I-(W^2+\Delta_I)J_I\right]
+\dfrac{1}{64\pi^2} \no \\
J_I & \equiv & J_I(W^2), \; \hat{J}_I \equiv J_I(W^2_q) \no \\
B^I_2 & \equiv & B^I_2(W^2), \; \hat{B}^I_2 \equiv B^I_2(W^2_q).
 \no \eeqa

The function $C(W^2_q,W^2,M^2_I,m^2_I)$ is given in App. \ref{loop}.
All the invariant amplitudes $V^+_1,\ldots,V^+_4$ are
real in the physical region. Of course, the same is true for the
$K_{l3}$ matrix element $F^+_\nu$.

The $K^0_{l3\gamma}$ amplitude has a very similar structure. Both the
$K^0_{l3}$ matrix element $F^0_\nu$ and the infrared finite
vector amplitude $\hat{V}^0_{\mu\nu}$ can be obtained from the
corresponding quantities $F^+_\nu$  and
$\hat{V}^+_{\mu\nu}$ by the following steps:
\bit
\item interchange $p'$ and $-p \:$ ;
\item replace $\dfrac{F_K}{F_\pi}$ by $\dfrac{F_\pi}{F_K}$ in $f^+_3$;
\item insert the appropriate coefficients $c^I_i$ for
$K^0_{l3\gamma}$ listed in table~\ref{tab:loop0};
\item multiply $F^+_\nu$ and $\hat{V}^+_{\mu\nu}$ by a
factor $-\sqrt{2}$.
\eit
\begin{table} \centering
\caption{Coefficients for the $K^0_{l3\gamma}$ loop amplitudes
corresponding to the diagrams $I=1,2,3$ in Fig. \protect\ref{fig42}. All
coefficients $c^I_i$ must be divided by $6 \protect\sqrt{2}F^2$.}
\label{tab:loop0} \vspace{.5cm}
$\ba{|l|l|l|r|r|c|}
\hline
I & M_I & m_I  & c^I_1
& c^I_2 & c^I_3 \\ \hline
1 & M_K & M_\pi & 0 & -3 & -3 M^2_K \\
2 & M_K & M_\eta & 6 & -3 &  M^2_K+2 M^2_\pi \\
3 & M_\pi & M_K & 4 & -2 & -2 M^2_K + 2 M^2_\pi \\
\hline \ea $
\end{table}

\vspace{.5cm}

\begin{subsub}{\bf
Numerical results}
\end{subsub}

In calculating the rates with the complete amplitudes of the previous
subsection, we use the same cuts as for the tree level rates in
Subsect.~\ref{subsec-tree}:
\beqa
E_\gamma & \geq & 30 MeV \\*
\theta_{l\gamma} & \geq & 20^\circ. \no \eeqa
The physical values of $M_\pi$ and $M_K$ are used in the amplitudes.
$M_\eta$ is calculated from the Gell-Mann--Okubo mass formula. The
values of the other parameters can be found in Ref. \cite{Intro}
 and in appendix \ref{notation}.

The results for $K^+_{l3\gamma}$ and $K^0_{l3\gamma}$ are displayed in
tables \ref{tab:+} and \ref{tab:0}, respectively. For comparison, the
tree level branching ratios of table \ref{tab:tree} and the rates for
the amplitudes without the loop contributions are also shown. The
separation between loop and counter\-term contributions is of course
scale dependent. This scale dependence is absorbed in the scale invariant
constants $\ol{L_9}, \ol{L_{10}}$ defined in Eqs.(\ref{eq:fi}),
(\ref{eq:loop}). In other words, the entries in tables \ref{tab:+},
\ref{tab:0} for the amplitudes without loops correspond to setting
all coefficients $c^I_i$ in tables \ref{tab:loop+},
\ref{tab:loop0} equal to zero.

\begin{table} \centering
\caption{Branching ratios and expected number of events at DA$\Phi$NE
for $K^+_{l3\gamma}$.} \label{tab:+} \vspace{.5cm}
$\ba{|l|c|c|}
\hline
\multicolumn{1}{|c|}{K^+_{e3\gamma}} &
$BR$  & \# $events/yr$ \\ \hline
$full $ O(p^4) $ amplitude$ & 3.0\times 10^{-4} & 2.7\times 10^6 \\
$tree level$ & 2.8\times 10^{-4} & 2.5\times 10^6 \\
O(p^4) $ without loops$ & 3.2\times 10^{-4} & 2.9\times 10^6 \\
\hline \ea$ \\*[.5cm] $\ba{|l|c|c|} \hline
\multicolumn{1}{|c|}{K^+_{\mu 3\gamma}}  &
$BR$ & \# $events/yr$ \\ \hline
$full $ O(p^4) $ amplitude$ & 2.0\times 10^{-5} & 1.8\times 10^5 \\
$tree level$ & 1.9\times 10^{-5} & 1.7\times 10^5 \\
O(p^4) $ without loops$ & 2.1\times 10^{-5} & 1.9\times 10^5 \\
\hline \ea$ \end{table}
\nopagebreak[2]
\begin{table} \centering
\caption{Branching ratios and expected number of events at DA$\Phi$NE
for $K^0_{l3\gamma}$.} \label{tab:0} \vspace{.5cm}
$\ba{|l|c|c|}
\hline
\multicolumn{1}{|c|}{K^0_{e3\gamma}}  &
$BR$  & \# $events/yr$ \\ \hline
$full $ O(p^4) $ amplitude$ & 3.8\times10^{-3} & 4.2\times 10^6 \\
$tree level$ & 3.6\times 10^{-3} & 4.0\times 10^6 \\
O(p^4) $ without loops$ & 4.0\times 10^{-3} & 4.4\times 10^6 \\
\hline \ea$ \\*[.5cm] $\ba{|l|c|c|} \hline
\multicolumn{1}{|c|}{K^0_{\mu 3\gamma}}  &
$BR$ & \#$events/yr$ \\ \hline
$full $ O(p^4) $ amplitude$ & 5.6\times 10^{-4} & 6.1\times 10^5 \\
$tree level$ & 5.2\times 10^{-4} & 5.7\times 10^5 \\
O(p^4) $ without loops$ & 5.9\times 10^{-4} & 6.5\times 10^5 \\
\hline \ea$ \end{table}

\subsection{Improvements at DA$\Phi$NE}
The numerical results given above
demonstrate very clearly that the non-trivial
CHPT effects of $O(p^4)$ can be detected at DA$\Phi$NE in all four channels
without any problem of statistics. Of course, the rates are bigger for
the electronic modes. On the other hand, the relative size of the
structure dependent terms is somewhat bigger in the muonic channels
(around 8\% for the chosen cuts). We observe that there is negative
interference between the loop and counterterm amplitudes.

The sensitivity
 to the counterterm coupling constants $L_9, L_{10}$ and to the
chiral anomaly can be expressed as the difference in the number of events
between the tree level and the $O(p^4)$ amplitudes (without loops).
In the optimal case of $K^0_{e3\gamma}$, this amounts to more than
$4 \times 10^5$ events/yr at DA$\Phi$NE. Almost all of this difference is
due to $L_9$. It will be very difficult to extract the coupling constant
$L_{10}$ from the total rates. A more detailed study is needed to
determine whether $L_{10}$ can be extracted from differential
distributions.

The chiral anomaly is more important for $K^+_{l3\gamma}$, but even there
it influences the total rates rather little. Once again, a dedicated
study of differential rates is necessary to locate the chiral anomaly,
if possible at all.

On the other hand, taking
into account that $L_9$ is already known to good accuracy (see
table 1 in Ref.  \cite{Intro}), $K_{l3\gamma}$
decays will certainly allow for precise and unambiguous tests
of the one-loop effects in CHPT \cite{beg}.

\newpage
\setcounter{equation}{0}
\setcounter{subsection}{0}
\setcounter{so}{1}
\setcounter{table}{0}
\setcounter{figure}{0}
\clearpage

         \section{$K_{l4}$ decays}
         In this subsection we discuss the decays
         \bearr
          K^+(p) & \rightarrow & \pi^+(p_1) \;\pi^-(p_2) \;l^+(p_l) \; \nu_l
(p_{\nu})          \label{h1} \\
          K^+(p) & \rightarrow & \pi^0(p_1)\; \pi^0(p_2) \;l^+(p_l) \;\nu_l
(p_{\nu})         \label{k2}\\
          K^0(p) & \rightarrow & \pi^0(p_1)\; \pi^-(p_2)\; l^+(p_l) \;\nu_l
(p_{\nu})         \label{h3}
        \eearr
        and their charge conjugate modes. The letter $l$ stands for
        $e$ or $\mu$. We do
        not consider isospin violating contributions and
        correspondingly set $m_u = m_d$, $\alpha = 0$.

        \subsection{Kinematics}

        We start with the process (\ref{h1}).
       The full kinematics of this decay requires five variables.
       We will use the
       ones introduced by Cabibbo and Maksymowicz \cite{cabmak}. It is
       convenient to
       consider three reference frames, namely the $K^+$ rest system
       $(\Sigma_K)$, the
       $\pi^+
       \pi^-$ center-of-mass system $(\Sigma_{2 \pi})$ and the
       $l^+\nu_l$ center-of-mass system $(\Sigma_{l \nu})$. Then
       the variables are

       \begin{enumerate}
       \item $s_\pi$, the effective mass squared of the dipion system,

       \item $s_l$, the effective mass squared of the dilepton system,

       \item $\theta_\pi$, the angle of the $\pi^+$ in $\Sigma_{2\pi}$
       with respect to the dipion line of flight in $\Sigma_K$,

       \item $\theta_l$, the angle of the $l^+$ in $\Sigma_{l\nu}$
       with  respect to the dilepton line of flight in
       $\Sigma_K$, and

       \item $\phi$, the angle between the plane formed by the pions
       in
       $\Sigma_K$ and the corresponding plane formed by the dileptons.

       \end{enumerate}

       The angles $\theta_\pi$, $\theta_l$ and $\phi$ are displayed in
       Fig.
       \ref{figkin}. In order to specify these variables more
\begin{figure}[t]
\vspace{6cm}
\caption{Kinematic variables for $K_{l4}$ decays. The angle $\theta_\pi$ is
defined in $\Sigma_{2\pi}, \theta_l$ in $\Sigma_{l\nu}$ and $\phi$ in
$\Sigma_K$.
\label{figkin}
         }
\end{figure}
       precisely, let
       $\vec{p}_1$ be the three-momentum of the $\pi^+$ in
       $\Sigma_{2\pi}$ and
       $\vec{p}_l$ the three-momentum of the $l^+$ in
       $\Sigma_{l\nu}$.
       Furthermore, let $\vec{v}$ be a unit vector along the direction
       of flight of the dipion in
       $\Sigma_K$, and $\vec{c}\,(\vec{d}\,)$ a unit vector along
       the  projection
       of $\vec{p}_1(\vec{p_l})$ perpendicular to $\vec{v}(-\vec{v})$,
       \bearr
       \vec{c} & = & (\vec{p}_1 - \vec{v} \vec{v} \cdot \vec{p}_1) /
       [\vec{p}^{\;2}_1 - (\vec{p}_1 \cdot \vec{v})^2]^{1/2}
       \nonumber \\
       \vec{d} & = & (\vec{p}_l - \vec{v} \vec{v} \cdot \vec{p}_l) /
       [\vec{p}^{\;2}_l - (\vec{p}_l \cdot \vec{v})^2]^{1/2} \; \; .
       \nonumber
       \eearr
       The vectors  $\vec{v}$, $\vec{c}$ and $\vec{d}$
       are indicated in Fig. \ref{figkin}. Then, one has
      \bearr
      s_\pi & = & (p_1 + p_2)^2 \; \;, \;\; s_l = (p_l + p_\nu)^2
      \nonumber \\
      \cos \theta_\pi & = & \vec{v} \cdot \vec{p}_1 / \mid \vec{p}_1
      \mid,\; \;  \cos \theta_l = - \vec{v} \cdot \vec{p}_l / \mid \vec{p}_l
\mid       \nonumber \\
      \cos \phi & = & \vec{c} \cdot \vec{d}, \; \; \sin \phi = (\vec{c}
      \times \vec{v}) \cdot \vec{d}.
      \label{h5}
      \eearr
      The range of the variables is
      \bearr
      4 M^2_\pi & \leq & s_\pi \leq (M_K - m_l)^2
      \nonumber \\
      m^2_l & \leq & s_l \leq (M_K - \sqrt{s_\pi})^2
      \nonumber \\
      0 & \leq & \theta_\pi, \theta_l \leq \pi, 0 \leq \phi \leq 2
      \pi. \label{h6}
      \eearr
      It is useful to furthermore introduce the following combinations
      of four vectors
      \be
      P = p_1 + p_2, \; \; Q = p_1 - p_2 \;,\; L = p_l + p_\nu \; , \;
      N = p_l - p_\nu
      \label{k6a}
      \ee
      together with the corresponding Lorentz invariant scalar
      products
      \bearr
      P^2 & = & s_\pi, \; \; Q^2 = 4 M^2_\pi - s_\pi,\; \;  L^2 = s_l, \; \;N^2
      = 2 m_l^2 - s_l,
      \nonumber \\
      PQ & = & 0, \nonumber \\
      PL & =& \frac{1}{2} (M_K^2 - s_\pi - s_l),
      \nonumber \\
      PN & = & z_l PL + (1-z_l) X \cos \theta_l,
      \nonumber \\
      QL & = & \sigma_\pi X \cos \theta_\pi,
      \nonumber \\
      QN & = & z_l QL + \sigma_\pi (1-z_l) \left[ PL \cos \theta_\pi \cos
      \theta_l \right.
      \nonumber \\
    & & -    \left. (s_\pi s_l)^{1/2} \sin \theta_\pi\sin \theta_l \cos
      \phi \right]
      \nonumber \\
      LN & = & m^2_l
      \nonumber \\
      <LNPQ> & \equiv & \epsilon_{\mu \nu \rho \sigma} L^\mu N^\nu
      P^\rho Q^\sigma
      \nonumber \\
      & = & - (s_\pi s_l)^{1/2}
      \sigma_\pi (1-z_l) X \sin \theta_\pi \sin \theta_l \sin
      \phi
      \label{k7}
      \eearr
      with
      \bearr
      X & = & ((PL)^2 -s_\pi s_l)^{1/2} = \frac{1}{2} \lambda^{1/2} (M_K^2,
s_\pi, s_l)       \nonumber \\
      z_l & = & m^2_l /s_l
      \nonumber \\
      \sigma_\pi & = & (1-4 M^2_\pi/s_\pi)^{1/2}.
      \label{k8}
      \eearr
      Below we will also use the variables
      \bearr
      t & = & (p_1 -p)^2
      \nonumber \\
      u & = & (p_2 - p)^2.
      \label{k9}
      \eearr
      These are related to $s_\pi, s_l$ and $\theta_\pi$ by
      \bearr
      t  +  u &=& 2 M^2_\pi + M^2_K + s_l - s_\pi
      \nonumber \\
      t  -  u& =& - 2 \sigma_\pi X \cos \theta_\pi \; \; .
      \label{k10}
      \eearr

      \subsection{Matrix elements}

      The matrix element for $K^+ \rightarrow \pi^+ \pi^- l^+ \nu_l$
      is
      \be
      T = \frac{G_F}{\sqrt{2}} V^\star_{us} \bar{u} (p_\nu) \gamma_\mu
      (1-\gamma_5) v(p_l) (V^\mu - A^\mu)
      \label{k11}
      \ee
      where
      \bearr
      I_\mu & = & < \pi^+ (p_1) \pi^- (p_2) \mbox{out}\mid
      I_\mu^{4-i5} (0) \mid K^+ (p)  >;\; I = V,A
      \nonumber \\
      V_\mu & = & - \frac{H}{M^3_K} \epsilon_{\mu \nu \rho \sigma} L^\nu
      P^\rho Q^\sigma
      \nonumber \\
      A_\mu & = & -i\frac{1}{M_K} \left [ P_\mu F +
      Q_\mu G + L_\mu R \right ]
      \label{k12}
      \eearr
and $\epsilon_{0123}=1$.
 The matrix elements for the other channels
(\ref{k2},\ref{h3}) may be obtained from (\ref{k11},\ref{k12}) by isospin
symmetry, see below.

      The form factors $F,G,R$ and $H$ are real analytic functions of the three
      variables $p_1 p_2$, $p_1 p$ and $p_2 p$. Below,
      we will use instead the variables $\{ s_\pi, s_l, \theta_\pi\}$ or
      $\{s_\pi, t, u \}$.

{\underline{Remark:}} In order to agree with the notation used by Pais and
Treiman \cite{paistr} and by Rosselet et al. \cite{ross}, we have changed our
previous  convention \cite{bijnenskl4,riggen} in the definition of the anomaly
form factor $H$. See also the comments after Eq. (\ref{h18}).

      \subsection{Decay rates}

      The partial decay rate for (\ref{h1}) is given by

      \be
      d\Gamma = \frac{1}{2M_K(2\pi)^8} \sum_{spins} \mid
      T \mid^2 d_{LIPS} (p; p_l, p_\nu, p_1, p_2).
      \label{k13}
      \ee

The quantity $\sum_{spins} \mid T \mid ^2$ is a Lorentz invariant quadratic
form in $F,G,R$ and $H$. All scalar products can be expressed
 in the 5 independent variables $s_\pi,s_l,\theta_\pi,\theta_l$ and $\phi$,
 such that
\be
\sum_{spins} \mid T \mid ^2 = {2G_F^2 \mid V_{us} \mid ^2}{M_K^{-2}}
J_5(s_\pi,s_\l,\theta_\pi,\theta_l,\phi) \; \; .
\label{k13a}
\ee
Carrying out the integrations over the remaining $4 \cdot 3 - 5 = 7$ variables
in (\ref{k13})       gives \cite{cabmak}
 \be
      d \Gamma_5 = G^2_F \mid V_{us} \mid^2  N(s_\pi, s_l) J_5
      (s_\pi, s_l, \theta_\pi, \theta_l, \phi) ds_\pi ds_l d (\cos
      \theta_\pi) d(\cos \theta_l) d\phi
      \label{k14}
      \ee
      where
      \be
      N(s_\pi, s_l) = (1-z_l) \sigma_\pi X
      /(2^{13} \pi^6 M_K^5) \; \; .
      \label{h15}
      \ee
      The form factors $F,G,R$ and $H$ are independent of $\phi$ and
$\theta_l$. It is       therefore possible to carry out two more
      integrations in
      (\ref{k14}) with the result
      \be
      d\Gamma_3 = G_F^2 \mid V_{us} \mid^2  N(s_\pi, s_l)
      J_3 (s_\pi, s_l, \theta_\pi) ds_\pi ds_l d (\cos \theta_\pi).
      \label{k14a}
      \ee
      The explicit form of $J_5$ is
      \bearr
J_5 &=& |F|^2 \left[ (\pl)^2 - (\pn)^2 - \spi s_l + m_l^2 \spi \right]
\nonumber \\
&+& |G|^2 \left[ (\ql)^2 - (\qn)^2 - Q^2 s_l + m_l^2 Q^2 \right]
\nonumber \\
&+& |R|^2 \; m_l^2 \left[ s_l - m_l^2 \right] \frac{ }{ }
\nonumber \\
&+& \frac{1}{M_K^4} |H|^2 \left[ (m_l^2 - s_l)\left[
Q^2 X^2 +s_\pi (QL)^2 \right] -<LNPQ>^2 \right]
\nonumber \\
&+& (F^* G+F G^*)  \left[ (\pl)  (\ql)^{ } - (\pn) (\qn) \right]
\nonumber \\
&+& (F^*R+F R^*) \; m_l^2 \left[(\pl) -  (\pn) \right] \frac{ }{ }
\nonumber \\
&+& \frac{1}{M_K^2} (F^*H+FH^*) \left[ (\qn) (\pl)^2
- (\ql) (\pl) (\pn) -  \spi s_l(\qn)   + \mee \spi  (\ql) \right]
\nonumber \\
&+& (G^*R+G R^*) \; m_l^2 \left[ (\ql) - (\qn) \right]  \frac{ }{ }
\nonumber \\
&+& \frac{1}{M_K^2} (G^*H+GH^*) \left[  (\pl) (\ql) (\qn)
 -(\pn)  (\ql)^2 +  s_l (\pn)  \qq
- \mee (\pl) \qq \right]
\nonumber \\
&+&  \frac{i}{M_K^2}<LNPQ>\left[-(F^*G-FG^*) M_K^2
+ (F^*H-FH^*)  (\pn) \right.
\nonumber \\
&+&\left. (G^*H-GH^*)  (\qn)
+ (R^*H-RH^*)  m_l^2 \right]  \; \; .\nl
      \label{h15a}
      \eearr
      For data analysis it is useful to represent this result in a still
      different form which displays the $\theta_l$ and $\phi$
      dependence more clearly \cite{paistr}:
      \bearr
J_5 &=&2(1-z_l)\left[ I_1 + I_2 \cos 2 \theta_l + I_3 \sin^2 \theta_l \cdot
\cos 2
\phi + I_4 \sin 2 \theta_l \cdot \cos \phi + I_5 \sin \theta_l \cdot \cos \phi
\right. \nonumber \\
&+& \left. I_6 \cos \theta_l + I_7 \sin \theta_l \cdot \sin \phi + I_8 \sin 2
\theta_l \cdot \sin \phi + I_9 \sin^2 \theta_l \cdot \sin 2 \phi \right]\fs\nl
       \label{h16}
      \eearr
      One obtains
      \bearr
I_1 &=& \frac{1}{4}\left\{ (1 + z_l) |F_1|^2  + \frac{1}{2}
(3+z_l)\left(|F_2|^2 +
|F_3|^2 \right) \sin^2 \thp   + 2z_l  |F_4|^2 \right\}
\nonumber \\
I_2 &=& - \frac{1}{4} (1-z_l)\left\{ |F_1|^2 - \frac{1}{2}
\left( |F_2|^2 + |F_3|^2 \right)  \sin^2 \thp  \right\}
\nonumber \\
I_3 &=& - \frac{1}{4} (1-z_l) \left\{ |F_2|^2 - |F_3|^2 \right\}\sin^2 \thp
\nonumber \\
 I_4 &=& \frac{1}{2}(1-z_l)\mbox{ Re} (F_1^* F_2)  \sin \thp
\nonumber \\
I_5 &=& -\left\{ \mbox{ Re} (F_1^* F_3) +
z_l  \mbox{ Re} (F_4^* F_2) \right\}  \sin \thp
\nonumber \\
I_6 &=&
- \left\{\mbox{ Re} (F_2^* F_3)\sin^2 \thp - z_l \mbox{ Re} (F_1^* F_4)\right\}
\nonumber \\
I_7 &=& - \left\{ \mbox{ Im} (F_1^* F_2) +
z_l  \mbox{ Im} (F_4^* F_3) \right\} \sin \thp
\nonumber \\
I_8 &=& \frac{1}{2} (1-z_l) \mbox{ Im} (F_1^* F_3) \sin \thp
\nonumber \\
I_9 &=& -\frac{1}{2}(1-z_l) \mbox{ Im} (F_2^* F_3) \sin^2 \thp  \;,
      \label{h17}
      \eearr
      where
      \bearr
 F_1& =& X \cdot F + \sigma_\pi (P L) \cos \thp \cdot G
\nonumber \\
 F_2& =& \sigma_\pi \left( s_\pi s_l \right)^{1/2} G
\nonumber \\
 F_3& =&  \sigma_\pi X \left( s_\pi s_l \right)^{1/2}  \frac{H}{M_K^2}
\nonumber \\
 F_4& =&- (P L)  F - s_l R - \sigma_\pi X \cos \thp \cdot G \; \; .
      \label{h18}
      \eearr
      The definition of $F_1, \ldots, F_4$ in (\ref{h18}) corresponds
      to the combinations used by Pais and Treiman \cite{paistr} (the
      different sign in the terms $\sim PL$
 is due to our use of the  metric
      $\mbox{diag} (+ ---)$). The form factors $I_1, \ldots, I_9$
      agree with the expressions given in \cite{paistr}. We conclude that our
convention for the relative phase in the definition of the form factors
 in Eq. (\ref{k12}) agrees with the one used by Pais and Treiman.
 The comparison of
(\ref{h15a}) with
\cite[table II]{ross} shows furthermore that it also agrees with
this reference.

      The quantity $J_3$ can now easily be obtained from (\ref{h16})
      by integrating over $\phi$ and $\theta_l$,
      \be
      J_3 = \int d\phi \; d({\mbox{cos}}\theta_l) J_5 =8 \pi (1 - z_l) \left [
I_1 - \frac{1}{3} I_2 \right].       \label{h19}
      \ee

      \subsection{Isospin decomposition}

      The $K_{l4}$ decays (\ref{k2}) and (\ref{h3}) involve the same
      form factors as displayed in Eq. (\ref{k12}). We denote by
      $A_{+-}$, $A_{00}$ and $A_{0-}$ the current matrix elements of
      the processes (\ref{h1})-(\ref{h3}). These are related by
      isospin symmetry \footnote{We use the  Condon-Shortley phase
conventions.},       \be
      A_{+-} =\frac{A_{0-}}{\sqrt{2}} - A_{00} \; \; .
      \label{i1}
      \ee
      This relation also holds for the individual form factors, which
      may be decomposed into a symmetric and an antisymmetric part under
      $t \leftrightarrow u \; (p_1 \leftrightarrow p_2)$. Because of Bose
      symmetry and of the $\Delta I = \frac{1}{2}$ rule of the
      relevant  weak currents, one has
      \bearr
      (F,G,R,H)_{00} & = & - (F^+, G^-, R^+, H^-)_{+-}
      \nonumber \\
      (F,G,R,H)_{0-} & = & \sqrt{2} (F^-, G^+, R^-, H^
      +)_{+-}
      \label{i2}
      \eearr
      where
      \be
      F^\pm_{+-} = \frac{1}{2} [F(s_\pi, t, u) \pm F(s_\pi, u, t)]
      \label{i3}
      \ee
      and $F(s_\pi, t, u)$ is defined in Eq. (\ref{k12}).

The isospin relation for the decay rates is
\be
\Gamma(K^+\rightarrow \pi^+\pi^-l^+\nu_l)=\frac{1}{2}\Gamma(K_L\rightarrow
\pi^0\pi^\pm l^\mp\nu) +2 \Gamma(K^+\rightarrow \pi^0\pi^0l^+\nu_l) \; \; .
\label{iso}
\ee
Isospin violating contributions affect the matrix elements and phase
space, as a result of which this relation  is modified. In
order to illustrate the (substantial) effects from asymmetries in
phase space, we take
 constant form factors
$F,G$ and set $R=0,H=0$. Eq.  (\ref{iso}) then reads (with
physical
masses for $K^+\rightarrow \pi^+\pi^-l^+\nu_l, \pi^0\pi^0 l^+\nu_l$ and with
$M_{\pi^0}=M_{\pi^\pm}=137$ MeV in  $K_L\rightarrow \pi^0\pi^\mp l^\pm \nu$)
\bearr
(16.0 F^2 +3.1 G^2)\Gamma_0&=& (20.1F^2 +2.0 G^2)\Gamma_0
\nonumber \\
\Gamma_0 &=& V_{us}^2 \cdot 10^2 {\mbox{sec}}^{-1}
\eearr
in the electron mode and
\bearr
(1.79 F^2 +0.25 G^2)\Gamma_0 &=& (2.64 F^2 +0.20 G^2)\Gamma_0
\eearr
in the muon mode.

      \subsection{Partial wave expansion}

      The form factors may be written in a partial wave expansion in
      the variable $\theta_\pi$.
      We consider a definite isospin $\pi \pi$ state. Suppressing isospin
indices, one has \cite{berends,cecile}
      \bearr
      F &= & \sum^{\infty}_{l=0} P_l (\cos \theta_\pi) f_l - \frac{
      \sigma_\pi PL}{X} \cos
      \theta_\pi G
      \nonumber\\
      G & = & \sum^{\infty}_{l=1} P_l' (\cos \theta_\pi) g_l
      \nonumber \\
      R & = & \sum^{\infty}_{l=0} P_l (\cos \theta_\pi) r_l +
      \frac{ \sigma_\pi s_\pi}{X}
      \cos \theta_\pi G
      \nonumber \\
      H & = & \sum^{\infty}_{l= 0} P_l' (\cos \theta_\pi) h_l
      \label{i4}
      \eearr
      where
      \be
      P_l'(z) = \frac{d}{dz} P_l(z) \; \; .
      \label{i10}
      \ee

      The partial wave amplitudes $f_l, g_l, r_l$ and $h_l$ depend on
      $s_\pi$ and $s_l$. Their phase coincides with the phase shifts
      $\delta^I_l$ in elastic $\pi \pi$ scattering (angular momentum
      $l$, isospin $I$).
      More precisely, the quantities
      \bearr
     && e^{-i \delta^0_{2l}} X_{2l}
      \nonumber \\
     && e^{-i\delta^1_{2l+1}} X_{2l+1}\; ; \;l=0,1, \ldots \; ;
      \; X = f,g,r,h
      \label{i11}
      \eearr
      are real in the physical region of $K_{l4}$ decay. The form
      factors $F_1$ and $F_4$ therefore have a simple expansion
      \bearr
      F_1 &=& X \sum_{l} P_l
      ( \cos \theta_\pi) f_l
      \nonumber \\
      F_4& =& - \sum_{l} P_l (\cos \theta_\pi) (PL f_l + s_l r_l).
      \label{i12}
      \eearr

      On the other hand, the phase of the projected amplitudes
      \be
      F_{2l} = \int P_l (\cos \theta_\pi) F_2 d (\cos \theta_\pi) \; ;
      \; l=0,1,2,\ldots
      \label{i13}
      \ee
      is not given by $\delta_l^I$, e.g., $e^{-i\delta^1_1} F_{20}$ is
      \underline{not} real in the isospin one case. A similar remark applies to
$F_3$.

      \subsection{Previous experiments}

\begin{table}[t]
\protect
\begin{center}
\caption{\label{kl441} Rates of $K_{l4}$ decays \protect\cite{pdg}. The data
for ${K_L} \rightarrow \pi^0 \pi^{\pm} e^{\mp} \nu $
 is from ref.
\protect\cite{kld}.}

\vspace{1em}
\begin{tabular}{|c|c|c|c|c|} \hline
\multicolumn{2}{|c|}{} & \multicolumn{2}{c|}{$\sharp$ events} & \\ \hline
  & branching  & Particle Data
  & DA$\Phi$NE  & improve-                  \\
  & ratio & Group
  & 1 yr & ment \\ \hline
  &&&& \\
$ K^+ \rightarrow \pi^+\pi^- e^+ \nu_e $ &
 $3.91 \cdot 10^{-5} $ & $ 3 \cdot 10^4 $ & $3\cdot 10^5$ & $10$      \\
$ K^+ \rightarrow \pi^0\pi^0 e^+ \nu_e $ &
 $ 2.1 \cdot 10^{-5} $ & $ <50 $ & $2\cdot 10^5$ & $>4\cdot 10^3$      \\
${K_L} \rightarrow \pi^0 \pi^{\pm} e^{\mp} \nu $ &
$5.16 \cdot 10^{-5}$  & $ 729$  & $ 7\cdot 10^4$ &
$7 \cdot 10^2$      \\
$K^+ \rightarrow \pi^+ \pi^- \mu^+ \nu_{\mu}$&
 $10^{-6}$      & $7$   & $2 \cdot 10^4$& $3 \cdot 10^3$ \\
 &&&& \\ \hline
\end{tabular}
\end{center}
\end{table}

      We display in table \ref{kl441} the number of events collected
      so far.
      The data are obviously dominated by the work of Rosselet et al.
      \cite{ross}, which measures the $\pi^+\pi^-$ final state with
      good statistics. The authors parametrize the form factors as
      \bearr
      F & = & f_s e^{i\delta^0_0} + f_p e^{i \delta^1_1} \cos \theta_\pi
      + \mbox{D-wave}
      \nonumber \\
      G & = & g e^{i\delta^1_1} + \mbox{D-wave}
      \nonumber \\
      H & = & h e^{i\delta^1_1} + \mbox{D-wave}
      \label{e1}
      \eearr
      with $f_s, f_p, g$ and $h$ assumed to be real
      \footnote{Note that, according to what is said in the previous
subsection, the terms denoted
      by ''D-wave'' in Eq. (\ref{e1}) all contain (complex) contributions which
      are proportional to $P_l (\cos \theta_\pi), l \geq 0$.
                }
              .
      Furthermore, they put $m_e = 0$, such that
      the form factors $R$ and $F_4$ drop out in the decay distribution.
Despite       the good statistics, the experiment has not been able to
      separate out the full kinematic behaviour of the matrix
      elements. Therefore certain approximations/assumptions had to
      be made. For example, no dependence on $s_l$ was seen within the
      limits of the data, so that the results were quoted assuming
      that such a dependence is absent. Similarly, $f_p$ was found to
      be compatible with zero, and hence put equal to zero when the
      final result for $g$ was derived. A dependence on $s_\pi$ was
      seen, and found to be compatible with
      \bearr
      f_s (q^2) & = & f_s(0) [1+ \lambda_f q^2]
      \nonumber\\
      g (q^2) & = & g(0) [1+ \lambda_g q^2]
      \nonumber \\
      h (q^2) & = & h(0) [1+ \lambda_h q^2] \nonumber\\
      q^2 & = &  (s_\pi-4 M_\pi^2)/4M_\pi^2
      \label{e2}
      \eearr
      with
      \be
      \lambda_f = \lambda_g = \lambda_h = \lambda.
      \label{e3}
      \ee

      These approximations to the form factors do not agree completely
      with what is found in the theoretical predictions. Dependence on
      $s_l$ and non-zero values for higher partial waves all occur in
      the theoretical results.

      The experimental results for the threshold values and the slopes
      of the form factors are \cite{ross}
      \bearr
      f_s(0) & = & 5.59 \pm 0.14
      \nonumber \\
      g(0) & = & 4.77 \pm 0.27
      \nonumber\\
      h(0) & = & - 2.68 \pm 0.68
      \nonumber \\
      \lambda & = & 0.08 \pm 0.02.
      \label{e4}
      \eearr
      We have used \cite{pdg} $\mid V_{us} \mid = 0.22$ in
      transcribing these results. (We note that from Eqs. (\ref{e1} - \ref{e4})
and $f_p=0$ we
obtain  $\Gamma_{K_{e4}} = (2.94 \pm 0.16)\cdot 10^3$ sec$^{-1}$. This value
must
be compared with $\Gamma_{K_{e4}} = (3.26 \pm 0.15)\cdot 10^3$ sec$^{-1}$
obtained in the same experiment.)  In addition to the threshold
values       (\ref{e4}) of the form factors, the phase shift difference
      $\delta = \delta^0_0 - \delta^1_1$ was determined \cite{ross} in
      five energy bins. The S-wave scattering length $a^0_0$ was then
      extracted by using a model of Basdevant, Froggatt and Petersen
      \cite{bafropet}. This model is based on solutions to Roy
      equations. The result for the scattering length is
      \be
      a^0_0 = 0.28 \pm 0.05.
      \label{e5}
      \ee
      A study by \cite{pipet}, based on a more recent solution to Roy
      equations, gives
      \be
      a^0_0 = 0.26 \pm 0.05.
      \label{e6}
      \ee

Turning now to the $\pi^0 \pi^{\pm} e^{\mp} \nu_e$ channel, we
consider the following recent data (based on $729\pm15$ events) \cite{kld}:
\be
BR(K_L\rightarrow\pi^0\pi^\mp e^\pm \nu) =
(5.16 \pm 0.2 \pm 0.22) \cdot 10^{-5} \per
\label{k51}
\ee
The group  also measured the $G$ form factor. Defining
\be
 G_{0-} = G_0(1+\lambda_g q^2)e^{i\delta^1_1} \co
\ee
they find
\bearr
G_0&=&7.8\pm0.7\pm0.2  \co \nn
\lambda_g&=&0.014\pm0.087\pm0.070 \per
\eearr
The slope agrees within the errors with the value (\ref{e4}) found by
Rosselet et al.  \cite{ross}. To compare the value of the form factor at
threshold, we use the isospin
prediction
 \be
 |g(0)|=|G_{0}|/\sqrt{2}  = 5.5 \pm 0.5 \co
\ee
which is not incompatible with $g(0)=4.77 \pm 0.27$ in eq. (\ref{e4}).
 (Here
we have used
$|V_{us}|=0.22$ to transcribe the data. Furthermore, we assume that the form
factor $G_{0-}$ measured in Ref. \cite{kld} indeed has to be divided by
$\sqrt{2}$ for the comparison with \cite{ross}. This is not quite clear to
us reading \cite{kld}.)

Finally for the channel $\pi^0 \pi^0 e^+ \nu_e$, we consider the
rate \cite{pdg}
\bearr\label{k50}
 \Gamma_{K^+\rightarrow\pi^0\pi^0e^+\nu_e} &=&\left(
1.70^{+0.34}_{-0.29}\right) \cdot 10^3 \mbox{sec}^{-1} \;\; \;\per
\eearr
The kinematic dependence of the form factors on
the variables $s_\pi,s_l$ and $\theta_\pi$ has not yet been resolved
experimentally in this decay. In order to proceed, we assume that the
form factors in this channel are independent of $\theta_\pi$, e.g.,
$F_{00}=F_{00}(s_\pi,t+u)$ etc.
As a result of this assumption, $G_{00}$ and $H_{00}$ vanish
by Bose statistics.  The contribution from $R_{00}$ is completely
negligible in the electron mode, and the contribution from the anomaly form
factor to the decay (\ref{k50}) is tiny. We neglect it altogether,
as a result of which the above decay is fully determined by $F_{00}$.  We write
\be
F_{00}=F_0(1+\lambda q^2)e^{i\delta^0_0} \co
\label{k52}
\ee
and obtain for the rate
\be
2\Gamma_{K^+\rightarrow\pi^0\pi^0e^+\nu_e} = |F_0 V_{us}|^2(
2.01 +1.7 \lambda +O(\lambda^2)) \cdot 10^3 {\mbox{sec}^{-1}}
\label{k53} \\ \per
\ee
This finally gives with
$\lambda=0.08$
 \be
|F_0| =  5.72^{+0.57}_{-0.49}\co
\ee
which compares very well with the isospin prediction
\be
|F_0| =  |f_s(0)| = 5.59 \pm 0.14\fs
\ee

      \subsection{Theory}

      The theoretical predictions of $K_{l4}$ form factors have a long
      history which started in the sixties with the current algebra
      evaluation of $F$, $G$, $R$ and $H$. For an early review of the
      subject and for references to work prior to CHPT we refer the
      reader to \cite{chounet} (see also \cite{shabkl4}).
      Here we concentrate
on the evaluation of the form factors in the framework of CHPT
      \cite{bijnenskl4,riggen,stern,BCG}.

\vspace{.5cm}

\begin{subsub}{\bf
Form factors at tree level}
\end{subsub}

The chiral representation of the form factors at leading order  was
originally given by Weinberg \cite{weinkl4},
\bearr
F&=&G=\frac{M_K}{\sqrt{2}F_\pi}=3.74 \co \nonumber \\
R&=& \frac{M_K}{2\sqrt{2}F_\pi}\left(\frac{s_\pi+\nu}{s_l-M_K^2} +1
\right) \co \nonumber \\ H&=&0 \per
\label{t1}
\eearr
 The next-to-leading order corrections  are displayed
 below, and the later sections contain an estimate of yet higher-order
contributions. Here we note that the
 total decay rates which follow from Eq. (\ref{t1}) are typically
a factor of two
(ore  more) below the data. As an example, consider the channel $K^+\rightarrow
\pi^+\pi^-e^+\nu_e$. Using (\ref{t1}), the total decay rate
becomes\footnote{If not stated otherwise, we use $F_\pi=93.2$ MeV,
$|V_{us}|=0.22$ and
$(M_\pi,M_K)=(139.6,493.6)$ MeV, $(135,493.6)$ MeV and $(137,497.7)$ MeV for
the decays (\ref{h1}), (\ref{k2}) and (\ref{h3}), respectively.} 1297
sec$^{-1}$,
whereas the experimental value is 3160$\pm$140 sec$^{-1}$ \cite{pdg}.

\vspace{.5cm}

\begin{subsub}{\bf
Form factors at one loop}
\end{subsub}

 The one-loop result for $F$ \cite{bijnenskl4},\cite{riggen} may be
written in the form       \be
      F(s_\pi, t, u)  =  \frac{M_K}{\sqrt{2} F_\pi} \left\{ 1 +
      \frac{1}{F_\pi^2}(U_F +P_F +C_F) +O(E^4)\right\} \per
      \label{T2}
      \ee
      The contribution
      $U_F(s_\pi, t, u)$ denotes the unitarity correction generated
      by the one-loop graphs which appear at order $E^4$ in the low-energy
expansion. Its expression will be given in appendix \ref{UFGR}.

      The contribution $P_F(s_\pi, t, u)$ is a polynomial in
      $s_\pi, t, u$ obtained from the tree graphs at order $E^4$. We
      find
      \be
      P_F (s_\pi, t, u) =  \sum^{9}_{i=1}
      p_{i,F} (s_\pi, t, u) L^r_i \co
      \label{T6}
      \ee
      where
      \bearr
      p_{1,F} & = & 32 (s_\pi - 2M^2_\pi ) \co
      \nonumber \\
      p_{2,F} & = &8 (M_K^2 + s_\pi
      - s_l) \co
      \nonumber \\
      p_{3,F} & = & 4(M^2_K -3M_\pi^2 +2 s_\pi -t) \co
      \nonumber \\
      p_{4,F} & = & 32 M^2_\pi \co
      \nonumber \\
      p_{5,F} & = & 4 M^2_\pi \co
      \nonumber \\
      p_{9,F} & = & 2 s_l \per
      \label{T7}
      \eearr
      The remaining coefficients $p_{i,F}$ are zero.

      The contributions $C_F$  contain
      logarithmic terms, independent of $s_\pi, t$ and $u$:
      \bearr
      C_F & = & \frac{1}{256 \pi^2} \left[ 5 M^2_\pi \ln
      \frac{M^2_\pi}{\mu^2} - 2 M^2_K \ln \frac{M^2_K}{\mu^2} - 3
      M^2_\eta \ln \frac{M^2_\eta}{\mu^2} \right] \per
      \label{T8}
      \eearr

      The corresponding decomposition of the form factor $G$
      is \cite{bijnenskl4},\cite{riggen}
      \bearr
      G(s_\pi, t, u)  &=&  \frac{M_K}{\sqrt{2} F_\pi} \left\{ 1 +
      \frac{1}{F_\pi^2}(U_G +P_G +C_G) +O(E^4) \right\}\co\nn
      \eearr
      For the expression of $U_G$ see appendix \ref{UFGR}.
      The polynomials
      \be
      P_G =  \sum^{9}_{i=1} p_{i,G} (s_\pi,
      t, u) L^r_i
      \label{T13}
      \ee
      are
      \bearr
      p_{2,G} & = & 8(t - u) \co
\nonumber \\
      p_{3,G} & = & 4(t-M_K^2 -M_\pi^2) \co
      \nonumber \\
      p_{5,G} & = & 4M^2_\pi \co
      \nonumber \\
      p_{9,G} & = &  2 s_l \co
      \nonumber \\
      \label{T14}
      \eearr
      The remaining $p_{i,G}$ vanish. The logarithms contained in
      $C_G$ are
      \be
      C_G = - C_F.
      \label{T15}
      \ee

The form factor $R$ contains a pole part $Z(s_\pi,t,u)/(s_l-M_K^2)$ and a
regular piece $Q$. [Since the axial current acts as an interpolating
field for a kaon, the residue of the pole part is related to
the $KK\rightarrow \pi\pi$ amplitude in the standard manner.] We write
 \bearr\label{formr}
 R &=& \frac{M_K}{2\sqrt{2}F_\pi}\left\{\frac{Z}{s_l-M_K^2} +Q +O(E^4)\right\}
\co
\nonumber
\\ I &=& B_I +\frac{1}{F_\pi^2}(U_I +P_I +C_I) \; , \; \; I=Z,Q \; \; .
\eearr
According to (\ref{t1}), the Born terms $B_I$ are \cite{weinkl4}
\bearr
B_Z &=& s_\pi +\nu \nonumber \co \\
B_Q&=&1.
\eearr
The one-loop corrections have been worked out in Ref. \cite{BCG}.
The unitarity corrections $U_I$ are displayed in the appendix \ref{UFGR}.
The residues $P_Z$ and $C_Z$ are

\bearr
      P_Z (s_\pi, t, u) &=&  \sum^{9}_{i=1}
      p_{i,Z} (s_\pi, t, u) L^r_i \co
\eearr
with
\bearr
p_{1,Z} &=& 32 (s_\pi -2M_K^2)(s_\pi -2M_\pi^2) \co
\nonumber \\
p_{2,Z}&=& 8 (s_\pi^2 +\nu^2) \co
\nonumber \\
p_{3,Z}&=&-2\left[2(\nu +4\Sigma)s_\pi -5 s_\pi^2 -\nu^2
-16M_K^2M_\pi^2\right] \co \nonumber \\
p_{4,Z}&=& 32 \left[ \Sigma s_\pi -4M_K^2M_\pi^2\right] \co
\nonumber \\
p_{5,Z}&=&4\left[(s_\pi+\nu)\Sigma -8M_K^2M_\pi^2\right] \co
\nonumber \\
p_{6,Z} &=& 128 M_K^2M_\pi^2 \co
\nonumber \\
p_{8,Z} &=&  64 M_K^2M_\pi^2 \co\nl
\Sigma&=&M_K^2+M_\pi^2\fs
\eearr
The remaining $p_{i,Z}$ vanish. The logarithms in $C_Z$ are
\bearr
C_Z&=& -\frac{M_K^2 -M_\pi^2}{128 \pi^2} \left[ 3 M^2_\pi \ln
      \frac{M^2_\pi}{\mu^2} - 2 M^2_K \ln \frac{M^2_K}{\mu^2} -
      M^2_\eta \ln \frac{M^2_\eta}{\mu^2} \right]
 \; \; .
\eearr
For the nonpole part Q we find:

\bearr
      P_Q (s_\pi, t, u) &=&  \sum^{9}_{i=1}
      p_{i,Q} (s_\pi, t, u) L^r_i \co
\eearr
with
\bearr
p_{1,Q}&=& 32(s_\pi -2M_\pi^2) \co
\nonumber \\
p_{2,Q}&=& 8(M_K^2 -s_l) \co
\nonumber \\
p_{3,Q}&=&2\left[ 4(s_\pi -2M_\pi^2) +M_K^2-s_l\right] \co
\nonumber \\
p_{4,Q}&=& 32M_\pi^2 \co
\nonumber \\
p_{5,Q}&=& 4 \Sigma \co
\nonumber \\
p_{9,Q}&=& 2\left[(s_\pi +\nu) -(M_K^2 -s_l)\right]\per
\eearr
The remaining $p_{i,Q}$ vanish. The logarithms in $C_Q$ are
\bearr
C_Q&=&   \frac{1}{128 \pi^2} \left[ 5 M^2_\pi \ln
      \frac{M^2_\pi}{\mu^2} - 2 M^2_K \ln \frac{M^2_K}{\mu^2} - 3
      M^2_\eta \ln \frac{M^2_\eta}{\mu^2} \right]
 \; \; .
\eearr

The first nonvanishing contribution in the chiral expansion of the form factor
$H$ is   due
to the chiral anomaly \cite{wessz}.
 The prediction is  \cite{wesszkl4}
 \be
      H = -\frac{\sqrt{2} M^3_K}{8 \pi^2 F^3_\pi} = -2.66 \co
      \label{T16}
      \ee
      in excellent agreement with the experimental value \cite{ross} $H=-2.68
\pm 0.68$. The
next-to-leading order corrections to $H$ have also been calculated
\cite{llamet}.
 If the new low-energy parameters are
estimated using the vector mesons only, these corrections are small.

      The results for $F,G$ and $R$ must satisfy two nontrivial
      constraints: i) Unitarity requires that $F,G$ and $R$ contain, in
      the physical region $4M^2_\pi \leq s_\pi \leq (M_K-m_l)^2$, imaginary
      parts governed by $S$- and $P$-wave $\pi \pi$ scattering [these
      imaginary parts are contained in the functions
      $\Delta_0(s_\pi), \Delta_1(s_\pi)]$. ii) The scale dependence
      of the low-energy constants $L^r_i$ must be compensated  by
      the scale dependence of $U_{F,G,Z,Q}$ and $C_{F,G,Z,Q}$ for all values
      of
      $s_\pi, t, u, M^2_\pi, M^2_K$. [Since we work at order $E^4$ in the
chiral expansion, the
      meson masses appearing in the above expressions satisfy the
      Gell-Mann-Okubo mass formula.] We have checked that these
      constraints are satisfied.

Because the one-loop contributions are rather large, one expects still
substantial corrections from higher orders. In the following
section, we therefore
first estimate the effects from higher orders in
the chiral expansion, using then this improved representation for the form
factors in a comparison with the data.

\vspace{.5cm}

\begin{subsub}{\bf
Form factors beyond one loop}
\end{subsub}

To investigate the importance of higher-order terms, we employ the method
developed in Ref. \cite{dgl}. It amounts to writing a dispersive
representation of the partial wave amplitudes, fixing the subtraction
constants using chiral perturbation theory. Here, we estimate the
higher-order terms in the $S$-wave projection of the amplitude $F_1$,
\be
f(s_\pi,s_l) = (4\pi X)^{-1}\int d\Omega F_1(s_\pi,t,s_l)\co
\label{un2}
\ee
 because
this form factor plays a decisive role in the determination of $L_1^r,L_2^r$
and $L_3$, and it is influenced by $S$-wave $\pi\pi$ scattering which is known
 \cite{truongswave} to produce substantial corrections.

\vspace{.5cm}

\begin{subsub}{\bf
Analytic properties of partial waves}
\end{subsub}

Only the crossing-even part
\be
F_1^+=XF^+ +\sigma_\pi (PL)\cos \thp \cdot G^-
\ee
contributes in the projection (\ref{un2}).
 The partial wave $f$ has the following analytic properties:
\begin{enumerate}
\item
At fixed $s_l$, it is analytic in the complex $\spi$-plane, cut along the real
axis for Re $\spi \geq 4M_\pi^2$ and  Re $\spi \leq 0$.
\item
In the interval $0\leq s_\pi \leq 4M_\pi^2$, it is real.
\item
In $ 4M_\pi^2\leq s_\pi \leq 16 M_\pi^2$, its phase  coincides with the
isospin zero $S$-wave phase $\delta_0^0$ in elastic $\pi\pi$ scattering,
\be
f_+ = e^{2i\delta_0^0}f_- \co f_\pm = f(s_\pi \pm i\epsilon,s_l).
\label{un3a}
\ee
\end{enumerate}
The proof of these properties is standard \cite{spearman}. Here we only note
that the presence of the cut for $s_\pi \leq 0$ follows from the relations
\bearr
t=M_\pi^2 +\frac{M_K^2 +s_l - s_\pi}{2} -\sigma_\pi X \cos \theta _\pi \co \nn
\label{un4}
t(\cos \theta_\pi=-1,s_\pi < 0) \geq (M_K+M_\pi)^2.
\label{bo1}
\eearr
Since $F^+$ and $G^-$ have cuts at $t\geq (M_K+M_\pi)^2$,
 the claim is proven.

\vspace{.5cm}

\begin{subsub}{\bf
Unitarization}
\end{subsub}

We introduce the Omn\`{e}s function
\be \label{e100}
\Omega(s_\pi) = \exp \left[ \frac{\spi}{\pi}\int_{4M_\pi^2}^{\Lambda^2}
\frac{ds}{s}\frac{\delta_0^0(s)}{s-\spi}\right] \co
\ee
where $\Lambda$ will be chosen of the order of 1 GeV below.
According to (\ref{un3a}), multiplication  by $\Omega^{-1}$
removes the cut in $f$ for  $4M_\pi^2 \leq \spi\leq 16 M_\pi^2$.
Consider now
\be
f=f_L+f_R \co
\ee
where $f_L(f_R)$ has only the left-hand (right-hand) cut, and introduce
\be
v=\Omega^{-1} (f -f_L) \; .
\ee
Then $v$ has only a right-hand cut, and we may represent it in a dispersive
way,
\be
v=v_0 +v_1 \spi+\frac{\spi^2}{\pi} \int_{4M_\pi^2}^\infty
\frac{ds}{s^2}\frac{{\mbox{Im}}\Omega^{-1}(f-f_L)}{s-\spi} \; .
\ee
We expect the contributions from the integral beyond 1 GeV$^2$ to be small.
Furthermore, $\Omega^{-1}f$ is approximately real between $16M_\pi^2$ and
1 GeV$^2$, as a result of which one has
\be
v = v_0 + v_1 \spi -\frac{\spi^2}{\pi}\int_{4M_\pi^2}^{\Lambda^2}
 \frac{ds}{s^2}\frac{f_L{\mbox{Im}}\Omega^{-1}}{s-\spi} \per
\ee
For given $v_0,v_1,f_L$ and $\Omega$, the form factor $f$ is finally obtained
from
\be
f=f_L + \Omega v \per
\label{un5}
\ee
The behaviour of $f_L$ at  $\spi \rightarrow 0$ is
governed by the large-$|t|$ behaviour of $F^+$ and $G^-$, see (\ref{bo1}).
Therefore, instead
of using CHPT to model $f_L$, we approximate the left-hand cut by resonance
exchange. To pin down the
subtraction constants $v_0$ and
$v_1$, we require that the threshold expansion of $f$ and $f_\CHPT$
agree
up to and including terms of order $O(E^2)$. For a specific choice of $f_L$,
this fixes ${v_0,v_1}$ in terms of the quantities which occur in the one-loop
representation of $F^+$ and $G^-$.
In the case where $f_L=0$, $f$ has then a particularly simple form at $s_l=0$,
\bearr
f(s_\pi,s_l=0)|_{f_L=0} &=& \Omega(v_0 + v_1 s_{\pi}) \co \nn
v_0 &=& \frac{M_K}{\sqrt{2} F_\pi}\biggr\{ 1.05 +\frac{1}{F_\pi^2}
\biggr[-64 M_\pi^2 L^r_1 + 8 M_K^2 L^r_2 \nn
  & &  +2(M_K^2-8M_\pi^2)L_3
  +\frac{2}{3}(M_K^2-4M_\pi^2)(4L^r_2+L_3)\biggr]\biggr\} \co \nn
v_1 &=& \frac{M_K}{\sqrt{2} F_\pi}\biggr\{0.38+\frac{1}{F_\pi^2}
 \biggr[32L^r_1+8L^r_2+10L_3 \nn
 & & \qquad \qquad -\frac{2}{3}\frac{M_K^2-4M_\pi^2}{4M_\pi^2}(4L^r_2+L_3)
\biggr]\biggr\}\per
\label{un6}
\eearr
The details of the evaluation of $f_L,v_0$ and of $v_1$ may be found in Ref.
\cite{BCG}.

In the partial wave $f$, the effects of the final-state interactions are
substantial, because they are related to the $I=0,S$-wave $\pi \pi$ phase
shift. On the other hand, for the leading partial wave in
$G^+=g e^{{i\delta_p}}+\cdots$, these effects are reduced, because the phase
$\delta_p$ is small at low energies. We find it more difficult to assess an
estimate for the higher-order corrections in this case -- we come back to this
point in the following sections.

We add a remark concerning the choice of the subtraction point in the
Omn\`{e}s function (\ref{e100}). To remove the right-hand cut in $f$, the
modified Omn\`{e}s function
\bea
\Omega(s_\pi,s_0)=\exp{\frac{(s_\pi-s_0)}{\pi}\int_{4M_\pi^2}^{\Lambda^2}
\frac{ds}{s-s_0}\frac{\delta_0^0(s)}{s-s\pi}}\nonumber
\eea
would do as well, with any value of $s_0 \leq 4M_\pi^2$. To illustrate the
meaning of $s_0$, we consider for simplicity the form factor $f$
 in the case where $f_L=0$. The choice $s_0=4M_\pi^2$ then amounts
to the statement that, at threshold,  there are no contributions from two loops
and beyond by fiat. We
consider this to be rather unlikely. (For a different opinion see
\cite{omnesstern}.) On the other hand, we have checked that our results do not
vary significantly if $s_0$ is taken of the order of the pion mass
squared, $|s_0| \leq M_\pi^2$.

\subsection{Determination of $L_1,L_2$ and $L_3$}
To illustrate the usefulness of forthcoming more accurate $K_{l4}$ data, we
determine here the low-energy constants $L^r_1,L^r_2$
and $L_3$ from  data on $K^+\rightarrow \pi^+\pi^-e^+\nu_e$
decays and on $\pi\pi \rightarrow \pi\pi$ threshold parameters, using the
improved $S$-wave amplitude $f$ set up above.
For a comparison with earlier work \cite{riggen} we refer the reader
to Ref. \cite{BCG}.

We perform  fits
 to $f_s(0),\lambda_f,g(0)$ as given in (\ref{e4}) and to
the $\pi\pi$ threshold
parameters listed in table \ref{fits}. We introduce for this purpose the
quantities
 \bearr
\label{proj} \fff(s_{\pi},s_l) &=&\left| (4\pi X)^{-1} \int
d\Omega F_1(s_{\pi},t,
s_l) \right|=\left|f(s_\pi,s_l)\right|  \co \nn
\ffg(s_{\pi},s_l) &=& \left| \frac{3}{8 \pi} \int d\Omega \sin^2\thp
G(s_{\pi},t,s_l) \right|\co
\eearr
where the factor $3/2\sin^2\theta$ appears because $G$ is expanded
in derivatives of  Legendre polynomials.
Below, we identify [$f_s(0)$, $g(0)$] with [$\bar{f}(4M_\pi^2,s_l)$,
 $\bar{g}(4M_\pi^2,s_l)$], which depend on $s_l$. Furthermore, we compare
the slope $\lambda_f$ with
\be
\bar{\lambda}_f(\spi,s_l)=
\frac{\fff(\spi,s_l)-\fff(4M_\pi^2,sl)}{\fff(4M_\pi^2,s_l)}\frac{4M_\pi^2}
{s_\pi-4M_\pi^2} \co
 \ee
which depends on both $s_\pi$ and $s_l$. We use these dependences to
estimate systematic uncertainties in the determination of the low-energy
couplings. [In future high-statistics experiments, the $s_l$-dependence of the
form factors will presumably  be resolved. It will  be easy to adapt the
procedure to this case.]

We have used MINUIT \cite{minuit} to perform the fits. The results for the
choice $s_l=0,s_\pi=4.4M_\pi^2$ are given in table \ref{fits}.
\begin{table}\centering
\caption{\label{fits}
Results of  fits with  one-loop and unitarized  form factors, respectively. The
errors quoted for the $L^r_i$'s are statistical only. The $L^r_i$ are given in
units of $10^{-3}$ at the scale $\mu=M_\rho$, the scattering lengths $a^I_l$
and the slopes $b_l^I$ in appropriate powers of $M_{\pi^+}$.}
\begin{center}
\vspace{.5cm}
\begin{tabular}{|c|cc|cc|c|}
\hline
         & \multicolumn{2}{c|}{ $K_{e4}$ data
alone}&\multicolumn{2}{c|}{$K_{e4}$  and $ \pi \pi$ data} & experiment\\
         & one-loop        & unitarized & one-loop & unitarized
&\cite{nagels} \\ \hline
$L^r_1$& $0.65 \pm 0.27$&$0.36\pm0.26$  & $0.60\pm0.24$  &$0.37\pm0.23$ & \\
$L^r_2$& $1.63 \pm 0.28$ &$1.35\pm0.27$ &$1.50\pm0.23$    &$1.35\pm0.23$
&\\ $L_3$&$-3.4\pm1.0$   &$-3.4\pm1.0$   &$-3.3\pm0.86$ &$-3.5\pm0.85$  & \\
\hline

$ a_0^0$& $0.20$& $0.20$ &$ 0.20$ & $0.20$       &$0.26\pm0.05$\\

 $b_0^0$ & $0.26$ & $0.25$& $0.26$ & $0.25$&$0.25\pm0.03$\\

$-$10 $a_0^2$ &$0.40$ &$0.41$ &$0.40$ &$0.41$ &$0.28\pm0.12$\\

$-$10 $b_0^2$     &$0.67$ &$0.72$ &$0.68$ &$0.72$ &$0.82\pm0.08$\\

$10 a_1^1$     &$0.36$  &$0.37$  &$0.36$  &$0.37$  &$0.38\pm0.02$\\

$10^2 b_1^1$     &$0.44 $ &$0.47$  &$0.43$  &$0.48$  &\\

$10^2 a_2^0$     &$0.22 $  &$0.18$   &$0.21$   &$0.18$   &$0.17\pm0.03$ \\

$10^3 a_2^2$     &$0.39 $   &$0.21 $   &$0.37 $   &$0.20 $   &$0.13\pm0.3$\\
\hline
$\chi^2/N_\mtiny{DOF}$& $0/0$&$0/0$&$8.8/7$&$4.9/7$&  \\
\hline
\end{tabular}
\end{center}
\end{table}
In the columns denoted by ``one-loop", we have evaluated $\bar{f},\bar{g}$ and
 $\bar{\lambda}_f$ from the one-loop representation given above
\footnote{We
always use for  $L_4^r,\ldots,L_9^r$ the values quoted
in table 1 of Ref. \cite{Intro}.}.
  In the fit with the unitarized form
factor  (columns 3 and 5), we have evaluated $\bar{f}$ from
 Eqs. (\ref{un5}), inserting in the Omn\`{e}s function the
parametrization of the $\pi \pi$ $S$-wave phase shift proposed by Schenk
\cite[solution B]{schenk}. For the form factor $G$,
we have again used the one-loop representation.
 The statistical errors
quoted for the $L_i$'s are the ones generated by the procedure MINOS in MINUIT
and correspond to an increase of $\chi^2$ by one unit.

A few  remarks are in order at this place.
\begin{enumerate}
\item
 It is seen that the
overall description of the $\pi \pi$ scattering data is better using
the unitarized form factors, in particular so for
the $D$-wave scattering lengths.
\item
The errors quoted do not give account of the fact that the simultaneous
determination of the three constants produces a strong correlation between
them.
To illustrate this point we note that, while
the values of the $L_i$'s in
column 4 and 5 are apparently consistent with each other within one error bar,
the $\chi^2$ in column 5 increases from 4.9 to 30.7 if
the $L_i$'s from column 4 are used in the evaluation of $\chi^2$ in
column 5. (For a discussion about the interpretation of the errors see
\cite{minuit}).
\item
The low-energy constants $\bar{l}_1,\bar{l}_2$ which occur in $SU(2)_L\times
SU(2)_R$ analyses may be evaluated from a given set of $L_1^r,L_2^r$ and $L_3$
 \cite{galenp1}. Their value changes in a significant way by
using the unitarized amplitude instead of the one-loop formulae: the
 values for $(\bar{l}_1,\bar{l}_2)$ in column 4 and 5 are
$(-0.5\pm 0.88,6.4\pm 0.44)$ and $(-1.7\pm 0.85, 6.0\pm 0.4)$, respectively.

\item
 $L_1^r,L_2^r$ and $L_3$ are related to  $\pi \pi$
phase shifts through sum rules \cite{truongsum,sternsum}. In principle, one
could
take these constraints into account as well\footnote{We thank B. Moussallam for
pointing this out to us.}. We do not consider them here, because we find it
very difficult to assess a reliable error for the integrals over the
total $\pi\pi$ cross sections which occur in those relations.
\end{enumerate}

The {\it{statistical}} error in the data is only one source of the uncertainty
in the
low-energy constants, which are in addition affected by the ambiguities in the
estimate of the higher-order corrections. These {\it systematic}
uncertainties have several sources:
\begin{enumerate}
\item[i)]
Higher-order corrections to $\ffg$ have not been taken into account.
\item[ii)]
The determination of the contribution from the left-hand cut is not unique.

\item[iii)]
The quantities $\fff$ and $\ffg$ depend on $s_l$, and $\bar{\lambda}_f$ is a
function of both $s_l$ and $s_\pi$.
\item[iv)]
The Omn\`{e}s function depends on the
elastic $\pi\pi$ phase shift and on the cutoff $\Lambda$ used.
\end{enumerate}

We have considered carefully these effects \cite{BCG}, and found that the best
determination of $L^r_1, L^r_2$, and $L_3$ which takes them into
account  is

\bearr
\begin{array}{rrr}
10^3L^r_1(M_\rho)&=& 0.4 \pm 0.3 \co\\
10^3L^r_2(M_\rho)&=&1.35\pm0.3 \co\\
10^3L_3&=&-3.5 \pm 1.1\per
\label{dl3}
\end{array}
\eearr
These values are the ones quoted in table 1 in Ref. \cite{Intro}.
For $ SU(2)_L\times SU(2)_R$ analyses it is useful to know the corresponding
values for the constants $\bar{l}_1$ and $\bar{l}_2$,
\bearr
\begin{array}{rrr}
\bar{l}_1&=&-1.7 \pm 1.0  \co\\
\bar{l}_2&=&6.1 \pm 0.5 \per
\label{dl4}
\end{array}
\eearr

The value and uncertainties in these couplings play a decisive role in a
planned experiment \cite{nemenov} to measure the lifetime of $\pi^+ \pi^-$
atoms,
which will provide a completely independent measurement of the $\pi \pi$
scattering lengths $|a_0^0-a_0^2|$.

One motivation for the analysis in \cite{bijnenskl4,riggen} was to test the
large-$N_C$
prediction $L_2^r=2L_1^r$. The above result shows that a small non-zero
value is preferred.
 To obtain a clean  error analysis, we have repeated the fitting
procedure using the variables
\bearr\label{dl5}
X_1&=&L_2^r-2L_1^r-L_3\co \nn
X_2^r&=&L_2^r\co\nn
X_3&=&(L_2^r-2L_1^r)/L_3  \per
\eearr
We
performed a fit to $K_{e4}$ and $\pi\pi$ data, including the theoretical error
in $G$ as discussed above, and found
\bearr
X_1&=&(4.8 \pm 0.8)\cdot 10^{-3} \co \nn
X_3&=&-0.17^{+0.12}_{-0.22}\per
\label{dl6}
\eearr
The result is that the large-$N_C$ prediction works remarkably well.

\subsection{Predictions}

In this section we make several predictions using the $L^r_i$'s from
table 1 in Ref. \cite{Intro}.
 It is clear that new and more accurate data on
$K_{e4}$ will allow for a better determination of $L^r_1, L^r_2$ and $L_3$,
and may correspondingly modify our predictions. However, unless a
dramatic change in the values of these constants occurs, the
modified predictions will be within the errors that we give.

 Whereas the slope $\lambda_g$  was assumed to coincide
with the slope $\lambda_f$ in the final analysis of the data in Ref.
\cite{ross}, these two quantities may differ in the chiral representation.
Furthermore, our amplitudes allow us to evaluate partial and total decay
rates. In this section, we consider the slope $\lambda_g$ and the total rates.

\begin{table}[t]\centering
\caption{\label{ratapp}
Approximations used to evaluate the total rates
in table \protect\ref{totrat}. Use of
$\bar{f}=\bar{f}_\protect{\mtiny{CHPT}}$,
$\bar{g}=\bar{g}_\protect{\mtiny{CHPT}}$
reproduces the one-loop results in table \protect\ref{totrat} to about $1\%$.}

\begin{center}
\vspace{.5cm}
a) $K^+$ decays

\vspace{.2cm}
\begin{tabular}{|c|c|c|c|}
\hline
& $\pi^0\pi^0 e^+ \nu_e$&$\pi^+\pi^-\mu^+
\nu_\mu$&$\pi^0\pi^0 \mu^+ \nu_\mu$
\\ \hline
&&&\\
$F_1$&$-X\bar{f}$&
$X\bar{f}+\sigma_\pi (PL)\cos\thp \bar{g}$&$-X\bar{f}$
\\
$F_2$&$0$&
$\sigma_\pi (s_\pi s_l)^{1/2}\bar{g}$&$0$
\\
$F_3$&0&0&0
\\
$F_4$&$(PL)\bar{f}$&
$-\left\{(PL)\bar{f}+s_lR_\mtiny{CHPT}+\sigma_\pi X\cos\thp \bar{g}\right\}$&
$\left\{(PL)\bar{f}+s_lR^+_\mtiny{CHPT}\right\}$
\\ \hline
\end{tabular}

\vspace{1cm}
b) $K^0$ decays. Shown are the amplitudes divided by $\sqrt{2}$.

\vspace{.2cm}
\begin{tabular}{|c|c|c|}
\hline
& $\pi^0\pi^-e^+ \nu_e$&$\pi^0\pi^- \mu^+ \nu_\mu$
\\ \hline
&&\\
$F_1$&$XF^-_\mtiny{CHPT}+\sigma_\pi (PL)\cos\thp \bar{g}$&
      $XF^-_\mtiny{CHPT}+\sigma_\pi (PL)\cos\thp \bar{g}$
\\
$F_2$&$\sigma_\pi (s_\pi s_l)^{1/2}\bar{g}$&
      $\sigma_\pi (s_\pi s_l)^{1/2}\bar{g}$
\\
$F_3$&0&0
\\
$F_4$&$-\left\{(PL)F^-_\mtiny{CHPT}+\sigma_\pi X\cos\thp \bar{g}\right\}$&
$      -\left\{(PL)F^-_\mtiny{CHPT} +s_lR^-_\mtiny{CHPT}+\sigma_\pi X\cos\thp
\bar{g}\right\}$ \\ \hline
\end{tabular}
\end{center}
\end{table}

\begin{table}[t]\centering
\protect
\caption{\label{totrat}
Total  decay rates in sec$^{-1}$. To evaluate the rates at one-loop accuracy,
we have used the $L_i^r$'s from table 1 in  Ref. \protect\cite{Intro}.
 The final predictions
are evaluated with the amplitudes shown in table \protect\ref{ratapp}, using
$\bar{f}=5.59(1+0.08q^2),\bar{g}=4.77(1+0.08q^2)$. For the evaluation of the
uncertainties in the
rates see text.}

\begin{center}
\vspace{.5cm}
a) $K^+$ decays

\vspace{.2cm}
\begin{tabular}{|c|cccc|}
\hline
 &$\pi^+\pi^- e^+\nu_e$&$\pi^0\pi^0 e^+ \nu_e$&$\pi^+\pi^-\mu^+
\nu_\mu$&$\pi^0\pi^0 \mu^+ \nu_\mu$
\\ \hline
{\small tree}&$1297$& $683$&$155$&$102$\\

{\small one-loop}&$2447$&$1301$&$288$&$189$\\
\hline
{\small final}&{\small input}&
$1625$&$333$&$225$\\
{\small prediction} & &$\pm90$&$\pm15$&$\pm11$\\ \hline

experiment &$3160$& $1700$&$1130$&  \\
\cite{pdg}&$\pm140$&$\pm320$&$\pm730$&\\
\hline
\end{tabular}

\vspace{1cm}
b) $K^0$ decays

\vspace{.2cm}
\begin{tabular}{|c|cc|}
\hline
 &$\pi^0\pi^-e^+ \nu_e$&$\pi^0\pi^- \mu^+ \nu_\mu$ \\ \hline
{\small tree}& $561$&$55$\\

{\small one-loop}&$953$&$94$\\
\hline
{\small final}&
$917$&$88$\\
{\small prediction} &$\pm170$&$\pm22$\\ \hline

experiment & $998$&  \\
\cite{kld}&$\pm39 \pm 43$&\\
\hline
\end{tabular}
\end{center}
\end{table}

\vspace{1cm}

\noindent
{\bf The slope $\lambda_g$}

\vspace{.5cm}
\noindent
We consider the form factor $\bar{g}$ introduced in  (\ref{proj}) and
determine its slope
 $\lambda_g$
\be
\bar{g}(s_\pi,s_l)=\bar{g}(4M_\pi^2,sl)(1+\lambda_g(s_l) q^2 +O(q^4))
\ee
from the one-loop expression for $G$. The result is
$\lambda_g(0)=0.08$. As the slope is a one-loop effect,
higher-order corrections may affect its value substantially. For this reason,
we have evaluated $\lambda_g$
 also from the modified form factor obtained by using the complete
resonance propagators (and the corresponding $L_i$'s), see Ref. \cite{BCG}.
The change is $\Delta \lambda_g =0.025$. We believe this to be a generous error
estimate and obtain in this manner
\be
\lambda_g(0)=0.08\pm 0.025 \per
\label{eslope}
\ee
The central value indeed agrees  with the slope $\lambda$ in (\ref{e4}).

\vspace{1cm}

\noindent
{\bf Total rates}

\vspace{.5cm}
\noindent
Once the leading partial waves $\bar{f}$ and $\bar{g}$ are known from e.g.
$K^+\rightarrow \pi^+\pi^-e^+\nu_e$ decays, the chiral representation allows
one to predict the remaining rates within rather small uncertainties. We
illustrate the procedure for $K^+\rightarrow \pi^0\pi^0e^+\nu_e$.
According to Eq. (\ref{i2}),
 the relevant amplitude is determined by
$F^+,G^-,R^+$ and $H^-$. The contribution from $H$  is kinematically
strongly suppressed and completely negligible in all total
rates, whereas the contribution from $R$ is
 negligible in the electron modes. Using the chiral representation
of the  amplitudes $F^+$ and $G^-$, we find that the rate is
 reproduced to about 1\%,
if one neglects $G^-$ altogether and uses only the leading partial wave
in
the remaining amplitude, $F_1^+\simeq -X\bar{f}$. From the measured \cite{ross}
form factor $\bar{f}=5.59(1+0.08q^2)$ we then find $\Gamma_{K^+\rightarrow
\pi^0\pi^0e^+\nu_e}=1625$sec$^{-1}$. Finally, we estimate the error from
\bearr
\Delta\Gamma&=& \left\{
[\Gamma(f_s(0)+\Delta f_s,\lambda_f)-
     \Gamma(f_s(0),\lambda_f)]^2 +\right.\nn
&&\left.[\Gamma(f_s(0),\lambda_f+\Delta \lambda_f)-
 \Gamma(f_s(0),\lambda_f)]^2
\right\}^{1/2}=90 {\mbox{sec}}^{-1}\co\nn
\eearr
where $\Delta f_s=0.14,\Delta \lambda_f=0.02$.
The final result for the rate is shown in the row ``final prediction" in
table \ref{totrat}, where we have also listed the tree and the
one-loop result, together
with the experimental data. The evaluation of the remaining rates is done in a
similar manner -- see table \ref{ratapp} for the simplifications used
and table \ref{totrat} for the
corresponding predictions.

 We have assessed an uncertainty due to contributions
from
$F^-_\mtiny{CHPT},R_\mtiny{CHPT}$ in the following manner.
i) We have checked that the results barely change by using the tree level
expression for $R_\mtiny{CHPT}$ instead of its one-loop representation. We
conclude
from this that the uncertainties in $R_\mtiny{CHPT}$ do not matter. ii) The
uncertainty from $F^-_\mtiny{CHPT}$
 is taken into account by adding to $\Delta \Gamma$ in quadrature the
change obtained
by evaluating $F^-_\mtiny{CHPT}$ at $L_3=-3.5+1.1=-2.4$. iii) In $K^0$ decays,
we have also added in quadrature the difference generated by evaluating the
rate with $M_\pi=135$ MeV.

The decay $K^0\rightarrow \pi^0\pi^-e^+\nu_e$  has recently been
measured \cite{kld} with considerably higher statistics than before \cite{pdg}.
We display the result for the rate in the first column of table
\ref{totrat}b.
The quoted errors correspond to the errors in the branching
ratio \cite{kld} and do not include the uncertainty  in the total
decay rate  quoted by the PDG \cite{pdg}.
Notice that the  value for $L_3$ determined in \cite{kld} should be multiplied
with $-1$ \cite{makpriv}.

\subsection{Improvements at DA$\Phi$NE}

      The chiral analysis of $K_{l4}$ decays has been used so far for
      three purposes:

      \begin{enumerate}

      \item  The $K_{e4}$ data from Ref. \cite{ross} allows one to make
      predictions for the slope of the $G$ form factor, for the total rates
in all the channels and for
the       $\pi \pi$ scattering lengths. These are given in Eq.
(\ref{eslope}), in table \ref{totrat} and in table  \ref{fits}, respectively.

      \item  The same $K_{e4}$ data allow one to test the large-$N_C$
      prediction, see Eqs. (\ref{dl5}) and (\ref{dl6}).

      \item The full set of $K_{e4}$ and $\pi \pi$ scattering data
      allows the best determination  of the coefficients $L_1, L_2$
and       $L_3$ in the chiral Lagrangian, see (\ref{dl3}).

      \end{enumerate}

      In the next generation of $K_{l4}$ decay experiments, there is the
      opportunity to improve the phenomenology of $K_{l4}$ (see table
      \ref{kl441}):

      \begin{enumerate}

      \item A very useful innovation would be to analyze the
      experimental data with a modified chiral representation. In
the
latter, the full $S$- and $P$- wave parts of $F_1$ and $F_2$ could be
inserted,
using the chiral representation solely to describe the small background effects
due to higher partial waves $l\ge 2$.
To  be more precise, one would take for $R$ and $H$ the one-loop chiral
representation, whereas for $G$ one writes
\bearr
G&=&g(s_\pi,s_l) e^{{i\delta_p}} + \Delta G\co \nn
\Delta G&=& G_\mtiny{CHPT} -\frac{3}{8\pi}\int d\Omega \sin^2\thp
G_\mtiny{CHPT}\co
\eearr
and similarly for $F$. The unknown amplitudes $g(s_\pi,s_l),f_s(s_\pi,s_l)$
and the phases $\delta_p,\delta_s$ would then be determined from the data.
We have checked that, if the errors in the form factors determined in this
manner can be reduced by e.g. a factor 3
 with respect to the
ones shown in (\ref{e4}), one could pin down particular combinations of
$L_1^r,L_2^r$ and $L_3$ to considerably better precision than was shown above.
This is true independently of an eventual improvement in the theoretical
determination of the higher-order corrections in the form factor $G$ -- which
is a theoretical challenge in any case.

      \item The present experimental uncertainty on $G$ is still too
      large to provide a precise value for the large-$N_C$ parameter
      $(L_2^r - 2 L_1^r)/L_3$. $(K^0 \rightarrow \pi^0 \pi^- e^+
 \nu_e$ decays are mainly sensitive to $G^+_{+-}$ which in turn can be
      used to pin down $L_3$. $K^+ \rightarrow \pi^0 \pi^0 e^+ \nu_e$
      is mainly sensitive to $F^+_{+-}$ which contains $L_1, L_2$ and
      $L_3$.)

      \item The observation of all $K_{l4}$ reactions with high
      statistics could provide a cleaner separation of the various
      isospin amplitudes.

      \item Finally, we come to a most important point. As we
      mentioned already, $K^+ \rightarrow \pi^+ \pi^- e^+ \nu_e$ has
      been used \cite{pipet} to determine the isoscalar $S$-wave
      scattering length with the result $a^0_0 = 0.26 \pm 0.05$. This
      value must be compared with the SU(2)$\times$SU(2) prediction
      \cite{galean,galepl} $a^0_0 = 0.20 \pm 0.01$. Low-energy $\pi \pi$
      scattering is one of the few places where chiral symmetry allows
      one to make a precise prediction within the framework of QCD. In
      their article, Rosselet et al. comment about the discrepancy
      between $a^0_0 = 0.26 \pm 0.05$ and the leading-order result
      \cite{weinpi} $a^0_0 = 0.16$ in the following manner: ``... it
      appears that this prediction can be revised without any
      fundamental change in current algebra or in the partial
      conservation of axial-vector current \cite{bonnier,franklin}."
      Today, we know that this is not the case: The standard picture of the
vacuum structure in QCD \cite{renner} would have to be revised,
 should the central value $a^0_0 = 0.26$ be
      confirmed with a substantially smaller error. For recent work which
supports this scenario see the contribution of Knecht and Stern in this
Handbook \cite{knechtstern}.

      $K_{l4}$ decays are -- at present \cite{nemenov} -- the only
      available source of clean information on $\pi \pi$ $S$-wave
      scattering near threshold. We refer the reader to Ref.
\cite{kl4swave} for a detailed analysis of the issue.

      \end{enumerate}

\newpage
\setcounter{equation}{0}
\setcounter{subsection}{0}
\setcounter{so}{1}
\setcounter{table}{0}
\setcounter{figure}{0}
\clearpage
\section{$K_{e5}$ decays}

In this subsection we discuss
\footnote{The material in this section is  taken from  Ref. \cite{blaser}.}
the decays
\bea
K^+&\rightarrow& \pi^+\pi^-\pi^0e^+\nu_e\nl
K^+&\rightarrow& \pi^0\pi^0\pi^0e^+\nu_e\nl
K^0&\rightarrow& \pi^0\pi^0\pi^-e^+\nu_e\nl
K^0&\rightarrow& \pi^+\pi^-\pi^-e^+\nu_e\fs
\eea
We do not consider isospin violating contributions and correspondingly set
$m_u=m_d$, $\alpha=0$.

\subsection{Matrix elements and decay rates}
The matrix element for  $K\rightarrow \pi\pi\pi e^+\nu_e$ is
\be
T=\frac{G_F}{\sqrt{2}}V_{us}^\star\bar{u}(p_\nu)\gamma_\mu(1-\gamma_5)
v(p_e)
(V^\mu-A^\mu)\co
\ee
where
\be
I_\mu=<\pi(p_1)\pi(p_2)\pi(p_3)\mbox{out}|I_\mu^{4-i5}(0)|K(p)>\;;\;I=V,A.
\ee
The decay rate is calculated from
\bdm
d\Gamma=\frac{1}{2M_K(2\pi)^{11}}\sum_{spins}|T|^2d_{LIPS}(p;p_e,p_\nu,
p_1,p_2,p_3)\fs
\edm
\subsection{Previous experiments}
The Particle Data Group \cite{pdg} quotes the upper bound
\bdm
BR(K^+\rightarrow \pi^0\pi^0\pi^0e^+\nu_e) < 3.5\cdot 10^{-6}.
\edm

\subsection{Theory}
In CHPT, the leading-order contribution is given by the
matrix element of the {\it vector}  current. The corresponding rates are
displayed
in table (\ref{tabkl5}). The smallness of these rates is due to the suppression
of phase space. Indeed, consider the ratio of the four- and five-dimensional
phase space volumes in the neutral pion channel,
\bdm
\frac{M_K^2\int
d_{LISP}(p;p_e,p_\nu,p_1,p_2)/2!}{(2\pi)^{12}}\frac{(2\pi)^{15}}{\int
d_{LISP}(p;p_e,p_\nu,p_1,p_2,p_3)/3!} \simeq 2.3\cdot 10^6\fs
\edm
\begin{table}[t]\centering
\caption{ Rates of $K_{e5}$ decays, evaluated from the leading-order term.
 \label{tabkl5}}
\vspace{.5cm}
\begin{center}
\begin{tabular}{|l|r|}\hline
&branching ratio\\
 \hline
$K^+\rightarrow \pi^+\pi^-\pi^0e^+\nu_e$&$ 3\cdot 10^{-12}$\\
$K^+\rightarrow \pi^0\pi^0\pi^0e^+\nu_e$&$ 2.5\cdot 10^{-12}$\\
$K^0\rightarrow \pi^0\pi^0\pi^-e^+\nu_e$&$12\cdot10^{-12} $\\
$K^0\rightarrow \pi^+\pi^-\pi^-e^+\nu_e$&$33 \cdot 10^{-12}$\\ \hline
\end{tabular}
\end{center}
\end{table}
It agrees well with the ratio of the corresponding
rates at tree level,
 \bdm
\frac{\Gamma(K^+\rightarrow \pi^0\pi^0e^+\nu_e)_{tree}}{\Gamma(
K^+\rightarrow \pi^0\pi^0\pi^0e^+\nu_e)_{tree}}\simeq 3.4\cdot
10^6\fs
 \edm
(The corresponding ratios for $K^+\rightarrow \pi^0e^+\nu_e/K^+\rightarrow
\pi^0\pi^0e^+\nu_e$ are $1.4\cdot 10^4$ and $0.53\cdot 10^4$ for phase space
volumes and decay rates, respectively.)
The contributions at order $p^4$ are due to i) the corrections to the matrix
element of the vector current,  and to ii)
 the matrix element of the {\it{axial}} current.
The latter stems from  the Wess-Zumino-Witten Lagrangian ${\cal L}_{WZW}$. Both
the
local and nonlocal term in the anomalous action contribute. The nonlocal part
is suppressed by the factor $m_e$ in the matrix element (in addition to the
phase space suppression just mentioned).

 Based on our experience  with
$K_{e4}$ decays, we expect the terms of order $p^4$ to enhance the above rates
by roughly a factor of two to three.

\subsection{Improvements at DA$\Phi$NE}
According to the standard model, $K_{e5}$ decays are invisible at DA$\Phi$NE,
but the existing upper limits can be improved significantly.

 \vspace{2cm}
{\bf{Acknowledgements}}

We thank G. Pancheri for the perfect organization of the DA$\Phi$NE Workshops
and the INFN for the hospitality in Frascati.
We have enjoyed numerous interesting discussions
with the members of the working groups.
 We thank J. Beringer for
checking traces in $K_{l2\gamma}$, M. Candusso for doing the contour plots
and C. Riggenbach for providing us with numerical routines for $K_{l4}$.
S. Blaser kindly allowed us to include his work \cite{blaser} on $K_{e5}$
decays.
 \appendix

\newcounter{zahler}
\renewcommand{\thesection}{\Alph{zahler}}
\renewcommand{\theequation}{\Alph{zahler}.\arabic{equation}}
\setcounter{zahler}{0}

\newpage
\setcounter{equation}{0}
\addtocounter{zahler}{1}

\section{Notation}
\label{notation}

The notation for phase space is the one without the factors of $2\pi$.
For the decay rate of a
particle with four momentum $p$ into $n$ particles
with momenta $p_1,\ldots,p_n$ this is
\begin{equation}
d_{LIPS}(p;p_1,\ldots,p_n) = \delta^4(p-\sum_{i=1}^n p_i )
\prod_{i=1}^n \frac{d^3 p_i}{2 p_i^0} ~.
\end{equation}
We use a covariant normalization of one-particle states,
\be
<{\vec{p}} \; '|\vec{p}> = (2\pi)^3 2p^0 \delta^3({\vec{p}} \; '-\vec{p})
 \;
\; , \ee
together with the spinor normalization
\be
\bar{u}(p,r)u(p,s) = 2m\delta_{rs} \; \; .
\ee
The kinematical function $\lambda(x,y,z)$ is defined as
\begin{equation}
\lambda(x,y,z) = x^2 + y^2 + z^2 - 2 (xy + yz + zx)~.
\end{equation}
We take the standard model in the current $\times$ current form, i.e.,
we neglect the momentum dependence of the $W$-propagator. The currents
used in the text are :
\begin{eqnarray}
V_\mu^{4-i5} &=& \bar{q} \gamma_\mu \frac{1}{2}(\lambda_4 - i\lambda_5)q
  ~=~ \overline{s}\gamma_\mu u
\nonumber\\
A_\mu^{4-i5} &=& \bar{q} \gamma_\mu \gamma_5 \frac{1}{2}
(\lambda_4 - i\lambda_5) q
  ~=~ \overline{s}\gamma_\mu \gamma_5 u
\nonumber\\
V^{em}_\mu &=& \bar{q}\gamma_\mu Q q
\nonumber\\
Q&=&\mbox{diag}(2/3, -1/3 ,-1/3)~.
\end{eqnarray}
The numerical values used in the programs are the physical masses
for the particles as given by the Particle Data Group \cite{pdg}.
In addition we have used the values for the decay constants derived
from the most recent measured charged pion and kaon
semileptonic decay rates\cite{pdg,lroos} :
\begin{eqnarray}
F_\pi &=& 93.2 ~MeV\nonumber\\
F_K &=& 113.6~MeV.
\end{eqnarray}
We do not need values for the quark masses. For the processes considered
in this report we can always use the lowest order relations to rewrite
them in   terms of the
pseudoscalar meson masses (see Ref. \cite{Intro}). For the
KM matrix element
we used the central value $|V_{us}|=0.220$ of Ref. \cite{pdg}.
The numerical values for the $L_i^r(M_\rho)$ are those
given in table 1 in  Ref. \cite{Intro}.

The number of events quoted for DA$\Phi$NE are based on a
luminosity of $5\cdot 10^{32}~ cm^{-2}s^{-1}$,
which is equivalent \cite{franzini}
to an annual rate of $9\cdot 10^9$ $(1.1\cdot 10^9)$ tagged
$K^{\pm}$ $(K_L)$ (1 year = $10^7~ s$ assumed).

Whenever we quote a branching ratio for a semileptonic
$K^0$ decay, it stands for the branching ratio of the corresponding
$K_L$ decay, e.g.,
\begin{equation}
BR(K^0 \to \pi^- l^+ \nu ) \equiv BR(K_L \to \pi^{\pm} l^{\mp} \nu )~.
\end{equation}
We use the Condon-Shortley phase conventions throughout.

\newpage
\setcounter{equation}{0}
\addtocounter{zahler}{1}
\section{Loop integrals}
\label{loop}

In this appendix we define the functions appearing in the loop integrals
used in the text.
First we define the functions needed for loops with two propagators,
mainly in the form given in Ref. \cite{galenp1}.
We consider a loop with two masses, $M$ and $m$.
 All needed functions can be given
in terms of the subtracted scalar integral $\bar{J}(t) = J(t) - J(0)$,
\begin{equation}
J(t) = ~  -i
\int \frac{d^dp}{(2\pi)^d} \frac{1}{((p+k)^2 - M^2)(p^2 - m^2)}
\end{equation}
with $t = k^2$.
The functions used in the text are then :
\begin{eqnarray}
\bar{J}(t)&=&-\frac{1}{16\pi^2}\int_0^1 dx~
\log\frac{M^2 - t x(1-x) - \Delta x}{M^2 - \Delta x}
\nonumber\\&=&
\frac{1}{32\pi^2}\left\{
2 + \frac{\Delta}{t}\log\frac{m^2}{M^2} -\frac{\Sigma}{\Delta}
\log\frac{m^2}{M^2} - \frac{\sqrt{\lambda}}{t}
\log\frac{(t+\sqrt{\lambda})^2 -
\Delta^2}{(t-\sqrt{\lambda})^2-\Delta^2}\right\}
{}~,
\nonumber\\
J^r(t) &=& \bar{J}(t) - 2k~,
\nonumber\\
M^r(t) &=& \frac{1}{12t}\left\{ t - 2 \Sigma \right\} \bar{J}(t)
+ \frac{\Delta^2}{3 t^2} \bar{J}(t)
+ \frac{1}{288\pi^2} -\frac{k}{6}
\nonumber\\&&
                               - \frac{1}{96\pi^2 t} \left\{
      \Sigma + 2 \frac{M^2 m^2}{\Delta}
     \log\frac{m^2}{M^2} \right\} ~,
\nonumber\\
L(t)&=& \frac{\Delta^2}{4t} \bar{J}(t)~,
\nonumber\\
K(t)&=&\frac{\Delta}{2t}\bar{J}(t)    ~,
\nonumber\\
H(t)&=&\frac{2}{3} \frac{L_9^r}{F^2} t + \frac{1}{F^2}[t M^r(t) - L(t)],
\nonumber\\
\Delta &=& M^2 - m^2~,
\nonumber\\
\Sigma &=& M^2 + m^2  ~,
\nonumber\\
\lambda&=&\lambda(t,M^2,m^2) ~=~ (t+\Delta)^2 - 4tM^2  ~.
\end{eqnarray}
In the text these are used with subscripts,
\begin{equation}
\bar{J}_{ij}(t)  =  \bar{J}(t)~~~\mbox{with}~~~M = M_i , m = M_j~
\end{equation}
and similarly for the other symbols.
The subtraction point
dependent part
is contained in the constant $k$
\begin{equation}
k = \frac{1}{32\pi^2} \frac{M^2 \log \left( \frac{M^2}{\mu^2} \right)
                       - m^2 \log\left(\frac{m^2}{\mu^2}\right)}
   {M^2 - m^2},
\end{equation}
where $\mu$ is the subtraction scale.

In addition, in subsection \ref{section4} these functions and symbols
appear in a  summation
over loops $I$
with
\begin{eqnarray}
J_I(t) &=& \bar{J}(t) ~~~\mbox{with}~~~M = M_I , m = m_I       ~;
\nonumber\\
\Sigma_I&=&M_I^2 + m_I^2
\end{eqnarray}
and again similarly for the others.
There the combination $B_2$ appears  as well :
\begin{eqnarray}
B_2(t,M^2,m^2)&=&B_2 (t,m^2,M^2)
\nl
&=&\frac{1}{288\pi^2} \left( 3\Sigma - t\right)
-\frac{\lambda(t,M^2,m^2) \bar{J}(t)}{12t}
+\frac{t\Sigma - 8M^2 m^2}{384\pi^2 \Delta} \log\frac{M^2}{m^2} ~.\nonumber
\end{eqnarray}

The last formula to be defined is the three propagator loop integral
function $C(t_1,t_2,M^2,$$ m^2)$ where one of the three external momenta
has zero mass and two of the propagators have the same mass $M$.
Here $t_1 = ( q_1 + q_2)^2$, $t_2 = q^2_2$ and $q_1^2 = 0$.
\begin{eqnarray}
C(t_1,t_2,M^2,m^2)&=& - i \int \frac{d^4p}{(2\pi)^d}
\frac{1}{(p^2 - M^2 ) ((p+q_1)^2 - M^2) ((p+q_1+q_2)^2 - m^2)}
\nonumber\\ &=&
-\frac{1}{16\pi^2}\int_0^1 dx \int_0^{1-x} dy
\frac{1}{M^2 - y ( \Delta + t_1 ) +  xy (t_1 - t_2) + y^2 t_1}
\nonumber\\
&=& \frac{1}{(4\pi)^2 (t_1 - t_2)}
\left\{ Li_2 \left( \frac{1}{y_+(t_2)}\right) +
  Li_2\left(\frac{1}{y_- (t_2)}\right)\right.
\nonumber\\& &  \left.
  - Li_2\left(\frac{1}{y_+(t_1)}\right)
  - Li_2\left(\frac{1}{y_-(t_2)}\right)\right\}~,
\nonumber\\
y_{\pm}(t)&=&\frac{1}{2t}\left\{ t + \Delta \pm
\sqrt{\lambda(t,M^2,m^2)}\right\}
\end{eqnarray}
where $Li_2$ is the dilogarithm
\begin{equation}
Li_2(x) = - \int_0^1    \frac{dy}{y}\log(1-xy)   ~.
\end{equation}

\newpage
\setcounter{equation}{0}
\addtocounter{zahler}{1}

\section{Decomposition of the hadronic tensors $I^{\mu \nu}$}
\label{kl2g}
Here we consider the tensors
\be
I^{\mu \nu} =
\int dx e^{iqx+iWy} < 0 \mid T V^\mu_{em} (x) I^\nu_{4-i5}(y) \mid K^+(p)>
\; \;, \; \; I=V,A \; \;
\ee
and detail their connection with the matrix element (\ref{k3}).

The general decomposition of $A^{\mu\nu}, V^{\mu\nu}$
 in terms of Lorentz invariant amplitudes reads \cite{BARDIN,beg} for $q^2
\neq 0$
\bearr
\frac{1}{\sqrt{2}} A^{\mu\nu} &=& - F_K \left \{ \frac{(2W^\mu + q^\mu)
W^\nu}{M_K^2 - W^2} + g^{\mu\nu} \right \}
\nonumber \\
&+& A_1 (q W g^{\mu\nu} - W^\mu q^\nu) + A_2 (q^2 g^{\mu\nu} - q^\mu
q^\nu)
\nonumber \\
&+& \left \{ \frac{2 F_K (F_V^K(q^2)-1)}{(M_K^2-W^2)q^2}
+ A_3 \right\}
 (qWq^\mu - q^2 W^\mu) W^\nu
\label{A6}
\eearr
and
\be
\frac{1}{\sqrt{2}} V^{\mu \nu} = iV_1 \epsilon^{\mu \nu \alpha \beta}
q_\alpha p_\beta
\label{A7}
\ee
where the form factors $A_i(q^2,W^2)$ and $V_1(q^2,W^2)$ are analytic functions
of $q^2$ and $W^2$. $F^K_V (q^2)$ denotes the electromagnetic form factor of
the kaon $(F_V^K (0) = 1)$. $A^{\mu\nu}$ satisfies the Ward identity
\be
q_\mu A^{\mu \nu} = - \sqrt{2} F_K p^\nu.
\label{A8}
\ee

In the process (\ref{k1}) the photon is real. As a consequence of this, only
the two form factors $A_1(0,W^2)$ and $ V_1(0,W^2)$ contribute. We set
\bearr
A(W^2)& = &A_1(0,W^2)
\nonumber \\
V(W^2) &=& V_1 (0, W^2)
\label{A10}
\eearr
and obtain for the matrix element (\ref{k3})
\be
T= -iG_F/\sqrt{2} e {V_{us}}^\star \epsilon^\star_\mu \left \{ \sqrt{2} F_K
l_1^\mu - (V^{\mu\nu} - A^{\mu\nu}) l_\nu \right \}_{ \mid_{q^2=0}} \; \;
, \label{A2}
\ee
with
\bearr
l^{\mu} & =& \bar{u} (p_\nu)\gamma^\mu  (1-\gamma_5) v(p_l)
\nonumber \\
l_1^\mu &=& l^\mu + m_l \bar{u} (p_\nu) (1+\gamma_5) \frac{2p_l^\mu + \not
\!{q} \gamma^\mu}{m_l^2 - (p_l + q)^2} v(p_l) \; \; .
\label{A3}
\eearr
Grouping terms into an IB and a SD piece gives  (\ref{k3},\ref{k4}).
As a consequence of (\ref{A8}), $T$  is
 invariant
under the gauge transformation $\epsilon_\mu \rightarrow \epsilon_\mu +
q_\mu$.

The amplitudes $A_1, A_2$ and $V_1$ are related to the
corresponding quantities $F_A,R$ and $F_V$ used by the PDG \cite{pdg} by
\be
- \sqrt{2} M_K (A_1, A_2, V_1) = (F_A, R, F_V).
\label{A9}
\ee
The last term in (\ref{A6}) is omitted in \cite{pdg}. It
contributes to
processes with a virtual photon, $K^\pm \rightarrow l^\pm \nu_l l'^+ l'^-$.

Finally, the relation to  the notation used in
\cite{ke22,km21} is
\bearr
2 (A \pm V)^2 &=& (a_k \pm v_k)^2 \; \; \; \cite{ke22} \nonumber \\
\sqrt{2} (A,V)& =& (F_A,F_V) \; \; \; \cite{km21} \; \; .
\label{A11}
\eearr

\newpage
\setcounter{equation}{0}
\addtocounter{zahler}{1}

\section{One-loop corrections to $K_{l4}$ form factors}
\label{UFGR}
In this appendix we give the expression of the unitarity corrections
to the form factors $F$, $G$, and $R$
(\cite{bijnenskl4},\cite{riggen},\cite{BCG}).
      \bearr
     U_F (s_\pi, t, u) & = &   \Delta_0 (s_\pi) +
      A_F(t) + B(t,u) \co
      \label{T3}
      \eearr
      with
      \bearr
      \Delta_0(s_\pi) & = & \frac{1}{2} (2 s_\pi - M^2_\pi) J^r_{\pi
      \pi} (s_\pi) + \frac{3s_\pi}{4} J^r_{KK} (s_\pi) +
      \frac{M^2_\pi}{2} J^r_{\eta \eta} (s_\pi) \co
      \nonumber \\
      A_F (t) & = & \frac{1}{16} \left[(14 M^2_K + 14 M^2_\pi - 19t)
      J^r_{K\pi} (t) + (2M^2_K + 2 M^2_\pi - 3t)
      J^r_{\eta K} (t) \right]
      \nonumber \\
      & + & \frac{1}{8} \left[ (3M^2_K - 7 M^2_\pi + 5 t) K_{K \pi}
      (t) + (M^2_K - 5 M^2_\pi + 3t) K_{\eta K} (t) \right]
      \nonumber \\
      & - & \frac{1}{4} \left[ 9 (L_{K\pi} (t) + L_{\eta K} (t)) +
      (3M^2_K - 3 M^2_\pi - 9 t) (M^r_{K\pi} (t) + M^r_{\eta K}
      (t))\right] \co
      \nonumber \\
      B(t,u) & = &  - \frac{1}{2} (M^2_K + M^2_\pi -t)
      J^r_{K\pi} (t) - ( t \leftrightarrow u).
      \label{T4}
      \eearr
      The loop integrals $J^r_{\pi \pi} (s_\pi), \ldots$ which occur
      in these expressions are listed in appendix \ref{loop}. The
      functions
      $J^r_{PQ}$ and $M^r_{PQ}$ depend on the scale $\mu$ at which the
      loops are renormalized. The scale drops out in the expression
      for the full amplitude.

      The imaginary part of  $F_\pi^{-2} \Delta_0 (s_\pi)$ contains
      the $I = 0$, $S$-wave $\pi \pi$ phase shift
      \be
      \delta^0_0 (s_\pi) = (32 \pi F^2_\pi)^{-1} (2s_\pi - M^2_\pi) \sigma_\pi
     + O(E^4) \co
      \label{T5}
      \ee
      as well as contributions from $K \bar{K}$ and $\eta \eta$
      intermediate states. The functions $A_F(t)$ and $B(t,u)$ are
      real in the physical region.

       \bearr
      U_G (s_\pi, t, u) & = &  \Delta_1 (s_\pi) +
      A_G(t) + B(t,u) \co
      \label{T10}
      \eearr
      with
      \bearr
      \Delta_1 (s_\pi) & = & 2 s_\pi \left\{ M^r_{\pi \pi} (s_\pi)+
      \frac{1}{2} M^r_{KK}(s_\pi) \right\} \co
      \nonumber \\
      A_G(t) & = & \frac{1}{16} \left[ (2M^2_K + 2M^2_\pi + 3t)
      J^r_{K\pi}(t) - (2M^2_K + 2M^2_\pi -3 t) J^r_{\eta
      K}(t) \right]
      \nonumber \\
      & + & \frac{1}{8} \left[ (-3 M^2_K + 7 M^2_\pi - 5 t) K_{K \pi}
      (t) + (-M^2_K + 5 M^2_\pi - 3 t) K_{\eta K} (t) \right]
      \nonumber \\
      & - & \frac{3}{4} \left[ L_{K\pi}(t) + L_{\eta K} (t) - (M^2_K -
      M^2_\pi + t) (M^r_{K\pi} (t) + M^r_{\eta K} (t)) \right]
      \nonumber \per \\
      \label{T11}
      \eearr
      The imaginary part of $F_\pi^{-2} \Delta_1(s_\pi)$ contains the
      $I=1$, $P$-wave phase shift
      \be
      \delta^1_1 (s_\pi) = (96 \pi F_\pi^2)^{-1} s_\pi
      \sigma_\pi^3 +  O(E^4) \per
      \label{T12}
      \ee
      as well as contributions from $K\bar{K}$ intermediate states.
      The function $A_G$ is real in the physical region.

The unitarity corrections $U_Z,U_Q$ in the form factor $R$ in
(\ref{formr}) are

\bearr
U_Z &=& s_\pi \Delta_0(s_\pi) + \nu \Delta_1(s_\pi) -\frac{4}{9}M_K^2 M_\pi^2
J^r_{\eta \eta}(s_\pi)
\nonumber \\
&+& \frac{1}{32}\left[ 11(s_\pi-\nu)^2 -20\Sigma (s_\pi -\nu) +12
{\Sigma}^2\right] J^r_{K\pi}(t)
\nonumber \\
&+& \frac{1}{96} \left[ 3(s_\pi -\nu) -2\Sigma\right]^2 J^r_{\eta K}(t)
\nonumber \\
&+& \frac{1}{4}(s_\pi +\nu)^2 J^r_{K \pi}(u)
\nonumber \\
&+& \frac{1}{4}(M_K^2 -M_\pi^2)\left[5(s_\pi -\nu) -6\Sigma\right] K_{K\pi}(t)
\nonumber \\
&+& \frac{1}{4}(M_K^2 -M_\pi^2)\left[3(s_\pi -\nu) -2\Sigma\right] K_{\eta
 K}(t)
\nonumber \\
&+& \frac{3}{8} \left[2s_\pi (\nu +4\Sigma) -3 s_\pi^2 + \nu^2 -16 M_\pi^2
M_K^2\right] \left[M^r_{K \pi}(t) + M^r_{\eta K} (t)\right]
\nonumber \\
&-& \frac{3}{4} (3 s_\pi +\nu -2 \Sigma)(L_{\eta K}(t) +L_{K \pi}(t)) \co
\eearr

\bearr
U_Q&=& \Delta_0(s_\pi) +\frac{M_K^2-s_l}{32}\left\{11 J^r_{K \pi}(t) +8J^r_{K
\pi}(u) +3 J^r_{\eta K}(t)\right\}
\nonumber \\
&-& \frac{1}{8}(5(s_\pi -\nu) +5(M_K^2-s_l)-6\Sigma)K_{K\pi}(t)
\nonumber \\
&-& \frac{1}{8}(3(s_\pi -\nu) +3(M_K^2-s_l)-2\Sigma)K_{\eta K}(t)
\nonumber \\
&-&\frac{9}{4}(L_{\eta K}(t) + L_{K \pi}(t))
\nonumber \\
&+&\frac{3}{8}(4(\nu +2M_\pi^2) -3(M_K^2 -s_l))(M^r_{K \pi}(t) + M^r_{\eta
K}(t)) \co
\eearr
with
\bdm
\Sigma=M_K^2+M_\pi^2\fs
\edm

\newpage
\addcontentsline{toc}{section}{\hspace{1cm}Bibliography}

\end{document}